\documentclass[useAMS,usenatbib]{mn2e}
\usepackage{times}
\usepackage{natbib}
\usepackage{aas_macros}
\usepackage{amsmath}
\usepackage{amssymb}
\usepackage[pdftex]{graphicx}
\usepackage{listings}

\title[]{Magnetic Field Evolution in Giant Radio Relics using the example of CIZA J2242.8+5301}
\author[J. M. F. Donnert]{
	J. M. F. Donnert$^{1,2,3}$\thanks{ERC Marie Curie fellow, donnert@ira.inaf.it}, A. Stroe$^{4,1}\thanks{ESO fellow}$,  G. Brunetti$^{2}$, D. Hoang$^{1}$, H. Roettgering$^{1}$  \\
$1$ Leiden Observatory, Leiden University, P.O. Box 9513, NL-2300 RA Leiden, The Netherlands\\
$2$ INAF-Istituto di Radioastronomia, via P. Gobetti 101, I-40129 Bologna, Italy\\
$3$ School of Physics and Astronomy, University of Minnesota, Minneapolis, MN 55455, USA\\
$4$ European Southern Observatory, Karl-Schwarzschild-Str. 2, 85748, Garching, Germany}

\begin{document}

\date{Accepted ???. Received ???; in original form ???}

\pagerange{\pageref{firstpage}--\pageref{lastpage}} \pubyear{2015}

\maketitle 

\label{firstpage}

\begin{abstract}
	Giant radio relics are the arc-shaped diffuse radio emission regions observed in the outskirts of some merging galaxy clusters. They are believed to trace shock-waves in the intra-cluster medium. Recent observations  demonstrated that some prominent radio relics exhibit a steepening above 2 GHz in their radio spectrum. This challenges standard theoretical models because shock acceleration is expected to accelerate electrons to very high energies with a power-law distribution in momentum. In this work we attempt to reconcile these data with the shock-acceleration scenario. We propose that the spectral steepening may be  caused by the highest energy electrons emitting preferentially in lower magnetic fields than the bulk of synchrotron bright electrons in relics. Here, we focus on a model with an increasing magnetic field behind the shock front, which quickly saturates and then declines. We derive the time-evolution of cosmic-ray electron spectra in time variable magnetic fields and an expanding medium. We then apply the formalism on the large radio relic in the cluster CIZA J2242.8+5301 (the Sausage relic). We show that under favourable circumstances of magnetic field amplification downstream, our model can explain the observed radio spectrum, the brightness profile and the spectral index profile of the relic. 
A possible interpretation for the required amplification of the magnetic field downstream is a dynamo acting behind the shock with an injection scale of magnetic turbulence of about 10 kpc. Our models require injection efficiencies of CRe - which are in tension with simple diffusive shock acceleration from the thermal pool. We show that this problem can likely be alleviated considering pre-existing CRe.
\end{abstract}


\section{Introduction}\label{sect.intro}

Galaxy clusters are the largest bound structures in the cosmological matter distribution, with masses of a up to few $10^{15} \,\mathrm{M}_\mathrm{sol}$ \citep{2012ARA&A..50..353K}. In their massive Dark Matter dominated gravitational potential, baryons accumulate to form the intra-cluster-medium (ICM).  The ICM is a hot, thin plasma and makes clusters bright X-ray sources \citep{1971Natur.231..107M,1988xrec.book.....S}. Observations of these sources reveal a medium behaving collisional on small scales, with complex features such as steep density gradients, cold fronts and shocks  \citep{2007PhR...443....1M}. The features are probably caused by cluster mergers, which dissipate potential energy in the ICM and drive shocks and turbulence in the otherwise nearly isothermal cluster atmosphere. \par
 
The intra-cluster-medium also hosts non-thermal components: cosmic-ray electrons (CRe) and magnetic fields at the $\mu\mathrm{G}$ level.  In some clusters, this is prominently evidenced by diffuse steep spectrum radio synchrotron sources \citep{2012A&ARv..20...54F,2014IJMPD..2330007B}. These radio sources are commonly classified as giant radio halos and giant radio relics: \emph{radio halos} are characterised as diffuse unpolarised Mpc-sized emission centered on the X-ray emission from the ICM. In contrast, \emph{radio relics} are strongly polarised elongated radio sources at the outskirts of clusters. Their formation mechanism is believed to be very different: radio halos are probably associated with merger-driven turbulence and the subsequent re-acceleration of a cluster-wide radio dark cosmic-ray electron population \citep{2014IJMPD..2330007B}. On the other hand, radio relics are probably connected to merger-driven low Mach number shocks and the subsequent acceleration of CRe by the shock \citep{2012SSRv..166..187B,2014IJMPD..2330007B}.  \par

Giant radio relics have been a subject of intense study. There are about 40 radio relics known to date, all connected to merging clusters \citep{2012A&ARv..20...54F}. Some of them occur in pairs as double relics \citep{2012MNRAS.426...40B,2015MNRAS.453.3483D}, some are connected to radio haloes \citep{2015MNRAS.449.1486S,2012A&A...546A.124V,2005ApJ...627..733M,2016arXiv160106029V}, many are connected to radio galaxies and show an irregular, complex morphology \citep[e.g.][]{2013ApJ...769..101V}.  \par

Theoretical models for radio relics assume particle acceleration at the shock to inject cosmic-ray electrons (CRe) into the ICM, which subsequently lose energy through inverse Compton and synchrotron losses and emit synchrotron radiation \citep{1998A&A...332..395E}. A similar model, based on diffusive shock acceleration (DSA) from the thermal pool, successfully explains the non-thermal emission in supernova remnant (SNR) shocks, however at much higher Mach numbers of about 100 \citep{1949PhRv...75.1169F, 1987PhR...154....1B}. Observations of SNR show that high-Mach number shocks in these environment accelerates many more protons than electrons \citep{2012A&A...538A..81M,2015ApJ...798L..28C}. The energy density of accelerated CRp commonly exceeds the CRe energy density by a factor of ten to 100. \par
 
In clusters however, this picture is challenged by the large efficiencies required to reproduce the total radio synchrotron brightness of many relics \citep[see ][ for a review]{2014IJMPD..2330007B}. In cluster relics this scenario would predict that CRp induced $\gamma$-ray emission exceeds the tight upper limits set by the FERMI satellite \citep{2014MNRAS.437.2291V,2014ApJ...787...18A}.  \par
To alleviate the large requirements for acceleration efficiencies of CRe, recent models for cluster relics assume shock re-acceleration of a pre-existing population of CRe \citep{2012ApJ...756...97K,2013ApJ...764...95K,2015MNRAS.451.2198V,2013MNRAS.435.1061P,2014ApJ...788..142K,2005ApJ...627..733M}. However, the absence of CRp-induced $\gamma$-ray emission in clusters remains puzzling \citep{2014IJMPD..2330007B}. CRp are expected to be subject to shock re-acceleration as well and hence should be $\gamma$-ray bright. Relic models commonly employ comparatively large magnetic fields in the emission region, usually from $1-7\,\mu\mathrm{G}$. Lower limits to the magnetic field strength were found from IC limits of up to $4\,\mu\mathrm{G}$ in some clusters \citep{2010ApJ...715.1143F}. Observations now also suggest a filamentary structure of the post-shock field on scales above 100 kpc \citep{2016arXiv160106029V,2014ApJ...794...24O}. Furthermore, the polarization of many relics has lead to the view that the required magnetic field amplification originates from shock compression. However, magnetic field amplification has not been studied in great detail \citep[see however][]{2012MNRAS.423.2781I}, and downstream magnetic field decay has never been discussed in the literature.  \par

The prototype of radio relics is probably the northern relic in CIZA J2242.8+5301 \citep{2010Sci...330..347V}. Its large extension of 2 Mpc, very narrow brightness profile with an FWHM of 55 kpc at 610 MHz and homogeneous brightness distribution have coined the term ''Sausage relic''. The data provide direct evidence for a large shock and subsequent CRe cooling at the outskirts of the cluster. Its simplicity has made the cluster a preferred target for the study of shock acceleration in the ICM. The cluster hosts another morphologically irregular counter relic in the south that is possibly connected to a few radio galaxies \citep{2013A&A...555A.110S}.     \par

Recently, \citet{2014MNRAS.441L..41S,2016MNRAS.455.2402S} discovered spectral steepening beyond 2 GHz in the radio spectrum of the northern relic and the ''Toothbrush'' relic, the same was also found in A2256 by \citet{2015A&A...575A..45T}. This new constraint appears to \emph{fundamentally contradict} the standard picture for the formation of radio relics, which predicts a power-law up to very high frequencies \citep{1998A&A...332..395E, 2015ApJ...809..186K}. In case of the northern relic in CIZA J2242.8+5301, the observed emission at 30 GHz is more than a factor 15 below the simple power-law model used to fit the low frequencies \citep{2010Sci...330..347V}.   \citet{2015ApJ...809..186K} conclude that the steepening is unlikely caused by the re-acceleration of a pre-existing CRe population. \citet{2015arXiv151103245B} suggest that the steepening can be attributed to the SZ decrement, however only for steep radio spectra corresponding to Mach numbers below three. \citet{2016arXiv160203278K} considered a scenario where a shock has crossed a cloud of finite extend and the relic emission originates from aged CR electrons. They conclude that this could lead to the observed steepening in the spectrum. In any case, our understanding of radio relics has been shaken by the Sausage relic alone: Neither exists a satisfactory model for its radio spectrum, nor do we understand the mechanism leading to the high magnetic field value required in current models. \par

In this article, we attempt to solve this discrepancy by exploring the possibility that the steepening in the spectrum of relics is indicative of higher energy CRe emitting in a lower magnetic field, hence significantly reducing the total emission at high frequencies. This could be caused by a number of physical situations, among them slow magnetic field amplification directly behind the shock or diffusion of preferentially high energy CRe into regions with low magnetic field.  \par
 
We focus this work on effects from time-dependent magnetic fields and adiabatic expansion of the thermal gas behind the shock. We neglect other effects like CR diffusion, reconnection and the possibly pre-existing CRe distribution \citep{2015ApJ...809..186K}, for simplicity. We show that under favourable conditions for the evolution of the magnetic field downstream, it is possible to model the spectral properties and brightness distribution observed in this relic. Magnetic fields and their amplification in or behind collisionless shocks in the outskirts of clusters are not well understood or constrained, in fact the problem has been largely ignored in the literature \citep[see however][]{2012MNRAS.423.2781I}. Hence an empirical model can yield constrains on the magnetic field evolution at shocks at the cluster outskirts. As motivated above, the simple structure of the northern radio relic in CIZA J2242.8+5301, its potentially high Mach number and the excellent available radio data make it the ideal target for this investigation.\par
 
This work will require to calculate the synchrotron brightness of CRe cooling in a time-variable magnetic field and adiabatic expansion. We develop this formalism in the next section.  In section \ref{sect:model}, we review the recent state-of-the-art observations of the cluster to set reasonable parameters for the shock. We also describe four magnetic field models in the downstream region, which we chose \emph{empirically to fit the data} (we explore the parameter space of these models in appendix \ref{app.parameters}). We compare the model with observations in section \ref{sect:results}.  The discussion and interpretation of the results is put forward in section \ref{sect:discussion}. Our summary is reported in section \ref{sect.conlusions}.  
 
Throughout the paper, we assume a standard $\Lambda$CDM concordance cosmology, with $H_0 = 70\,\mathrm{km}\, \mathrm{s}^{-1} \mathrm{Mpc}^{-1}$, $\Omega_\mathrm{M} = 0.3$ and $\Omega_\Lambda = 0.7$. Hence, at the redshift of the cluster, $1 \,\mathrm{arcmin} = 192\,\mathrm{kpc}$, this means a luminosity distance of $d_\mathrm{lum} \approx 930 \,\mathrm{Mpc}$.
 
\section{Scenario \& Formalism} \label{sect:formalism}

Here we explore an extension of the standard theoretical scenario for radio relics evaluating the changes in the observed radio spectrum induced by the evolution of the magnetic field in the downstream region. The goal is to scrutinize whether these effects can explain the steepening observed at high frequencies in the integrated spectrum of some radio relics. \\
We assume the following steps :
\begin{enumerate}
	\item Cosmic-ray electrons are accelerated or re-accelerated at the shock via DSA with a power-law spectrum. We assume the usual dependency of acceleration efficiency with Mach number \citep{1987PhR...154....1B}.
	\item Cosmic-ray electrons evolve and cool downstream, also due to adiabatic expansion.
	\item The magnetic field is gradually amplified downstream and then declines with increasing distance from the shock. Amplification by compression in the shock is relevant only so far, as it sets the initial field in our model: $B_\mathrm{min}$.
\end{enumerate}

\subsection{Adiabatic Expansion behind an ICM Shock}\label{sect.adexp}

Behind a shock in the ICM, the thermal plasma with number density $n_\mathrm{th}$ is going to settle into the gravitational potential of the cluster atmosphere, subsequently expanding. As a simple first model for this process, we assume that this expansion is only one-dimensional along the shock normal with velocity $v_\mathrm{exp}$ away from the shock on a time scale of $t_\mathrm{exp}$, which in turn increases the downstream velocity ($v_\mathrm{dw}$) as a function of time and distance from the shock ($r(t)$): 
\begin{align}
	v_\mathrm{dw}(t) &= v_\mathrm{dw,0} e^{\frac{t}{t_\mathrm{exp}}} \\
	v_\mathrm{exp}(t) &=  v_\mathrm{dw,0} \left(e^{\frac{t}{t_\mathrm{exp}}} - 1 \right) \\
	r(t) &= v_\mathrm{dw,0} t_\mathrm{exp} \left(e^{\frac{t}{t_\mathrm{exp}}} - 1\right) \\
	n_\mathrm{th}(t) &= n_\mathrm{th,0} e^{-\frac{t}{t_\mathrm{exp}}}, \label{eq.AdExNth}
\end{align}
where the downstream velocity at $t=0$ is $v_\mathrm{dw}(0) = c_\mathrm{s,up} M / \sigma$, with the upstream sound speed $c_\mathrm{s,up}$, Mach number $M$ and the compression factor $\sigma$ (see table \ref{tab.sausage}). The expansion formally lasts until $t = t_\mathrm{exp} \mathrm{ln}\left( v_\mathrm{up} / v_\mathrm{dw,0} \right)$, i.e. $v_\mathrm{dw} = v_\mathrm{up}$. However, as the ICM density is stratified, expansion will likely seize long before this time. As usual, for a frozen in magnetic field: $B(t) \propto n_\mathrm{th}^{2/3}$ during the expansion.

\subsection{Cosmic-ray Electron Evolution} \label{sect:lifetimes}

In the downstream region, a population of CR electrons can be described by its \emph{isotropic} spectrum in number density $n(p,t)\,\mathrm{d}p$ over momentum  $p = E/c$ (in the ultra-relativistic limit).  The time-evolution of this spectrum is  given by the diffusion-loss equation \citep[see e.g.][ for a pedagogic introduction]{2011hea..book.....L}:
\begin{align}
	\frac{\mathrm{d}n(p,t)}{\mathrm{d}p} &= \frac{\partial}{\partial p} \left[ \left( \left. \frac{\mathrm{d}p}{\mathrm{d}t}\right|_\mathrm{loss} + \left. \frac{\mathrm{d}p}{\mathrm{d}t}\right|_\mathrm{AE} \right) n(p,t) \right] + Q(p,t), \label{eq:cre_spectrum} 
\end{align}
where  we neglect spatial diffusion and non-linear momentum diffusion as well as escape. \par
CRe are subject to a number of energy losses. In radio relics, radiative losses through inverse Compton scattering with CMB photons and synchrotron emission due to the ambient magnetic field $B(t)$ are dominant at synchrotron bright momenta (see also figure \ref{fig:lifetime}). These are given by \citep[e.g.][]{1962AZh....39..393K}:
\begin{align} 
	\left. \frac{\mathrm{d}p}{\mathrm{d}t}\right|_\mathrm{rad} &= -\frac{4}{9} \left( \frac{r_0}{m_\mathrm{c}c} \right)^2  p^2 \left( B(t)^2 + B_\mathrm{CMB}^2 (1+z)^4 \right), \label{eq.cooling}
\end{align}
where  $r_0 = e^2/m_\mathrm{e}c^2$ the classical electron radius, $B_\mathrm{CMB} = 3.2\,\mu\mathrm{G}$ the inverse Compton equivalent magnetic field and  $z$ the redshift of the cluster.  \par
Another source of systematic losses is due to adiabatic expansion \citep{1962AZh....39..393K}:
\begin{align}
	\left. \frac{\mathrm{d}p}{\mathrm{d}t}\right|_\mathrm{AE}  &= -\frac{p}{3} \frac{\dot{n}_\mathrm{th}(t)}{n_\mathrm{th}} \label{eq.AdEx}\\
\dot{n}_\mathrm{th}(t)/n_\mathrm{th} &= -1/t_\mathrm{exp} = X,
\end{align}
where we have assumed that the CRe number density evolves like the thermal number density $n_\mathrm{th}(t)$ (eq. \ref{eq.AdExNth}).
For simplicity, we assume that the shock instantaneously injects a power-law spectrum of CR electrons only at time $t=0$, i.e. the injection function in eq. \ref{eq:cre_spectrum}:
\begin{align}
	Q(p,t) &= n_0 p^{-s} \delta(t), \label{eq.IC}
\end{align}
where $n_0$ is the normalisation of the CRe spectrum\footnote{As we have defined the spectrum over $p$, the unit of this normalisation is $\mathrm{cm}^{-3} \left( \frac{\mathrm{sec}}{\mathrm{g cm}}\right)^s $, where s is the spectral index. Standard formulae are usually integrated over $E = pc$ missing a factor of $c^{s-1}$} and $s$ is the spectral index. According to standard DSA the spectral index is related to the Mach number $M$ of the shock by \citep[e.g.][]{1983RPPh...46..973D,1987PhR...154....1B}:
\begin{align}
	s &= 2 \frac{M^2 + 1}{M^2 - 1}.
\end{align}

\subsubsection{Solution}

Following \citet{1973ranp.book.....P}, chapter 6.3, we obtain the time-evolution of $n(p,t)$ in absence of additional injection, escape or any other effect except cooling and expansion from the sum of equations \ref{eq.cooling} and \ref{eq.AdEx}. This leads to:
\begin{align}
	p(t) &= \frac{p_0  e^{X t}}{1 + C_\mathrm{p} \beta(t) p_0}, \\
	C_\mathrm{p} &= \frac{4 r_0^2}{9 m_\mathrm{e}^2c^2}, \\
	\beta(t) &= \int\limits_{t_0}^t e^{X\tau} \left[ B^2(\tau) + B_\mathrm{IC}^2 \right] \,\mathrm{d}\tau . \label{eq.beta}
\end{align}
Here $\beta(t)$ is the time-integrated magnetic energy density, which parametrizes the cumulative effect of cooling and expansion on the spectrum. The time evolution of the CR electron spectrum can then be found from $n(p,t) \left . \,\mathrm{d}p\right|_{t=0} = n(p_0,0)\,\mathrm{d}p_0 $ and the injection function eq. \ref{eq.IC} as initial condition:
\begin{align}
	n(p,t) &= n_0 p^s e^{s X t} \left[ e^{X t} + C_\mathrm{p} \beta(t) p \right]^{s-2} e^{-\frac{t}{t_\mathrm{exp}}}. \label{eq.electron_spectrum}
\end{align}
where the last factor accounts for the change in volume of the gas until expansion stops, formally at $v_\mathrm{dw} = v_\mathrm{shock}$. The important quantity governing the cooling of the CRe population is $\beta(t)$, which for a constant magnetic field $B(t) = B_\mathrm{const}$ and no expansion ($t_\mathrm{exp} \rightarrow \infty$) becomes: 
\begin{align}
	\beta_\mathrm{JP}(t) &= \left(B_\mathrm{const}^2 + B_\mathrm{CMB}^2(1+z)^4 \right) t, \label{eq.beta_JP}
\end{align}
so equation \ref{eq.electron_spectrum} becomes the standard Jaffe-Perola model \citep{1973A&A....26..423J}, which leads to a cut-off when the term in brackets equals zero. This happens at a momentum of 
\begin{align}
	p_\mathrm{cut}(t) &= \frac{m_\mathrm{e}^2c^2}{C_\mathrm{p}\beta(t)}. \label{eq:pcut}
\end{align}
The additional effect of adiabatic expansion is then to shift this spectrum to lower momenta and reduce its normalisation (see figure \ref{fig:single_synchro_spectra}, left panel, red curve).

\subsection{Synchrotron Emission}
 
The synchrotron emissivity $j_\nu$ in $\mathrm{erg \, cm^{-3} sec^{-1} Hz^{-1}}$  of an isotropic CR electron population $n(p,t) \,\mathrm{d}p$ in a homogeneous magnetic field $B(t)$ with pitch angle $\theta$ at frequency $\nu$ is \citep{1965ARA&A...3..297G,1994hea2.book.....L}: 
\begin{align}
	j_\nu(t) &= \frac{e\sqrt{3}}{m_\mathrm{e} c^2} \int\limits^{\pi/2}_{0} B(t) \sin^2\theta  \int n(p,t) K(x) \,\mathrm{d}p\,\mathrm{d}\theta \label{eq.synchro}\\ 
	K(x) &= x \int\limits_x^\infty K_{5/3}(z) \,\mathrm{d}z \\
	x &= \frac{\nu}{C_\mathrm{Crit} B(t)\sin{\theta} p^2} \\
	C_\mathrm{Crit} &= \frac{3e}{4 \pi m_\mathrm{e}^3 c^3}
\end{align}
where $K(x)$ is the synchrotron kernel, $K_{5/3}$ is the Bessel function. The total synchrotron luminosity $L(\nu)$ in erg/sec/Hz can then be found by integrating eq. \ref{eq.synchro} over the volume of the relic. Under the assumption of slowly varying ICM properties and a downstream speed $v_\mathrm{dw}(t)$ to convert distance from the shock to time:
\begin{align}
	L(\nu) &=  \int \int \int j_\nu(t)v_\mathrm{dw}(t) \, \mathrm{d}t \,\mathrm{d}y\,\mathrm{d}z \label{eq:Inu} \\
		   &\approx 1.25 \times 10^{50} \mathrm{cm^2} \,\,  \int j_\nu(t) v_\mathrm{dw}(t) \, \mathrm{d}t .\nonumber
\end{align}
For the second equation we again used the parameters from section \ref{sect:model}. The integrations in $t$, $\theta$ and $p$ have to be done numerically, we use a midpoint rule for the former two and a Simpson rule for the latter\footnote{efficient IDL routines are available from the authors upon request.}. 

\subsection{Lifetimes and Basic Relic Physics}

\begin{figure}
	\centering
	\includegraphics[width=0.45\textwidth]{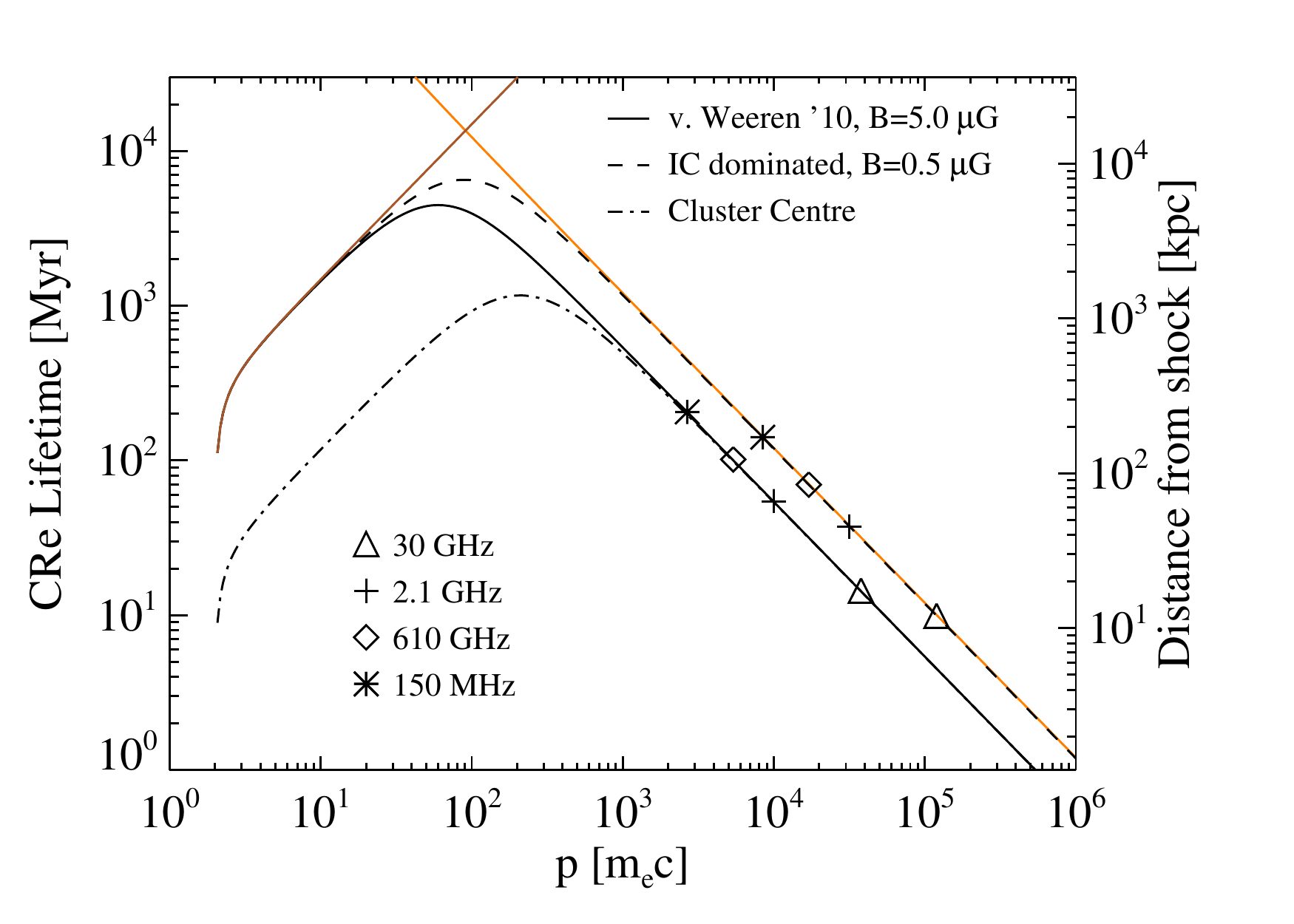}
	\caption{Lifetime of CR electrons over momentum due to IC and synchrotron cooling in a constant magnetic field of $0.5 \,\mu \mathrm{G}$ (dashed line), and $5 \,\mu \mathrm{G}$ (full line). We over-plot 4 times the sampling momentum $p_\mathrm{sample}$, eq. \ref{eq:psample} at four frequencies: 30 GHz (triangles), 2.1 GHz (crosses), 610 MHz (diamonds), 150 MHz (asterisk). We relate the lifetime to the distance of the shock in CIZA J2242.8+5301 with a downstream velocity of 1000 km/s.}
	\label{fig:lifetime}
\end{figure}

The lifetime of CRe is defined as $t_\mathrm{life} = p/\dot{p}$, where $p$ is the momentum of a CRe and $\dot{p}$ designates the systematic momentum losses of the CRe due to radiative and Coulomb losses (equation \ref{eq.cooling} and \citet{2014MNRAS.443.3564D}). \par
In figure \ref{fig:lifetime}, we show this lifetime of CRe over momentum at cluster outskirts assuming magnetic fields of 5 and 0.5 $\mu\mathrm{G}$, as full black line and dashed black line, respectively. We also show the lifetime in cluster centres as dot-dashed black line. We mark the smallest synchrotron bright momenta at 30 GHz (triangles), 2.1 GHz (crosses), 610 MHz (diamonds) and 150 MHz (asterisks). These momenta can be defined as: 
\begin{align}
	p_\mathrm{Sample} &= \sqrt{\frac{\nu}{ C_\mathrm{Crit} B(t) } } \label{eq:psample}
\end{align} 
where  one assumes $x = 1, \sin\theta = 1$. The lifetime can be converted into a distance from the injection of CRe, using the downwind shock speed (here we neglect expansion of the gas). This distance is shown on the right ordinate.  \par
We see that at high observed radio frequencies, lifetimes of the emitting CRe are as short as a few ten Myrs (triangles), which translates to a few ten kpc distance from the shock. That means, the emitting region becomes very thin at high frequencies ($>10$ GHz). \par
Lifetimes increase substantially at radio frequencies below 5 GHz (crosses, diamonds, asterisks), resulting in a relic thickness of several 100 kpc. Because of high IC cooling ($B_\mathrm{IC} = B_\mathrm{CMB} (1+z)^2 = 4.6 \,\mu\mathrm{G}$ at the cluster redshift), lifetimes are not strongly dependent on the ambient magnetic field. Under these conditions, an increase in the magnetic field by a factor of ten even leads to an increase of lifetime of the \emph{observed} CRe, because the sampling momentum increases as well (eq. \ref{eq:psample}).  It is in general not trivial to assess the effect of the changing field on the total spectrum of a relic from the lifetime alone. However, lifetimes of CRe at a given frequency is the maximized, if $B = B_\mathrm{IC}/\sqrt{3}$ \par
Lifetimes peak with 1 Gyr at 300 $m_\mathrm{e}\mathrm{c}$ for the cluster centre (thermal number density $n_\mathrm{th} = 10^{-3} \,\mathrm{cm}^3$) and 7 Gyr at 100 $m_\mathrm{e}\mathrm{c}$ at the location of the relic ($n_\mathrm{th} = 10^{-4} \,\mathrm{cm}^3$). Cooling is here dominated by Coulomb losses, hence dependent mostly on thermal density, not magnetic field strength \citep[see][ for details]{2014MNRAS.443.3564D}. This means CRe can accumulate at cluster outskirts from z = 1 onwards, with radio galaxies as a certain source of the CR injection (see also section \ref{sect.eff}).

\section{A Model for the Shock in CIZA J2242.8+5301}\label{sect:model}

\begin{table*}
	\centering
	\begin{tabular} {c|l|l|l}
		\hline
		Name & Value & Comment & Citation \\\hline
		$d_\mathrm{center}$ & 1.3 Mpc & Distance to center & \citet{2015MNRAS.450..646S} \\
		$z$ & 0.188 & Redshift & \citet{2015ApJ...805..143D} \\
		Length & 2 Mpc & Longest Extent on the sky &  \citet{2010Sci...330..347V} \\
		Width & 200 kpc & Extent of the brightness profile & \citet{2016MNRAS.455.2402S}  \\
		$\Psi$ &  $< 10 \,\mathrm{deg}$ & Angle into plane of sky & \citet{2010Sci...330..347V} \\ 
		M & 4.6 & Mach number  & \citet{2010Sci...330..347V} \\
		$T_\mathrm{up}$ & 3.0 keV & Upstream temperature & \citet[][ Akamatsu et al. 2015]{2014MNRAS.440.3416O} \\
		$n_\mathrm{th,up}$ & $1.6 \times 10^{-4}\,\mathrm{cm}^{-3}$ & Upstream thermal number density & Akamatsu priv. comm.  \\\hline
		$s$ & 2.1 & Injection spectral index of CRe & \\
		$v_\mathrm{dw}$ & 1184 km/s &  Downwind speed & \\
		$c_\mathrm{s,up}$ & 902 km/s & Sound speed of upstream medium & \\
		$v_\mathrm{shock}$ & 4144 km/s &  Shock speed in upstream medium &  \\
		$\sigma$ &  3.5 &  Compression ratio & \\
		$\sigma_\mathrm{Temp}$ &  7.5 &  Thermal compression ratio & \\
		$n_\mathrm{th,dw}$ & $5.6 \times 10^{-4}\,\mathrm{cm}^{-3}$ &  Downwind thermal number density &  \\
		$T_\mathrm{th,dw}$ & 22.4 keV &  Downwind temperature &  \\
		$B_\mathrm{IC}$ & $4.6 \,\mu \mathrm{G}$ & $B_\mathrm{CMB}$ at z=0.188 & \\
		\hline
	\end{tabular}
	\caption{Observed and derived parameters of the shock in CIZA J2242.8+5301 as assumed in this work.}
	\label{tab.sausage}
\end{table*}

In the discovery paper, \citet{2010Sci...330..347V} modelled the relic with a Mach number of 4.6, a shock speed of 1000 km/s and a magnetic field of $5 \mu\,\mathrm{G}$, using the standard formalism \citep[e.g. ][]{1998A&A...332..395E,2007MNRAS.375...77H}. \citet{2011MNRAS.418..230V} showed a simple numerical model for the shock and constrained the merger scenario to two colliding clusters with a mass ratio of one to two.  The shock related to the northern relic was discovered in the X-rays by \citet{2013PASJ...65...16A,2015A&A...582A..87A}, who find a Mach number of 2.7 for the shock. \citet{2013MNRAS.429.2617O} observed the cluster with XMM-Newton constraining the thermal ICM properties also in front of the shock wave. \citet{2015MNRAS.450..646S,2015MNRAS.450..630S} studied the interaction of the shock with the star-forming galaxies in CIZA J2242.8+5301. \citet{2014MNRAS.440.3416O} investigate the internal structure of the ICM and find several density discontinuities and shock candidates in the center of the cluster. \citet{2014MNRAS.445.1213S} conducted spatially resolved age modelling on the relic, finding a Mach number of 2.9 and an downstream speed of 905 km/s from aging arguments. \citet{2015ApJ...802...46J} constrained the Dark Matter distribution in the  merger, resulting in a total mass of $2 \times 10^{15} \,\mathrm{M}_\mathrm{sol}$. CIZA J2242.8+5301 is thereby one of the most massive clusters known to date.  \par

The redshift of the cluster CIZA J2242.8+5301 has been found to be $z\approx0.19$ \citep{2015ApJ...802...46J}. Observations show no significant difference in redshift between the two sub-clusters, hence we assume the cluster merger is roughly in the plane of the sky  \citep{2015ApJ...805..143D}, see also \citet{2012ApJ...756...97K}. \par
We assume the volume associated with the northern giant radio relic has the form of a cuboid with  a length of 2000 kpc, a cross-section area of $260\,\mathrm{kpc} \times 260\,\mathrm{kpc}$ and a distance from the centre of the cluster of 1500 kpc \citep{2010Sci...330..347V}. We assume a shock with a homogeneous Mach number $M$ of 4.6, this results in a compression ratio of $\sigma=3.5$.  We neglect projection effects, \citet{2010Sci...330..347V} showed that the relic extends less than 10 degrees into the plane of the sky \citep[see also][]{2012ApJ...756...97K}. We discuss this in section \ref{sect:discussion}. 

We note that the Mach number assumed here is inconsistent with Mach numbers inferred from X-ray observations. However, we anticipate that shocks with Mach numbers $M \le 4$ will not reproduce the observed spectral index profile of the relic, which require a downstream velocity of $v_\mathrm{dw} \ge 1200\,\mathrm{km}/\mathrm{s}$ (see \citet{2014MNRAS.445.1213S} and section \ref{sect.spix}). A discussion of our models with a Mach number of 3 can be found in appendix \ref{app.Mach}. This problem of inconsistency between the Mach numbers measured in the X-rays and those derived from radio observations of relics (assuming DSA) is well known \citep{2015A&A...582A..87A,2014IJMPD..2330007B}. \citet{2014ApJ...785..133H} discussed this problem using numerical simulations. They find that complex shocks contain a range of Mach numbers and radio observations are likely biased towards the highest Mach numbers in the distribution. \par
We set the upwind temperature to 3.0 keV, consistent with measurements from X-ray observatories, which found $T_\mathrm{up} = 2.7^{+1.2}_{-0.7}\,\mathrm{keV}$ \citep{2013PASJ...65...16A,2014MNRAS.440.3416O,2015A&A...582A..87A}. This leads to a downwind shock speed of 1184 km/s using the standard Mach number of 4.6. For this calculation we took a pre-shock density of $n_\mathrm{th,up} = 1.6 \times 10^{-4}\,\mathrm{cm}^{-3}$, which is based on Suzaku data (Akamatsu priv. comm.).

\subsection{Magnetic Field Models}

\begin{figure*}
	\centering
	\includegraphics[width=0.9\textwidth]{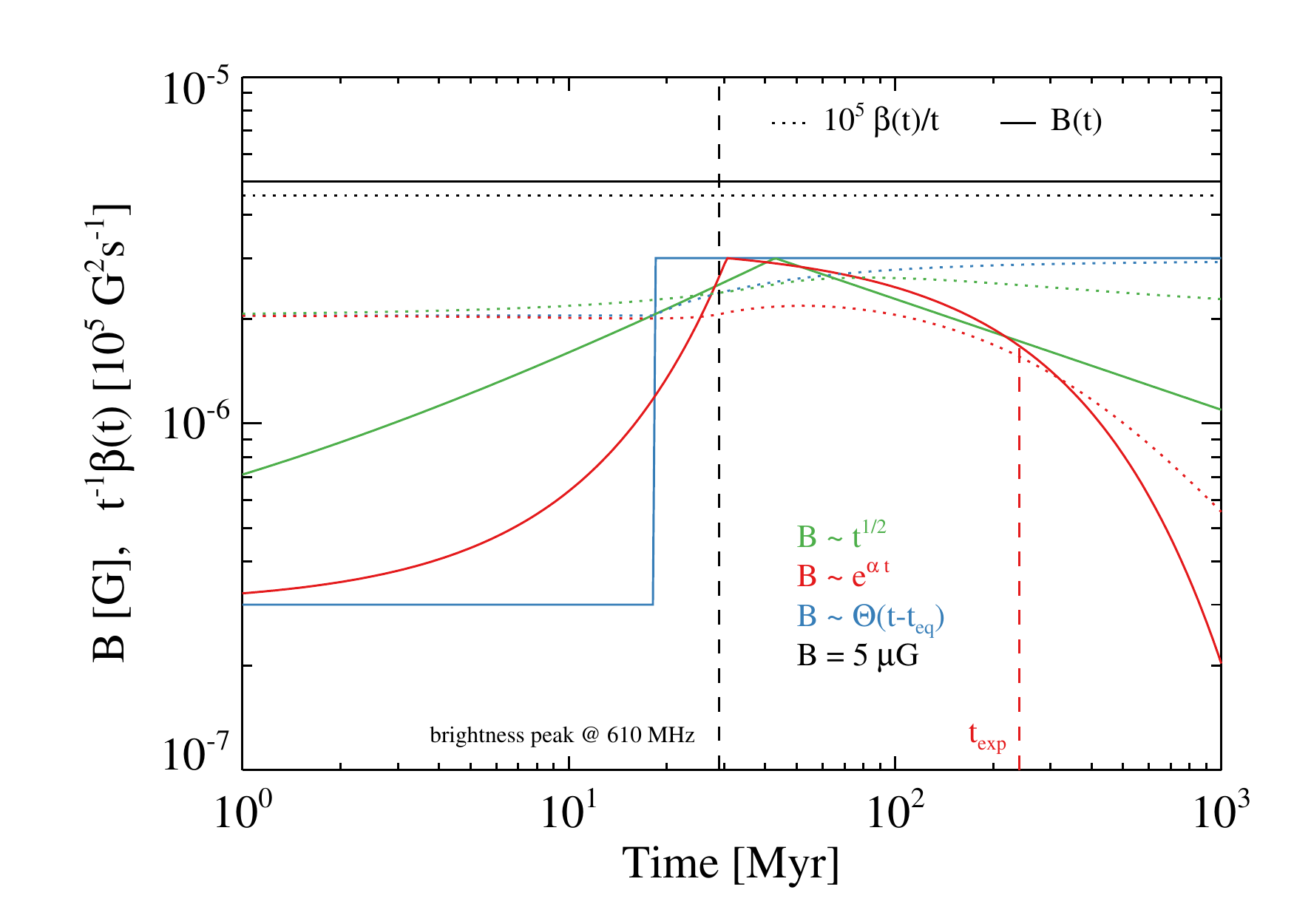}
	\caption{Time and spatial evolution of the four magnetic field models used in this work. We show the constant field model (JP) as black line, step function in blue, linear increase in green and exponential increase in red. Alongside we plot the corresponding value of $\beta(t)/t$ in units of $10^{5}\,\mathrm{G}^2\mathrm{s}^{-1}$ as dotted lines. We also add the location of the brightness peak at 610 MHz as vertical dashed line and the e-folding time scale of the exponential model as red vertical dashed line.}
	\label{fig:bfld}
\end{figure*}

Radio relics are commonly modelled with spatially constant downstream magnetic fields, in the case of CIZA J2242.8+5301, \citet{2010Sci...330..347V} model the relic with $B = 5 \,\mu \mathrm{G}$, which was subsequently used in other work \citep{2014MNRAS.445.1213S}. We adopt this value in a model we name our ''standard model'', which we will show as a black line in all figures.  
As mentioned before, a break in the total synchrotron spectrum motivates  that shorter lived CRe radiate in a smaller magnetic field. This can be realised by \emph{increasing} in the magnetic field just behind the shock and the subsequent injection of the CR electrons. This leads to a natural decoupling of the region of CRe injection/acceleration (at the shock) and the magnetic field amplification, which is gradually amplified in the downstream region. We heuristically chose three functional forms for this increase, linear increase of magnetic energy (green in all figures, henceforth ''linear model''), exponential increase of the field (red in all figures, henceforth ''exponential model'') and step function (blue in all figures, henceforth ''step model''). The former two can be physically motivated, which we discuss in section \ref{sect:discussion}, the step-function increase is physically not motivated, we chose it as an extreme case for comparison. In the linear and exponential case, we assume that the magnetic field reaches a maximum value at distance $d_\mathrm{eq}$, and then declines at larger distances. In the exponential model, the decline is assumed to result from the adiabatic expansion (section \ref{sect.adexp}) under the hypothesis of flux-freezing of the magnetic field into the thermal plasma, i.e. $B \propto n_\mathrm{th}^{2/3}$. We use an exponential decay of the density with an e-folding time of 240 Myrs in the exponential model (red). For the linear model (green), we use an exponent $\delta =  -0.3$, but do not follow the adiabatic expansion of the gas, i.e. leave velocity and density constant. Then the magnetic field models take the form: 
\begin{itemize}
	\item Standard model (black):
		\begin{align}
			B &= 5\,\mu\mathrm{G}
		\end{align}
	\item Step function model (blue):
		\begin{align}
			B_\mathrm{step} &= B_\mathrm{min} + \Theta(t-t_\mathrm{eq}) \left( B_\mathrm{max} - B_\mathrm{min}  \right) 
		\end{align}
	\item Exponential model (red):
		\begin{align}
			B_\mathrm{exp} &=  \begin{cases}
			B_\mathrm{min} e^{\alpha_0 t} & \Rightarrow t < t_\mathrm{eq} \\
			B_\mathrm{max} e^{-\frac{2t}{3t_\mathrm{exp}}}  & \Rightarrow t \ge t_\mathrm{eq}
			\end{cases} \\
		\alpha_0 &= \mathrm{log}\left( \frac{B_\mathrm{max}}{B_\mathrm{min}} \right) / ( t_\mathrm{eq}) \nonumber
		\end{align}
	\item Linear model (green):
		\begin{align}
			B_\mathrm{lin} &= \begin{cases}
			a \sqrt{t} + B_\mathrm{min} &  \Rightarrow t < t_\mathrm{eq} \\
			B_\mathrm{max}\left( \frac{t}{t_\mathrm{eq}}\right)^\delta &  \Rightarrow t \ge t_\mathrm{eq} \\
			\end{cases} \\
			a &= \frac{B_\mathrm{max}-B_\mathrm{min}}{\sqrt{t_\mathrm{eq}}}  \nonumber
		\end{align}
\end{itemize}
with $d_\mathrm{eq} = \int\limits_0^{t_\mathrm{eq}} v_\mathrm{dw}(t)$. For the standard model, $\beta(t)$ is then given by eq. \ref{eq.beta_JP}, for the other models it follows straight forward from eq. \ref{eq.beta}, in case of the step and linear model with $X=0$. \par
\begin{table*} 
	\centering
	\begin{tabular} {r|c|c|c|c|c}
		\hline 
		Model & Magnetic Field $[\mu\mathrm{G}]$ & $d_\mathrm{eq} \, [\mathrm{kpc}]$ & Decay index $\delta$ & $t_\mathrm{exp}[\mathrm{Myrs}]$  \\\hline
		Standard (black) & 5  & - & - & $\rightarrow \infty$ \\
		Step (blue) & $0.3 < B < 3.0$ & 20 & - & $\rightarrow \infty$ \\
		Exponential (red) & $0.3 < B < 3.0$ & 37 & - &  240 \\ 
		Linear (green) & $0.3 < B < 3.0$ & 50 & -0.3 & $\rightarrow \infty$ \\ 
		\hline 
	\end{tabular}
	\caption{Magnetic field parameters used in this work} \label{tab.bfldparam}
\end{table*} 
An overview of the parameters used in this work is given in table \ref{tab.bfldparam}. To fit the data, we empirically find a minimum magnetic field value roughly consistent with naive estimates at the outskirts of clusters $B_\mathrm{min} = 0.3 \,\mu \mathrm{G}$, and a maximum value of $B_\mathrm{max} = 3.0 \,\mu \mathrm{G}$, roughly consistent  with previous work \citep{2010Sci...330..347V,2014MNRAS.445.1213S,2015ApJ...809..186K}. A best fit model is found for saturation scales $d_\mathrm{eq}$ of 20, 35 and 50 kpc for step, exponential and linear model, respectively. An exploration of the parameter space of magnetic field values is given in appendix \ref{app.parameters}, where we conclude that at high frequencies the radio spectrum is most sensitive to the minimum magnetic field and the saturation scale. \par
We show all four magnetic field models over time and distance in figure \ref{fig:bfld}, where we also plot  $\beta(t)/10^9$ as dotted lines. In our new models, the cooling is IC dominated, so the cooling term $\beta$ is basically the same for the new models. Small differences arise only for  $d \gg d_\mathrm{eq}$.  This is different from the standard model, where the magnetic field about the same as the IC equivalent field $B_\mathrm{CMB}$. Here both mechanism contribute to the cooling, and \citet{2014MNRAS.445.1213S} showed that the resulting CRe cooling speed is close to its minimum with these values. \par
In our new models, the structure of the relic then motivates a different location of the shock, so the brightness peak of model and observed data coincide. In the standard model, the shock and the peak of the brightness profile coincide, i.e. $d_\mathrm{eq} = 0$. In contrast, in the new models we assume that the shock is located $d_\mathrm{eq} > 0$ in front of the brightness peak. The situation is further complicated by the finite resolution of the observed brightness profiles and projection effects, which broaden the rising flank of the emission and shift its peak, as shown in \citet{2010Sci...330..347V}.  In the standard model, the emission is consistent with 610 MHz data and the model in \citet{2010Sci...330..347V}. For the other models, we chose $d_\mathrm{eq}$ so the model fits the brightness peak, without considering projection effects. Simulations are required to further study the shock geometry, projection effects and constrain this aspect of the model.  

To ease computation, we assume instantaneous injection of the CR electrons in all models. 

\section{Results} \label{sect:results}

\subsection{Time-dependent CRe and Synchrotron Spectra}
 
\begin{figure*}
	\centering
	\includegraphics[width=0.45\textwidth]{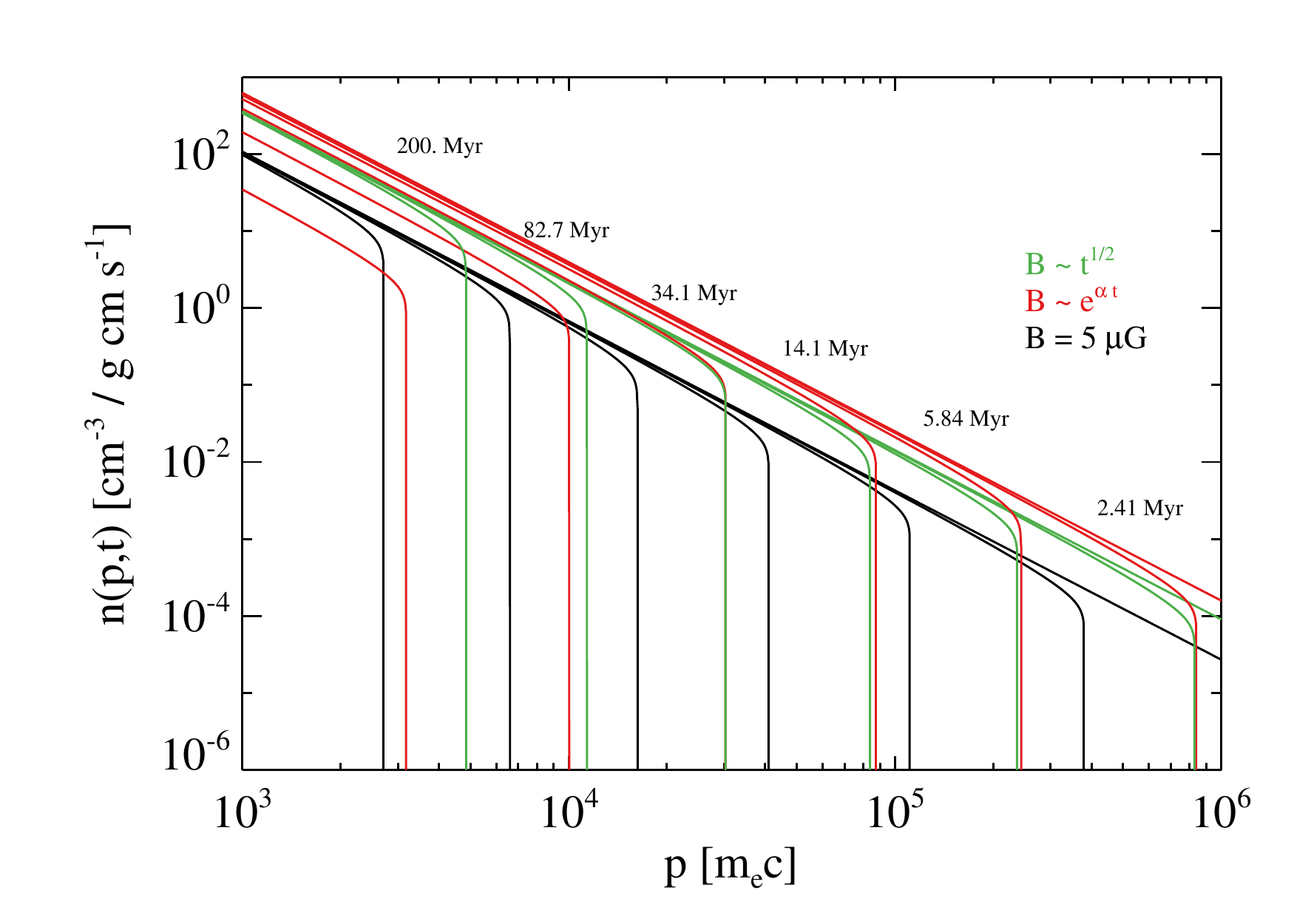}
	\includegraphics[width=0.45\textwidth]{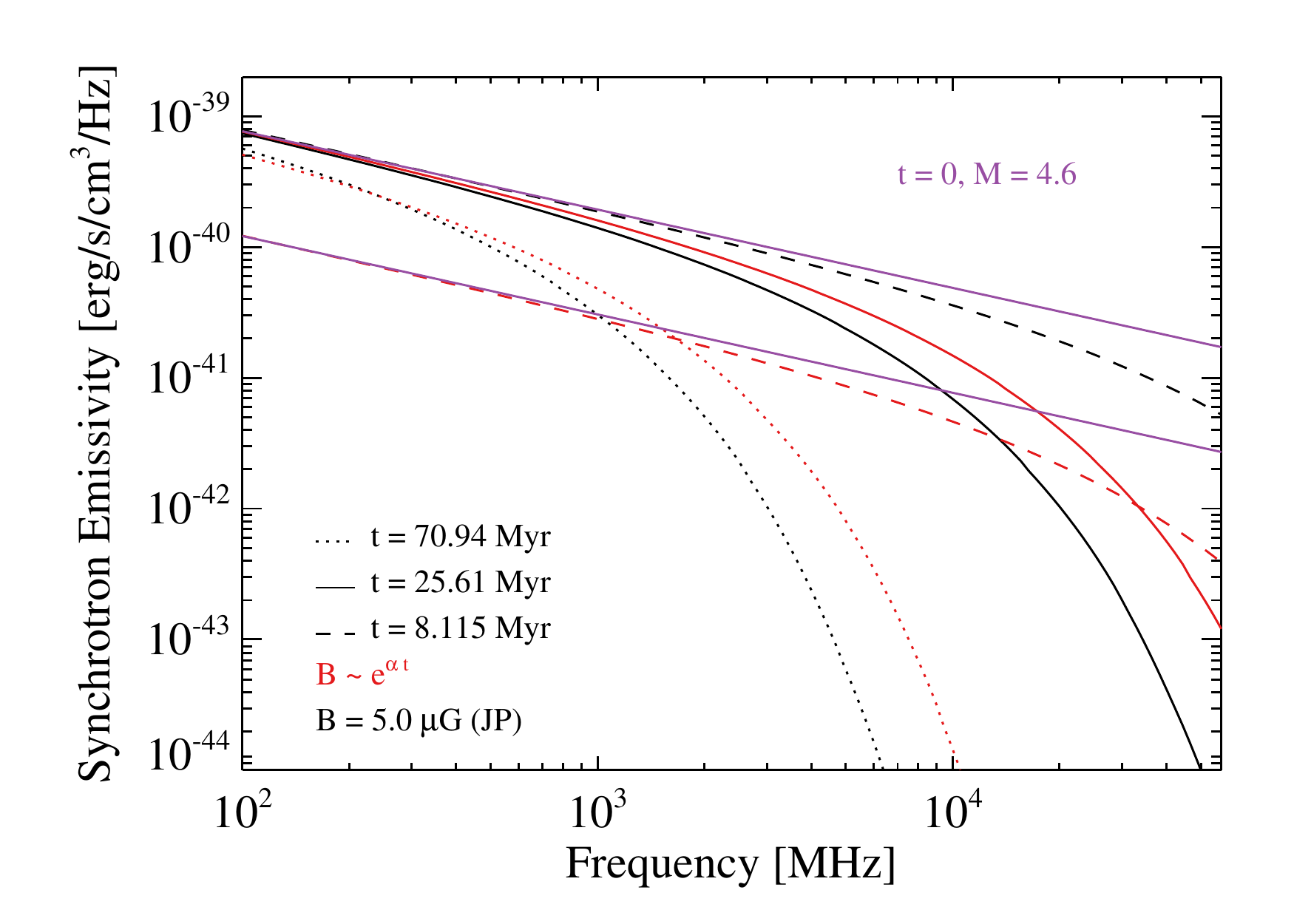}
	\caption{Left: Time evolution of cooling CRe spectra $n(p,t)$ in cgs units. The standard JP model is shown in black with the time in Myr above the break. The linear and exponential models are shown as in green and red, respectively. Right: Time evolution of synchrotron spectra  of the standard and the exponential model (standard, black; exponential, red) at three different times,  8, $t_\mathrm{eq}$ and 83 Myr as dashed, full and dotted line, respectively. The injection is marked as purple line. }
	\label{fig:single_synchro_spectra}
\end{figure*}

We begin by demonstrating the action of cooling and synchrotron sampling in the formalism. In figure \ref{fig:single_synchro_spectra} left, we show the time evolution of $n(p,t)$ from equation \ref{eq.electron_spectrum}, for three magnetic field models at 7 times (0, 2.41, 5.84, 14.1, 34.1, 82.7, 200 Myr) in cgs units. We do not show the step model, because it is not very instructive. We over-plot the age of the spectrum above the break momentum of the standard JP model with $5 \,\mu\mathrm{G}$ (black). The other two models (standard color scheme) show a delayed break in the spectrum, because the magnetic field contributes less to the cooling in the beginning, i.e. the cooling is completely IC dominated. The exponential model shows a shift in amplitude at late times, due to the expansion of the thermal plasma. This uniformity in cooling is a result of the high IC equivalent magnetic field of $B_\mathrm{IC} = 4.6 \,\mu\mathrm{G}$.\par
In figure \ref{fig:single_synchro_spectra}, right, we compare synchrotron spectra at 8, 26 and 83 Myr for the standard (black) and exponential model (red) from equation \ref{eq.synchro}. We add the emission from the injection, which is calculated fully analytically from the standard formulae in purple for both models. In contrast to the CRe spectra, the associated synchrotron spectrum of the exponential model increases in brightness by nearly a factor of ten. This is despite the expansion and due to the increasing magnetic field, which also leads to changes in sampling momenta (compare eq. \ref{eq:psample}).  

\subsection{Integrated Synchrotron Spectra} \label{sect:int_spectrum}

\begin{figure*}
	\centering
	\includegraphics[width=0.9\textwidth]{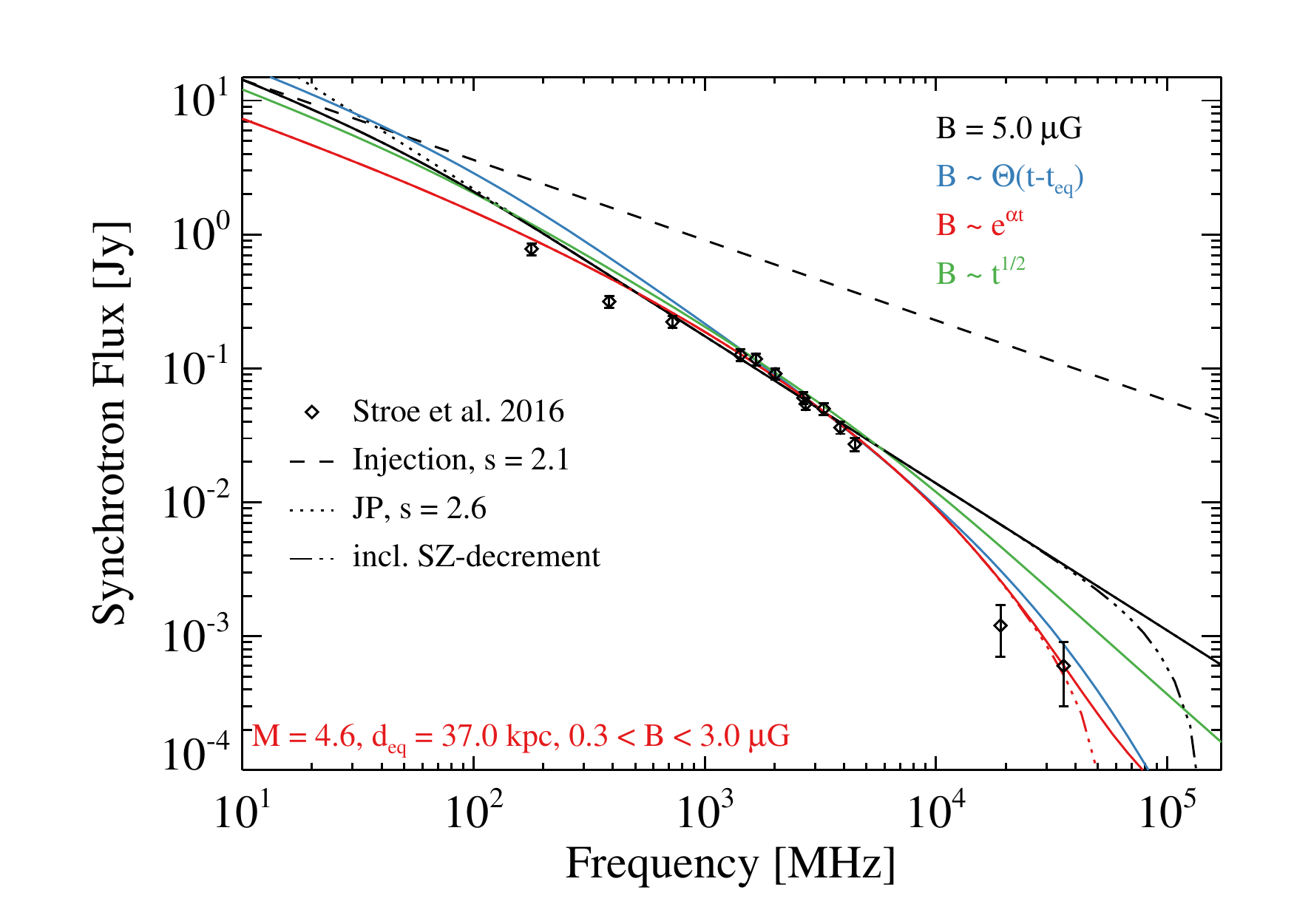}
	\caption{Total integrated synchrotron spectrum of the large relic in CIZA J2242.8+5301 from the standard model (black, \citet{2010Sci...330..347V}), and our three models in blue, red and green. We over-plot the recently observed spectrum with K-corrected frequencies from \citet{2016MNRAS.455.2402S} as black diamonds with error bars, and the injection power-law (dashed line) as well as the JP model (dotted line).}
	\label{fig:total_synchro_spectrum}
\end{figure*}

We solve equation \ref{eq:Inu} for the standard values given in table \ref{tab.sausage} for all four models. We chose a normalisation $n_0$ of the CRe spectrum equation \ref{eq:cre_spectrum}, so the total flux from the model roughly fits the observed spectrum around 1 GHz. The normalisation values are reported for all four models in table \ref{tab:cre_norm} \par
The resulting integrated radio synchrotron spectra over frequency are shown in figure \ref{fig:total_synchro_spectrum}. We mark the standard model as solid black line, the new model in the usual color scheme. We add the injection spectrum of the standard model and the cooled JP spectrum of the standard model as black dashed line and dotted line, respectively. Recent observations of the large relic in CIZA J2242.8+5301 by \citet{2016MNRAS.455.2402S} are added as black diamonds with error bars, where we convert the observed to rest frame frequencies by multiplying by $1+z$ for all frequencies. We also show the modification of the standard model by the SZ-decrement as dot-dashed black line, following  \citet{2013A&A...558A..52B}, eq. 1, assuming constant thermal pressure in the relic from the values shown in table \ref{tab.sausage} and considering a 150 kpc region behind the shock, consistent with the beam size at 30 GHz (see fig. \ref{fig:conv_brightness_profiles}). \par
As expected, the standard model shows curvature only at low frequencies and does not reproduce the steepening at the highest frequencies. Our new models are generally curved outside the frequency range of 610 to 2 GHz. In particular, they show a steepening at high frequencies. The exponential model is here roughly consistent with the observations of the relic. At intermediate frequencies, all models reproduce the power-law behaviour expected from the standard JP-model. At low frequencies all models flatten and the exponential and linear model are roughly consistent with the data. For our thermal model and a volume of 2 Mpc x 150 kpc x 260 kpc, the SZ-decrement does not play a role at 30 GHz.  

\subsection{Normalized Brightness Profiles} \label{sect:brightnes_profiles}

\begin{figure*}
	\centering
	\includegraphics[width=0.45\textwidth]{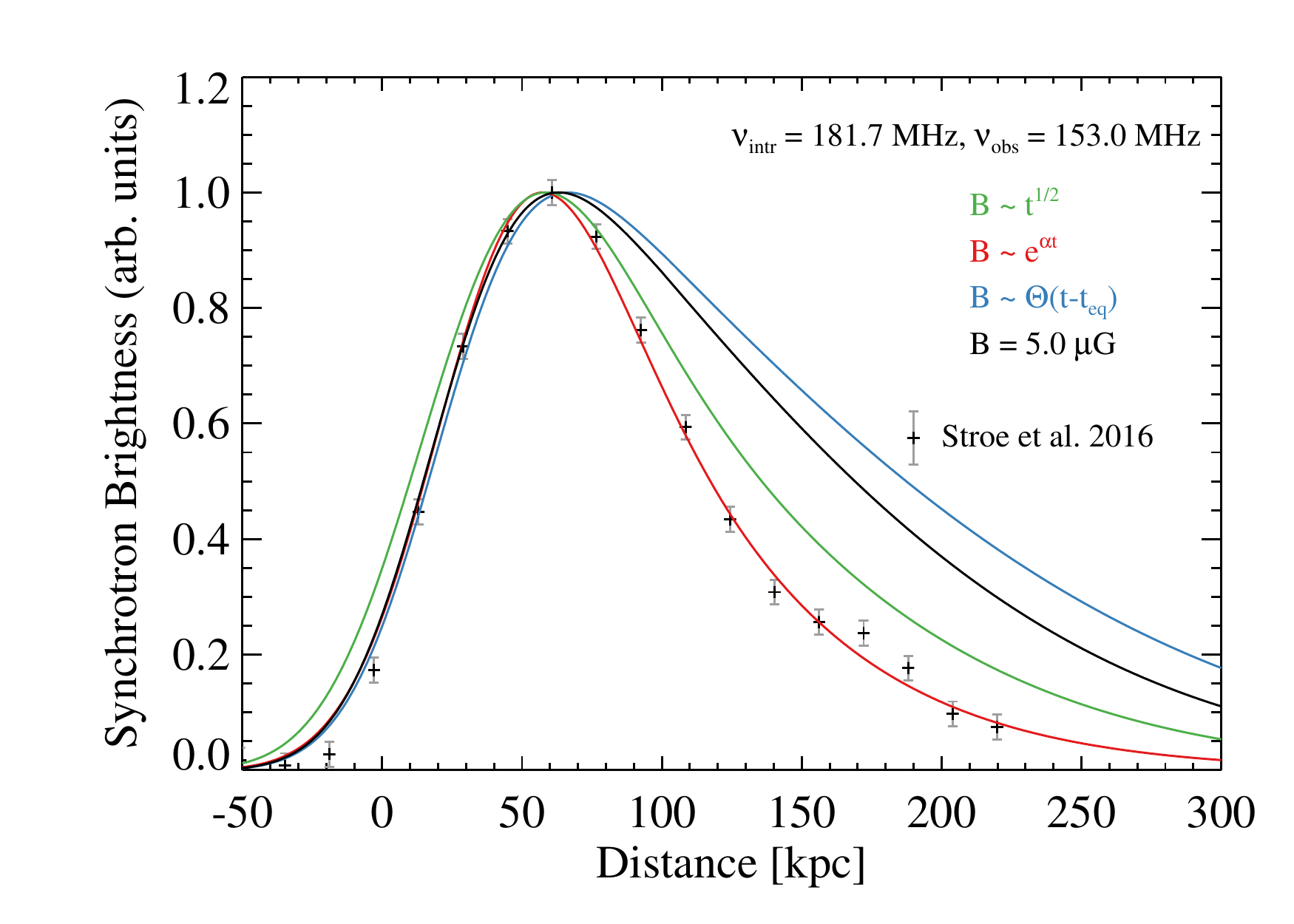}\\
	\includegraphics[width=0.45\textwidth]{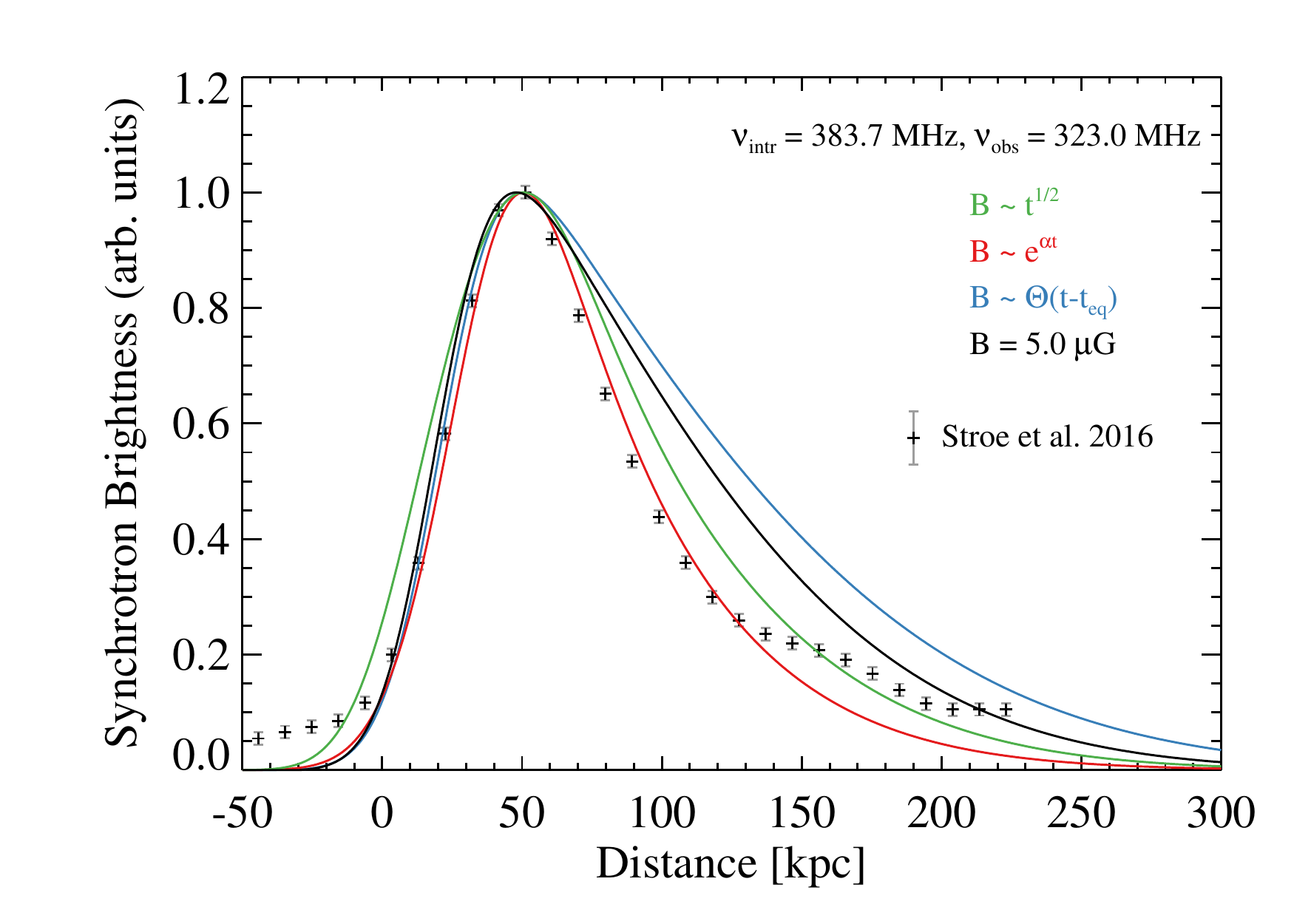}
	\includegraphics[width=0.45\textwidth]{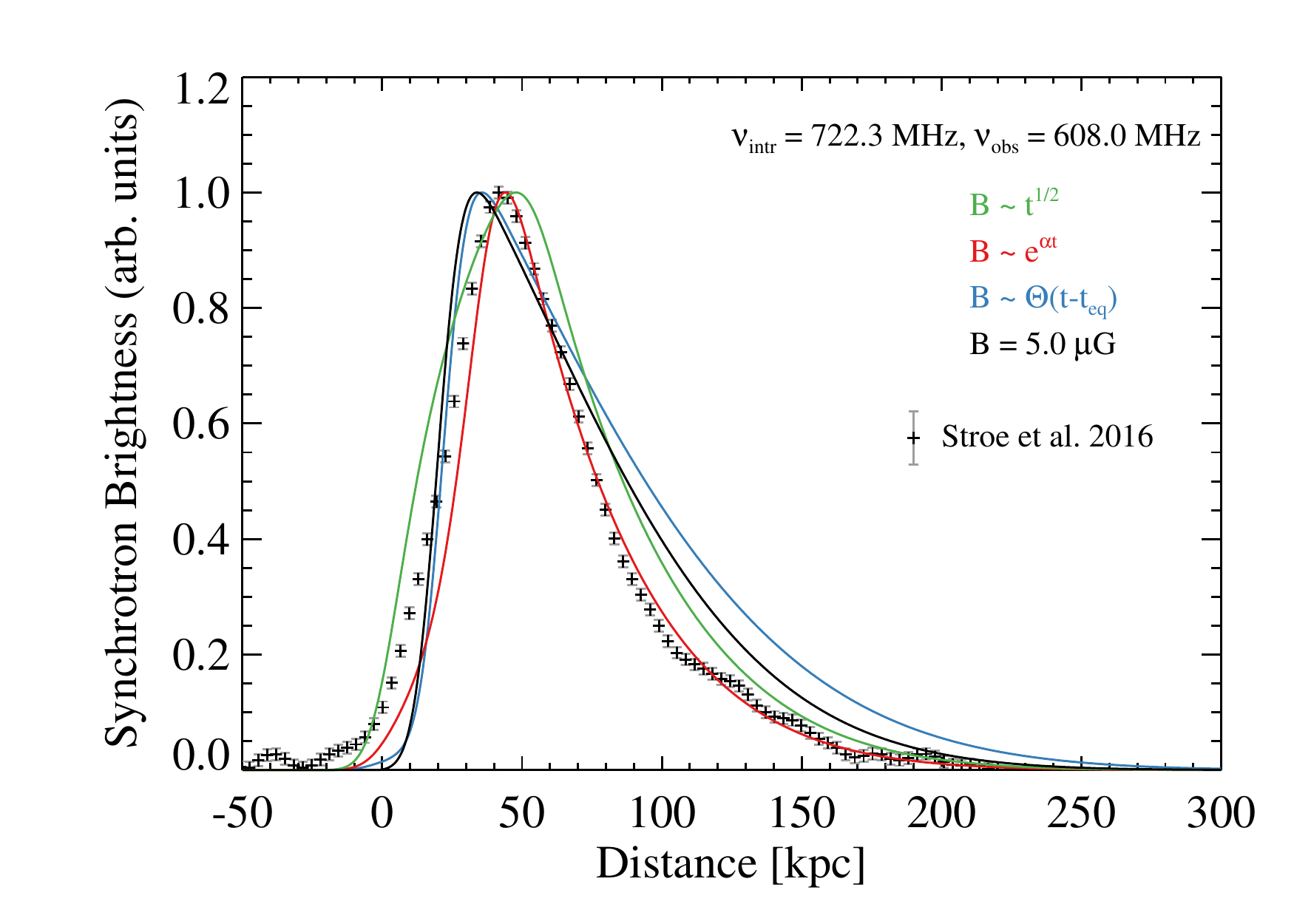}\\
	\includegraphics[width=0.45\textwidth]{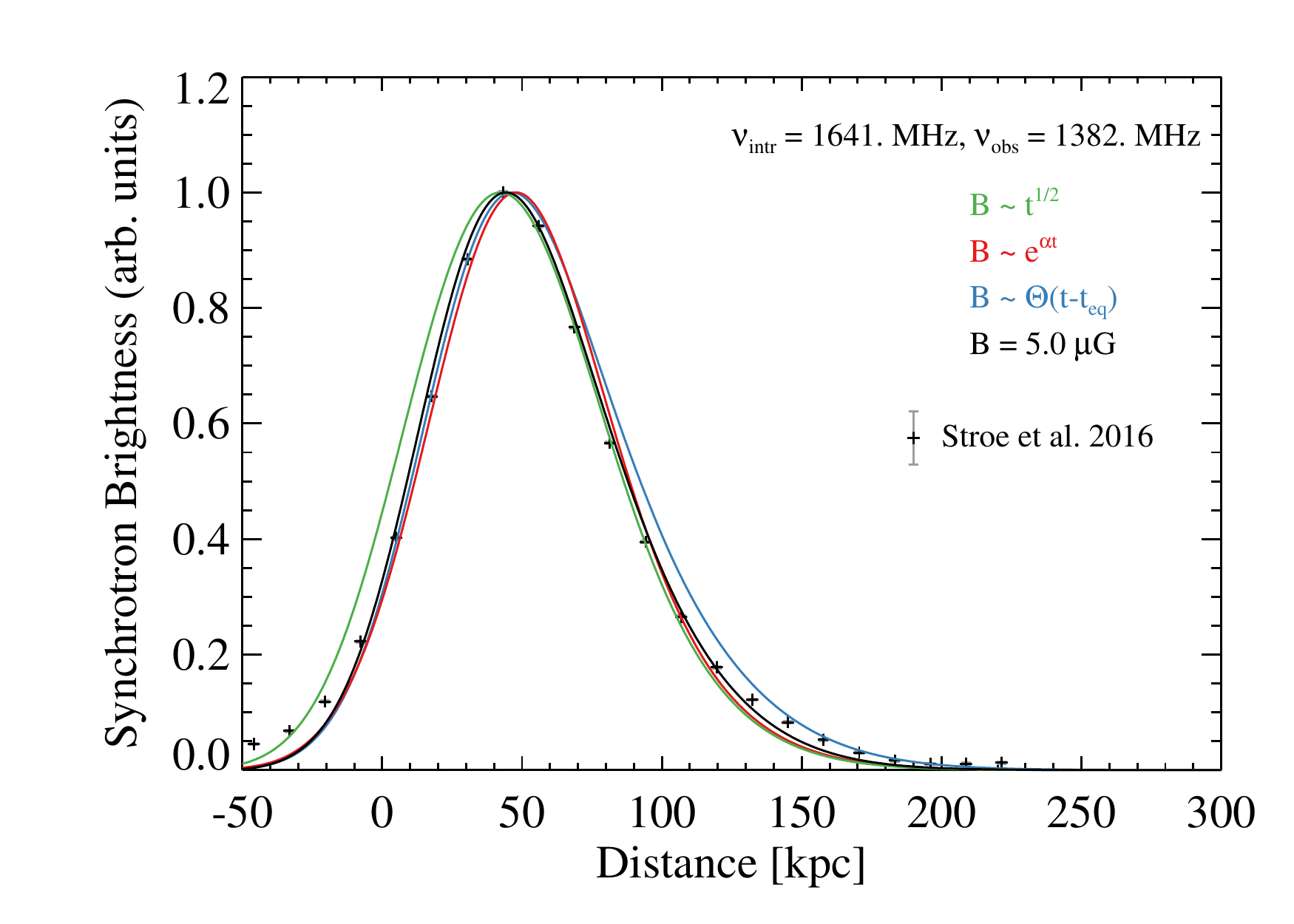}
	\includegraphics[width=0.45\textwidth]{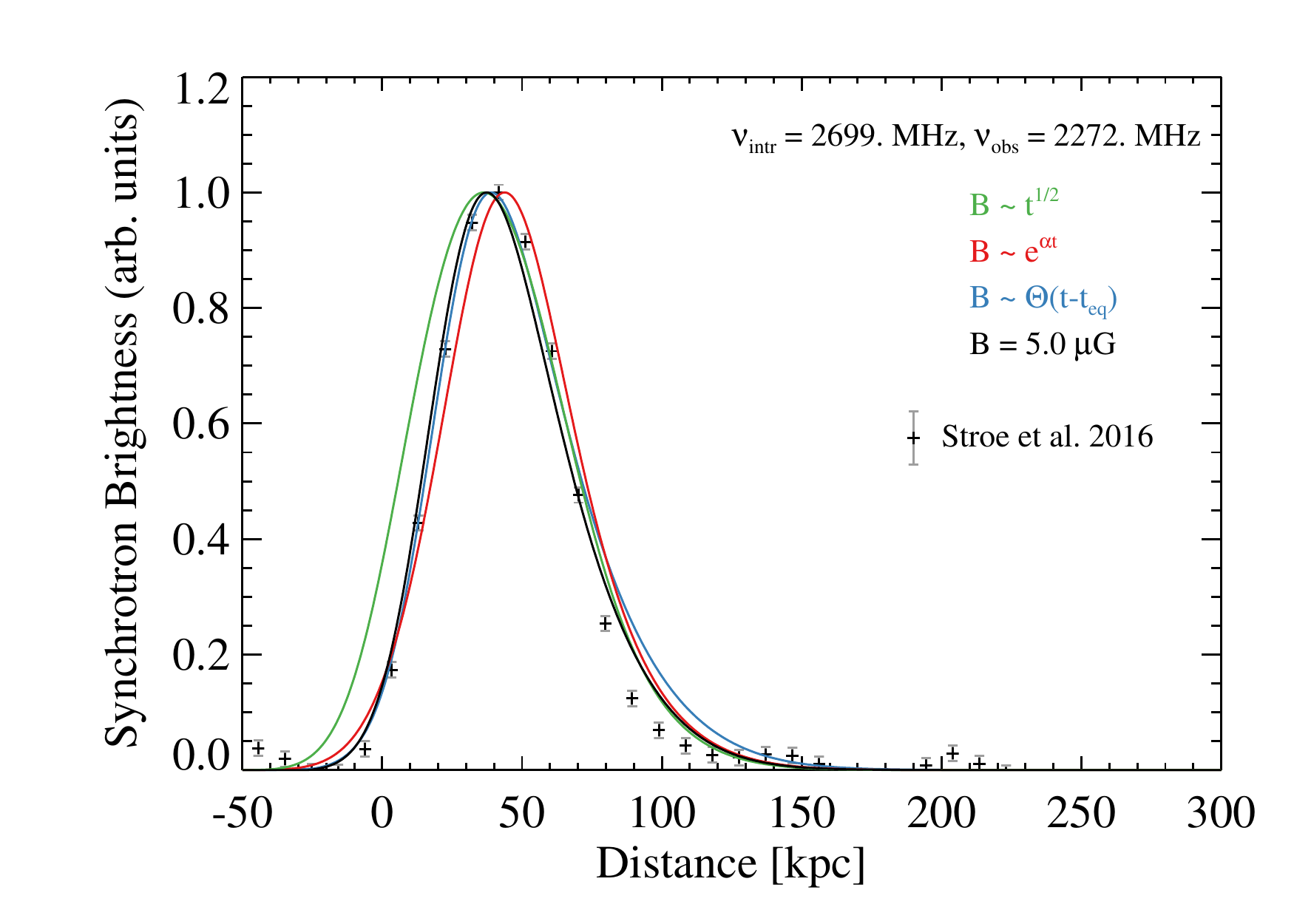}\\
	\includegraphics[width=0.45\textwidth]{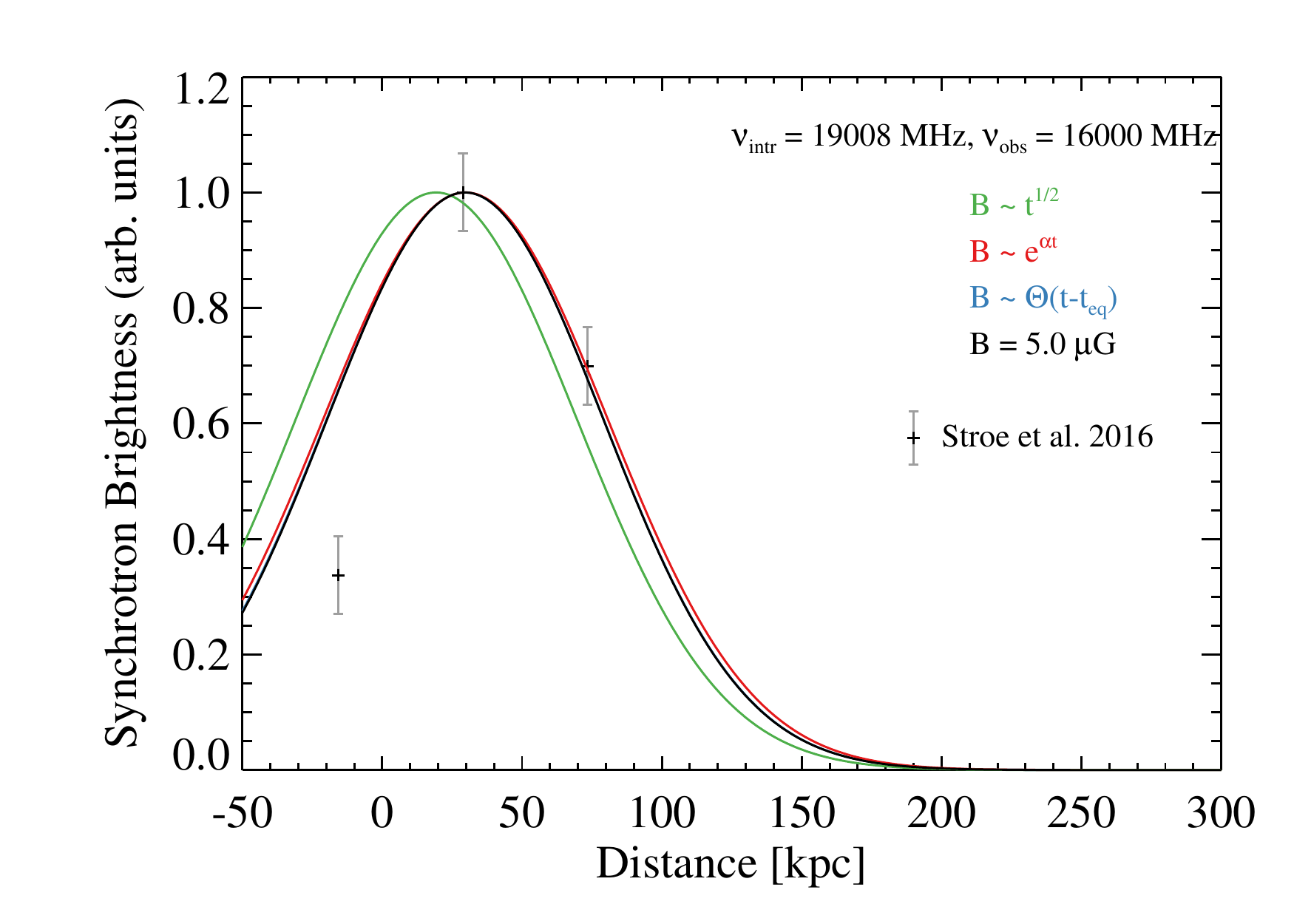}
	\includegraphics[width=0.45\textwidth]{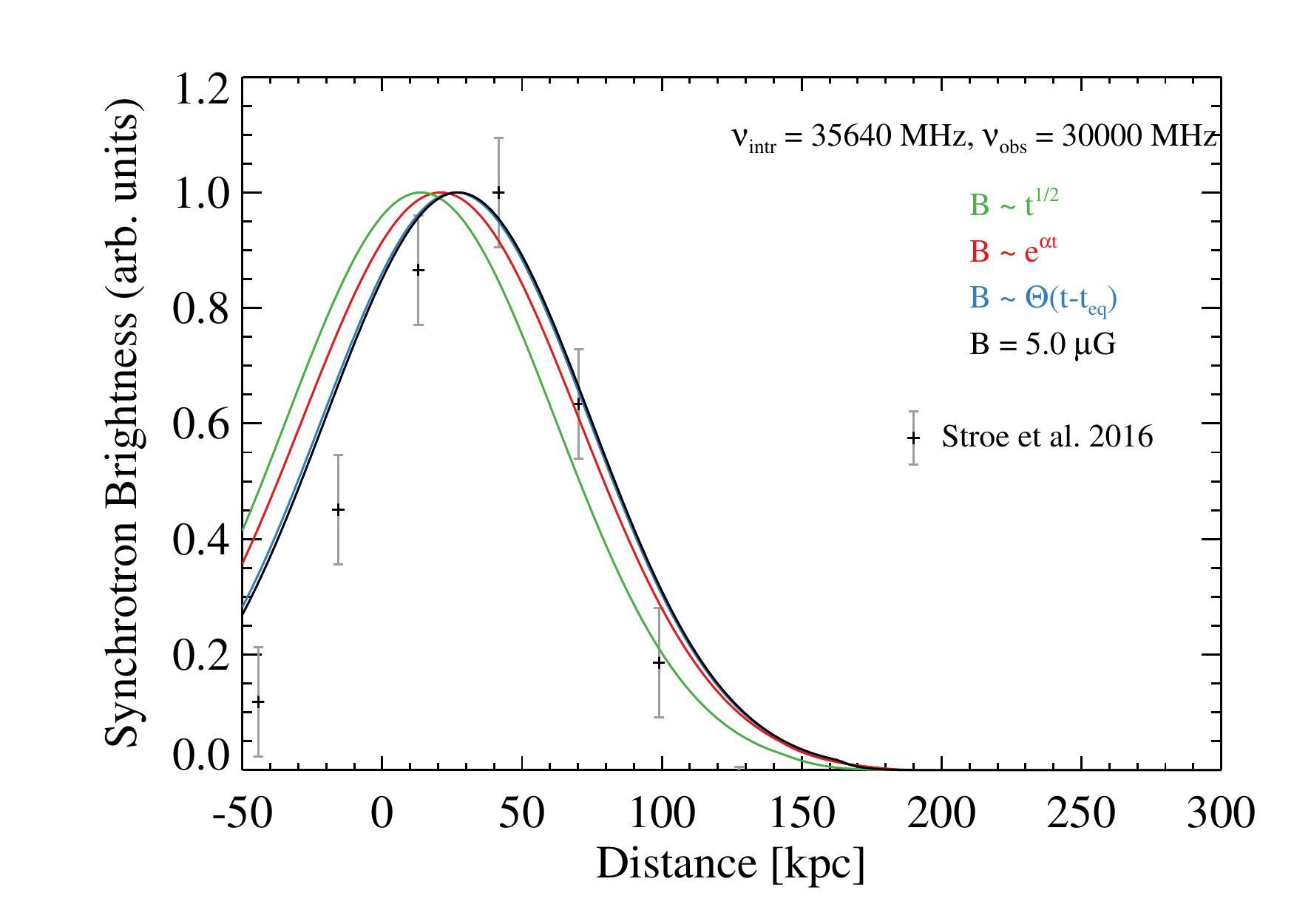}
	\caption{\emph{Beam convolved} normalised relic brightness from the models over distance perpendicular to the relic in kpc. The model was computed at eight intrinsic frequencies (top left to bottom right: 182, 384.3, 723.5, 1644, 2703, 19040, 35700 MHz), corresponding to observed frequencies: 153, 323, 608, 1382, 2272, 16000, 30000 MHz. The standard model is shown in black with grey errors, the step function model in blue, the exponential model in red and the linear model in green. Black crosses are profiles from the data presented in \citet{2016MNRAS.455.2402S}. }	\label{fig:conv_brightness_profiles}
\end{figure*}

For a more detailed comparison with the observed data, we now consider the radio brightness profiles of the relic. 
We obtain brightness profiles using the data available on the `Sausage' relic, which were previously described in \citet[][]{2013A&A...555A.110S,2014MNRAS.441L..41S,2016MNRAS.455.2402S}. We produce images using all the visibilities, using a uniform weighting to maximise the resolution. To increase signal to noise, we average along the relic, using circular caps aligned with the shock structure. This was done in a similar fashion to Figure 20 from \citet{2013A&A...555A.110S}. This average profile describes an average cut through the relic tracing the upstream, shock and downstream region. We note that the relic is, as expected, not perfectly traced by a circle, so a small amount of averaging across its width will happen.  We calculate the error at each position within the average profile as the error obtained by averaging over $N_\mathrm{beams}$ number of beams, each with noise $\sigma_\mathrm{RMS}$:
\begin{align}
	E_\mathrm{profile} &= \frac{ \sqrt{ \Sigma^{N_\mathrm{beams}}_1 \sigma^2_\mathrm{RMS}  } }{N_\mathrm{beams}}\label{eq:profileErr}.
\end{align}

In figure \ref{fig:conv_brightness_profiles}, we show the observed radio profiles from \citet{2016MNRAS.455.2402S} at eight observed frequencies, top left to bottom right: 50, 153, 323, 608, 1382, 2274, 16000, 30000 MHz. We convert the frequencies to the intrinsic frequencies at redshift $z = 0.19$: 59.50, 182, 384.3, 723.5, 1644, 2703, 19040, 35700 MHz. The standard model is shown as black line and our new models in the usual colors. We add the model profiles, convolved with the appropriate beam from the observations. We note that for the new models, the shock is located at a distance of 0 kpc. In contrast, for the standard model, the shock is located at 30 kpc. \par
We find a reasonable fit of the exponential model and linear model at all frequencies.  The standard and step model  predict excess emission at low frequencies and large distances. The linear model predicts a significant shift in the position of the brightness relic at the highest frequencies compared to the lower frequencies.  We note that \citet{2010Sci...330..347V} showed that a broadening at the rising flank of the profile is consistent with projection effects into the plane of the sky.  All but the exponential model exhibit excess emission at the lowest frequencies, indicating the best fit of an exponential decay of the magnetic field after the brightness peak. We conclude that the exponential and linear model are roughly consistent with the data, if the relic extends into the plane of the sky with an angle $\Psi < 10\deg$.

\subsection{Spectral Index Profiles} \label{sect.spix}

\begin{figure}
	\centering
	\includegraphics[width=0.45\textwidth]{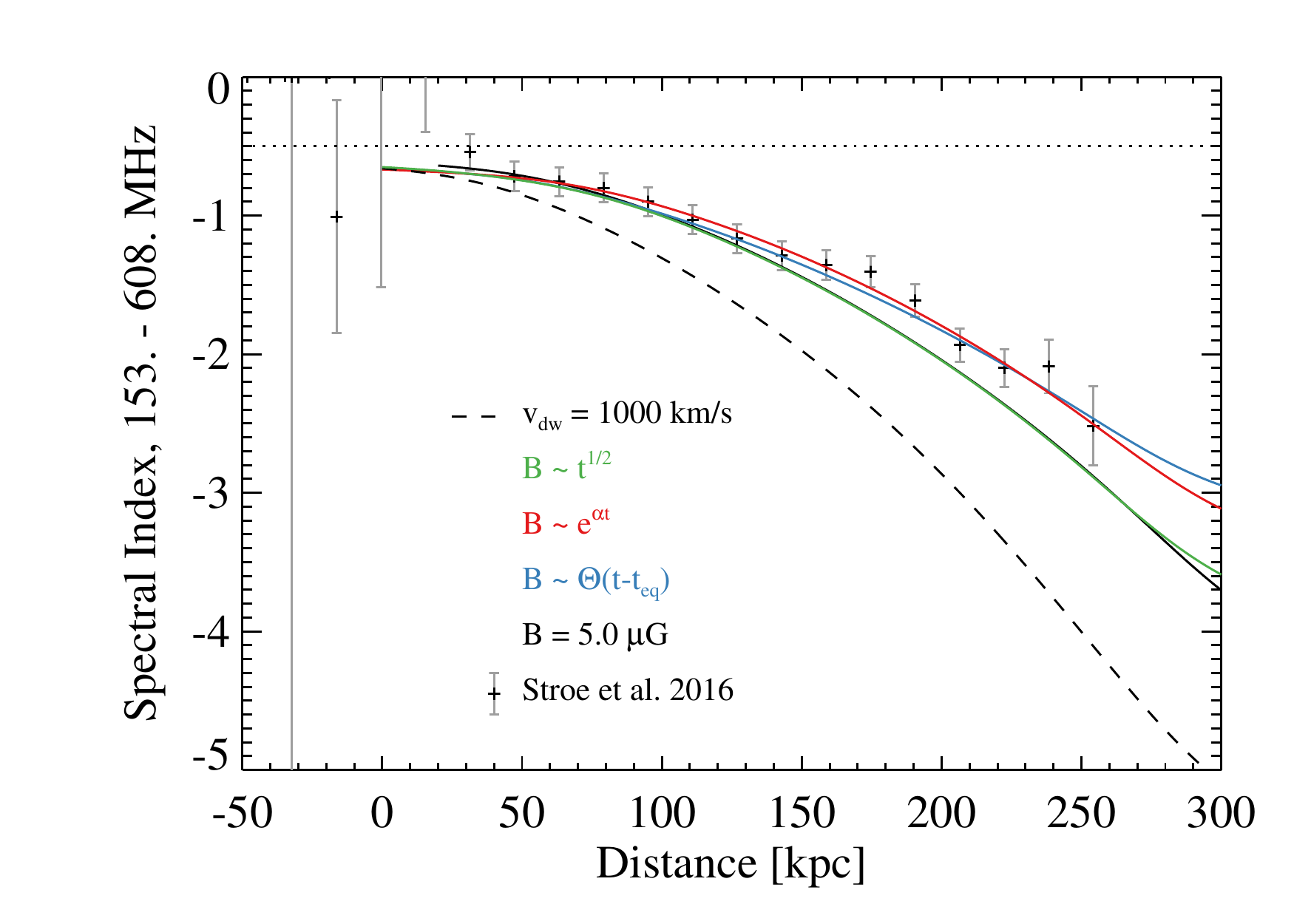}
	\caption{Spectral index over distance from the shock in kpc in the range  153-610 MHz. Black crosses are data derived from observations, with grey error bars \citep{2016MNRAS.455.2402S}. As usual, the standard model is shown in black, the step function model in blue, the exponential model in red and the linear energy model in green. All models imply a downwind speed of roughly 1200 km/s. For comparison we show the standard model with 1000 km/s as a black dashed line.}
	\label{fig:spectral_index_profile}
\end{figure}

In figure \ref{fig:spectral_index_profile}, we show the spectral index profiles obtained from GMRT data at 153 and 608 MHz \citep{2016MNRAS.455.2402S} and from the standard model (black line) and our three new models in the usual colors. We add the standard model with a downwind speed of 1000 km/s as a dashed black line. The error bars on the data include a $10\%$ error in flux scale. We convolve our models with a beam of  16 arcsec. We mark the flattest spectral index expected from simple DSA (-0.5) as a dotted horizontal line.   \par
Our models are well consistent with the data. The standard model with 1000 km/s downwind speed is not consistent with the data. We note that because the spectral index profiles are obtained only from two frequencies, they are very sensitive to errors in the flux scale and fluctuations/noise present in only one frequency.

For magnetic fields smaller than $B_\mathrm{IC}$ (IC dominated cooling), the shape of the spectral index profile primarily measures the downwind speed $v_\mathrm{dw}$. Decreasing this quantity will lead to a steepening in the spectral index profile. In our model, the spectral index profile strongly disfavours a downwind speed of less than 1000 km/s (black dashed line). This is consistent with models shown in \citet{2014MNRAS.445.1213S}, who conclude that the aging requires downwind speeds not smaller than 1200 km/s. In the absence of additional physics, recent measurements for the upstream temperature in the cluster disfavour a scenario with Mach numbers below 4: E.g. a Mach number of 3.5 would require an upstream temperature above 4 keV to lead to a downwind speed of 1200 km/s. This temperature is then not consistent with recent X-ray observations that measure 8 keV downstream \citep{2014MNRAS.440.3416O,2015A&A...582A..87A}. These problems could be alleviated by introducing CRe diffusion away from the shock or "in situ" re-acceleration downstream, this is however beyond the aim of our paper \citep{2016arXiv160207304F}. \par
At the same time, however, we note that if we assume the geometry of our (exponential) model, the temperature and density jump that are derived from X-ray observations are likely biased low, due to the expansion downstream, leading to a possible underestimation of the Mach number. Indeed, the downstream temperature of 8 keV has been measured by \citet{2015A&A...582A..87A} using an extraction region of the order of Mpc. According to the parameters that we have to assume in order to fit the radio properties of the relic, namely the e-folding time $\approx 240 Myr$, the downstream temperature is expected to decline from 15 to about 8 keV in about 100-150 kpc. 

\section{Discussion} \label{sect:discussion}

We find that the total integrated synchrotron spectrum is best fit by an \emph{exponential} increase in the magnetic field by a factor of 5-10 on a scale of $d_\mathrm{eq} = 40\,\mathrm{kpc}$ in front of the brightness peak of the relic at 2.1 GHz. This assumes a maximum thickness into the plane of the sky of $< 260 \,\mathrm{kpc}$, consistent with previous models. The exponential model is also roughly consistent with the brightness profiles and the spectral index profile. \par
The simple relic model, based on a constant magnetic field and diffusive shock acceleration (DSA) from the thermal pool, are ruled out given the observed spectrum and brightness profiles for our set of parameters. The linear model does not fit the data as well as the exponential model. \par 

The magnetic field strengths proposed here are roughly within theoretical expectations given the distance from the cluster centre: Faraday rotation measurements in clusters as well as cosmological MHD simulations find small fields at cluster outskirts. A correlation of the ICM magnetic field with the ICM thermal density $n_\mathrm{th}$ exists \citep{2001A&A...378..777D,2009MNRAS.392.1008D}:
\begin{align}
	B(r) &\propto  n_\mathrm{th}^\xi(r) \\
		&= B_0 \left(1 + \frac{r^2}{r_\mathrm{c}^2}\right)^{-\beta \xi}
\end{align}
where it is usually assumed that $\xi = 0.5$ or $\xi = 1$. Here $r$ is the radius from the cluster centre. For the standard $\beta$-model inferred from X-ray observations of clusters and magnetic field normalisation found in the Coma cluster \citep{2010A&A...513A..30B}, this motivates magnetic fields at the outskirts of the cluster ($d_\mathrm{center} = 1.5\, \mathrm{Mpc}$) of $0.1 - 1.2 \mu \mathrm{G}$. The amplification factor assuming a shock with Mach number 4.6 is $< 3.5$ (3.5 for purely perpendicular shocks where the upstream field is perpendicular to the shock normal, and about 2.3 considering isotropic compression). This means that additional mechanisms are necessary to explain the increase of the magnetic field downstream that is  required in our model to match radio data.

\subsection{Magnetic Field Increase}

In our new models, we assume an amplification of the magnetic field by a factor of 5-10 behind the shock on a length scale of roughly $d_\mathrm{eq} \approx 40$ kpc. Unfortunately, the amplification of magnetic fields behind shocks in the ICM is not well understood. This limits us to the following simple energy arguments. \par
We can compare the energy flux of the shock to the magnetic energy flux required for the amplification. Following  \citet{2010ApJ...715.1143F}, the energy flux available at the shock is given by: 
\begin{align}
	F_\mathrm{shock} &= \frac{1}{2} \rho_\mathrm{up} v_\mathrm{shock}^3 \left(1 - \frac{1}{\sigma^2}\right)\label{eq.Ediss} \\
					 &= 1 \times 10^{-2} \,\frac{\mathrm{erg}}{\mathrm{cm}^2 \mathrm{sec}},\nonumber
\end{align}
where $\rho_\mathrm{up} = 3.04\times 10^{-28} \,\mathrm{g}/\mathrm{cm}^3$ is the upstream density and $v_\mathrm{shock} = 4149 \,\mathrm{km}/\mathrm{s}$ is the shock speed in the upstream medium. The increase in magnetic energy in the exponential model results in an approximate increase in magnetic energy flux:
\begin{align}
	F_\mathrm{mag} &\approx v_\mathrm{dw}(0) \left(B_\mathrm{max}^2 - B_\mathrm{min}^2\right)/8\pi  \\
								&= 4.2\times10^{-5} \,\mathrm{erg}/\mathrm{cm^2 sec}, \nonumber
\end{align}
which is a factor 20 below the shock energy flux. Hence, in principle the field amplification is energetically possible. \par
If we for the moment assume the magnetic field increase is due to solenoidal turbulence and an MHD dynamo, we can derive combined constrains on the turbulent injection scale $l_0$, the dynamo efficiency $\eta_\mathrm{B}$  and the magnetic field increase $\Delta B^2 = B_\mathrm{max}^2 - B_\mathrm{min}^2$. The energy flux into turbulence is then:
\begin{align}
	\eta_\mathrm{t} \frac{1}{2} \rho_\mathrm{up} v_\mathrm{up}^3 \left( 1 - \frac{1}{\sigma^2} \right) &= \frac{1}{2} \rho_\mathrm{dw} v_\mathrm{dw} V_0^2, \label{eq:cascade}
\end{align}
where $V_0$ is the turbulent velocity at injection scale $k_0$. If we assume $\eta_\mathrm{t} = 0.05$ is the fraction of energy flux available at the shock that is converted into turbulence downstream:
\begin{align}
	V_0 &= \eta_\mathrm{t}^{1/2} v_\mathrm{up} \left( 1- \frac{1}{\sigma^2} \right)^{1/2} \label{eq.v0} \\
		&\approx 890 \,\mathrm{km}/\mathrm{s}\left( \frac{\eta_\mathrm{t}}{0.05} \right)^{1/2} \nonumber
\end{align}
With the Alven speed $v_\mathrm{A} = \Delta B / \sqrt{4\pi\rho_\mathrm{dw}} \approx 200 \,\mathrm{km}/\mathrm{s}$, we obtain the Alvenic Mach number $M_\mathrm{A} = V_0 /V_\mathrm{A}$:
\begin{align}
	M_\mathrm{A} &\approx 2.9 \left( \frac{\eta_\mathrm{t}}{0.05}\right)^{1/2} \left( \frac{\Delta B}{3 \,\mu\mathrm{G}} \right)^{-1}
\end{align}
Hence turbulence is super-Alfvenic at the shock and then gradually becomes quasi-Alfvenic with distance. For solenoidal turbulence, the energy flux transported by the turbulent cascade to smaller scales is given by:
\begin{align}
	\Phi_\mathrm{casc} &\approx \rho_\mathrm{dw} V_0^3 k_0.
\end{align}
A fraction of this flux, $\eta_\mathrm{B}$, can be converted into magnetic field amplification through turbulent dynamo:
\begin{align}
	\frac{\mathrm{d}E_\mathrm{B}}{\mathrm{d}t} \approx \frac{\Delta B^2}{8\pi\Delta t} &= \eta_\mathrm{B}  \rho_\mathrm{dw}V_0^3 k_0. \label{eq:dynamo}
\end{align}
If we assume that $\eta_\mathrm{B} = 0.05$ of the turbulent energy is converted into magnetic energy, combining eq. \ref{eq:dynamo} and \ref{eq.v0} leads to an injection scale $l_0$ of: 
\begin{align}
	k_0^{-1} \propto l_0 &= \eta_\mathrm{B} \eta_{t}^{3/2} \rho_\mathrm{dw} v_\mathrm{up}^3 \left(1-\frac{1}{\sigma^2} \right)^{3/2} \frac{8\pi\Delta t}{\Delta B^2} \label{eq.l0}\\
	&\approx 5 \,\mathrm{kpc} \left(\frac{\eta_b}{0.05}\right) \left(\frac{\eta_\mathrm{t}}{0.05}\right)^{3/2} \left[ \left(\frac{\Delta B}{3\,\mu\mathrm{G}} \right)^2/\Delta t_\mathrm{eq}\right]^{-1}, \nonumber
\end{align}
which is a reasonable value. Here $\Delta t_{eq} = d_\mathrm{eq}/v_\mathrm{dw}(0)$. In order to achieve efficient amplification we require that turbulent motions reach the MHD scale, $l_\mathrm{A} = l_0 M_\mathrm{A}^{-3}$, where $M_\mathrm{A}$ is the Alfvenic Mach number, i.e. the turbulent cascading time scale, $\sim l_0/V_0$, should be smaller than the crossing time-scale, $d_\mathrm{eq}/V_\mathrm{dw}$. This requires $ \frac{l_0}{V_0} \le \Delta t_{eq}$, implying (eq. \ref{eq.v0} and \ref{eq.l0}):
\begin{align}
	\frac{\eta_\mathrm{t}}{0.05} &\le 3.5 \left(\frac{\eta_\mathrm{B}}{0.05}\right)^{-1}.
\end{align}
This provides a condition between turbulent efficiency and dynamo that is satisfied under our assumptions. As soon as the energy density of magnetic fields reaches a substantial fraction of the energy associated with turbulent motions (i.e. for $M_A$ reaching unity) magnetic field tension prevents turbulent motions and the amplification process slows down. In our scenario this situation would be realised at distance $\sim d_\mathrm{eq}$, at these distances magnetic field energy density saturates and the field strength starts declining due to adiabatic expansion. Above estimates demonstrate that the magnetic field luminosity required by the model is not in conflict with basic energy constraints, turbulent scales and relevant time-scales. \par 
An interesting outcome of this scenario is that during the initial phases of amplification the magnetic field topology is highly turbulent, the Alfvenic Mach number being very large.\par
This implies that the synchrotron emission at higher frequencies is not expected to be strongly polarised. On the contrary, stronger polarization is expected when the magnetic field reaches quasi-equipartition, at scales $\approx d_\mathrm{eq}$, where the bulk of the emission at lower (GHz) frequencies is produced in a strong field that is advected and possibly stirred downstream. Hence in our model, intrinsic polarisation in the relic depends on distance from the shock and frequency. This prediction can be tested with modern radio interferometers.

\subsection{Magnetic Field Decrease}

The magnetic field decrease is constrained by the decline in brightness behind the peak of the emission in the relic. The effect of the magnetic field itself on the CRe cooling is rather small, hence the field is constrained well, especially at low frequencies, where CRe cool slowly. \par
Several mechanisms may contribute to this decline, dissipation of turbulent motions, magnetic reconnection or adiabatic expansion. We limit our considerations here to the latter, because it is physically unavoidable and easy to model. \par
Our best-fit exponential model self-consistently includes adiabatic expansion along the shock normal. This result suggests that expansion indeed plays an important role in shaping the relic emission. The linear model reproduces the brightness profiles nearly as well, however with more dissipation of magnetic energy.  The data at 150 MHz rule out the standard model with constant density and a constant magnetic field of $5 \,\mu\mathrm{G}$. 

\section{Acceleration Efficiency in the Northern Relic of CIZA J2242.8+5301} \label{sect.eff}

\begin{table}
	\centering
	\begin{tabular} {r|c|c|c|c}
		\hline 
		Model & $n_0$ & $n_\mathrm{CRe} $ & $\epsilon_\mathrm{CRe}/\epsilon_\mathrm{th,dw}$ & $\eta_\mathrm{KR}$ \\\hline
		Standard  		& 1.62 	& 0.61	& 0.002 &  0.0013 \\
		Step-function  	& 4.8 	& 1.79 	& 0.009 & 0.0039 \\
		Exponential  	& 9.6 	& 3.59	& 0.012 & 0.0079 \\
		Linear 			& 5.4 	& 2.02	& 0.007 & 0.0044 \\
		\hline 
	\end{tabular}
	\caption{Normalisation of the CRe spectrum $n_0$ in $ 10^{-28} \left(\frac{\mathrm{g\,cm}}{\mathrm{sec}}\right)^{1-s} \mathrm{cm}^{-3}$. We add the number density in $10^{-9}\,\mathrm{cm}^{-3}$, the energy density fraction relative to the downstream thermal energy density of $\epsilon_\mathrm{th,dw} = 6.0\times10^{-11}\mathrm{erg}/\mathrm{cm}^3$ and the injection efficiency without pre-existing CRe (eq. \ref{eq.eta_KR}). We use a minimum momentum of $0.1 \,m_\mathrm{e}c$ .} \label{tab:cre_norm}
\end{table}

We now derive the required CRe density at injection using the new models for the underlying magnetic field. We also set constrains on the pre-existing CRe population required for shock re-acceleration to be efficient. \par
We can estimate the injection efficiency of relativistic particles behind the shock directly from our models. To do so, we first have to define a minimum momentum of cosmic-rays to consider. We set this momentum to $p_0 = 0.1\,m_\mathrm{e}c$.
\begin{align}
	n_\mathrm{CRe} &= \int\limits_{p_0}^{\infty} n_0 p^{-s} \,\mathrm{d}p = \frac{n_0}{s-1} p_0^{1-s} \label{eq:ncre}\\
	&=  9.3\times 10^{20}  n_0  \left(\frac{\mathrm{g\,cm}}{\mathrm{sec}}\right)^{1-s} \, \nonumber \\
	\epsilon_\mathrm{CRe} &=  \int\limits_{p_0}^{\infty} n_0 p^{-s} E \,\mathrm{d}p = n_0\frac{m_\mathrm{e}c^2 }{s-1}  \times \nonumber\\
	 &\times \left[  -p^{1-s} {}_{2}F_1\left(-\frac{1}{2},\frac{1-s}{2}; \frac{3-s}{2}; -\frac{p^2}{m_\mathrm{e}^2c^2}\right) \right]^\infty_{p_0} \label{eq:epscre}\\\
	&= 7.6\times 10^{14} n_0  \left(\frac{\mathrm{g\,cm}}{\mathrm{sec}}\right)^{1-s} \mathrm{erg} \,   \nonumber
\end{align}
, where ${}_2F_1(a,b;c;x)$ is the hypergeometric function \citep{1970hmfw.book.....A}. From the shock properties we find a downstream thermal number density of $n_\mathrm{th,dw} = 5.6\times10^{-4} \,\mathrm{cm}^{-3}$ and a thermal energy density of $\epsilon_\mathrm{th,dw} = 3 n_\mathrm{dw}k_\mathrm{Boltz}T = 6 \times 10^{-11}\mathrm{erg}/\mathrm{cm}^3$.
The injection efficiency relative to energy flux of the shock $\eta_\mathrm{flux}$ then follows by multiplying the CRe energy density eq. \ref{eq:epscre} with the shock speed in the upstream medium. \par
We give our values for the normalisation, CRe number density, CRe energy density and injection efficiency before cooling in table \ref{tab:cre_norm}. The energy densities are larger than in scenarios of SNR, where proton to electron ratios of $X_\mathrm{p/e} < 0.01$ were found \citep{2012A&A...538A..81M}. 

\subsection{Shock Re-acceleration}

Adiabatic compression of a pre-existing CRe population can possibly solve the efficiency problems encountered in recent models for relics \citep[][]{2015ApJ...809..186K}. Here we derive constrains on the energy density of pre-existing CRes before the shock in the northern relic of CIZA J2242.8+5301. \par
In DSA simulations, the efficiency for CRe injection $\eta_\mathrm{KR}$ in the frame of the upstream medium is defined as \citep{2013ApJ...764...95K}:
\begin{align}
	\eta_\mathrm{KR}(M) &= \frac{2 v_\mathrm{dw}}{\rho_\mathrm{up}v_\mathrm{shock}^3} \left[ \epsilon_\mathrm{CRe,dw} - \epsilon_\mathrm{CRe,up} \left(\frac{\rho_\mathrm{up}}{\rho_\mathrm{dw}}\right)^{\gamma_\mathrm{CR}} \right] \label{eq.eta_KR}
\end{align}
where $\gamma_\mathrm{CR} = 4/3$ is the adiabatic index of CRe and $\epsilon_\mathrm{CRe,up}$ and $\epsilon_\mathrm{CRe,dw}$ are the upstream and downstream CRe energy densities, respectively.  \par

\begin{figure}
	\centering
	\includegraphics[width=0.45\textwidth]{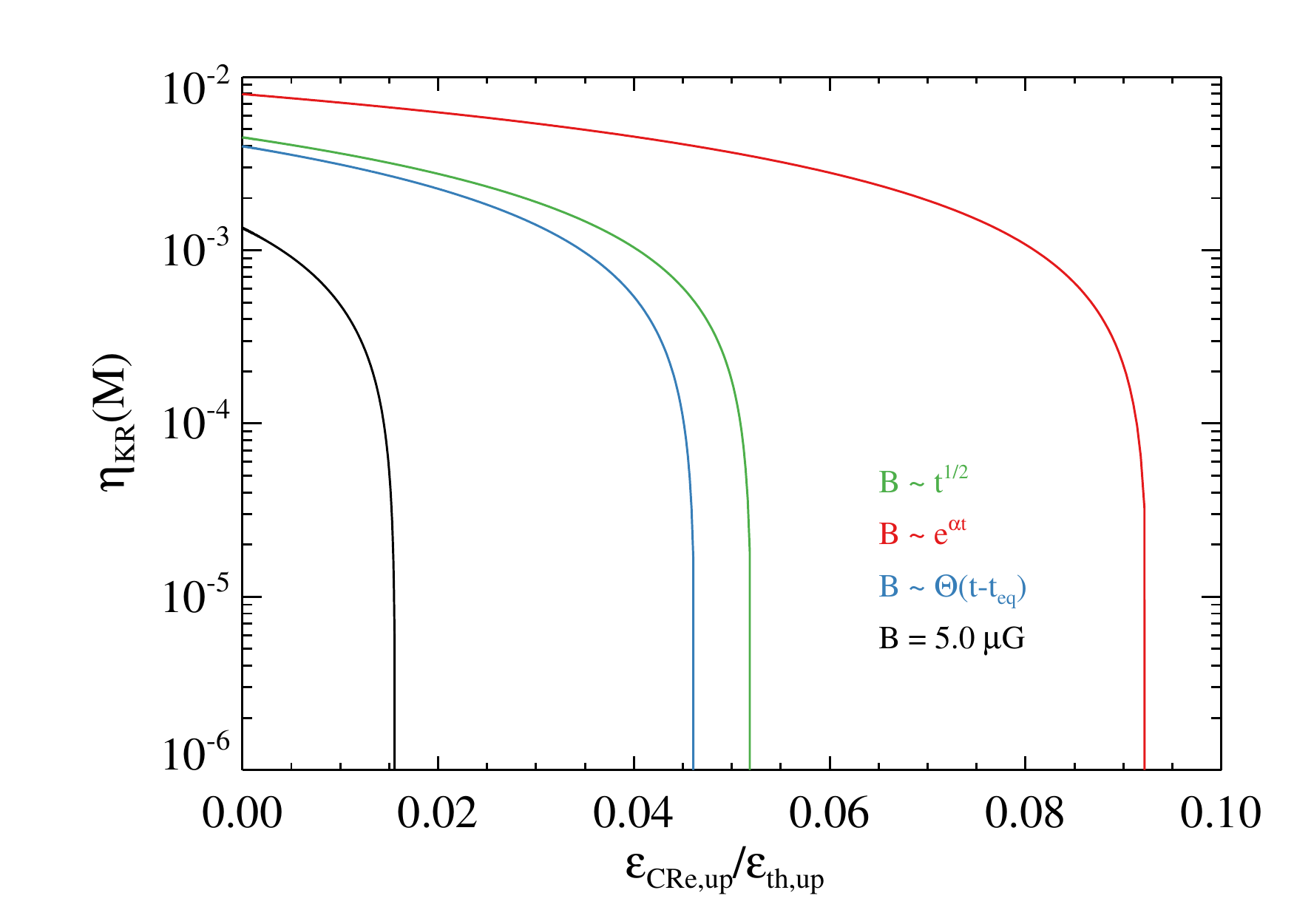}
	\caption{Required acceleration efficiency over pre-existing CRe energy density to match the observed relic. The  pre-existing CRe energy density is shown in units of the upstream thermal energy density. }
	\label{fig:eta_eps}
\end{figure}

In figure \ref{fig:eta_eps} we show the required acceleration efficiency $\eta_\mathrm{KR}(M)$ over the energy density in CRe ahead of the shock $\epsilon_\mathrm{CRe,up}$ for the relic. We plot the upstream CRe energy density in units of the upstream thermal energy density. For energy densities of 0.1 percent the required efficiency drops very quickly for all models, because of adiabatic gains in energy. This compression sets an upper limit on the pre-existing CRe energy density before the shock, less than one percent of the upstream thermal energy density all models. \par
This demonstrates that compression can reduce the required energy flux and solve the acceleration efficiency problem, even if only very moderate amounts of CRe are present in front of the shock \citep{2015ApJ...809..186K}. As shown in section \ref{sect:lifetimes}, CRe lifetimes at the location of the relic are around half a Hubble time, thus a radio dark population of CRe can accumulate in the outskirts of a cluster during its formation from z = 1. This pre-existing population of CRe is likely to have low momenta between 10 and 100 $\mathrm{m}_\mathrm{e}\mathrm{c}$, because at these momenta the lifetime shown in figure \ref{fig:lifetime} peaks. For the reasonable assumption of turbulent diffusion of CRe, this also means large diffusion lengths in the Mpc regime and smooth spatial distributions of the population. \par
This scenario differs from the scenario of localized ''CRe clouds'' deposited by radio galaxies and re-accelerated by shocks to form relics \citep[e.g.][]{2015ApJ...809..186K,2016arXiv160203278K}. The two scenarios could explain the morphological differences between the northern and the southern relic in CIZA J2242.8+5301: The northern source population of CRe was accumulated over Gyrs and hence is smooth and leads to the observed regular relic morphology, while the southern source population is younger and originates from outflows of nearby radio galaxies. \par
In the centre of clusters, lifetimes less than one Gyr imply that quiet or cool core clusters can in principle be rather depleted of CRe even at these low energies, potentially making relic formation from an old population more difficult.

\subsection{Comparison to Previous Models} \label{sect:alt_models}

\begin{figure}
	\centering
	\includegraphics[width=0.45\textwidth]{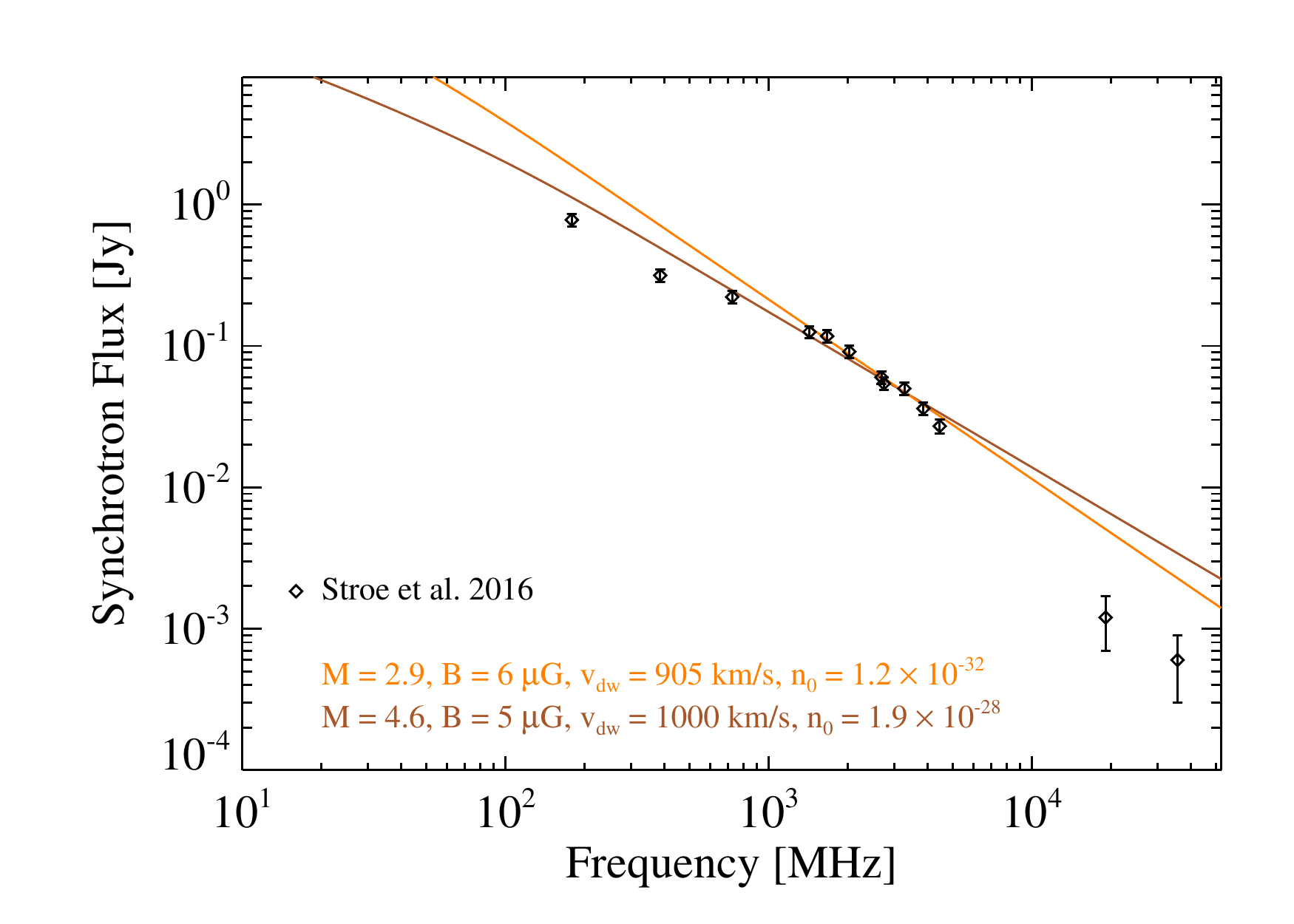}
	\caption{Total synchrotron spectra from our formalism using parameters of previous models from \citet[][ brown]{2010Sci...330..347V} and \citet[][ orange]{2014MNRAS.445.1213S}.}
	\label{fig:alt_synchro_spectrum}
\end{figure}

In their discovery paper, \citet{2010Sci...330..347V} modelled the relic with $5 \,\mu \mathrm{G}$, a shock speed of 1000 km/s and a Mach number of 4.6. In a more recent study, \citet{2014MNRAS.445.1213S} conducted detailed spectral age modelling of the relic, assuming a constant magnetic field of $6 \,\mu\mathrm{G}$. They find a best fit of a lower Mach number of 2.9 and aging speeds around $905 \,\mathrm{km}/\mathrm{sec}$, however leaving the CRe normalisation $n_0$ a free parameter. We stress that their formalism is \emph{fundamentally different} from the approach taken in this work, where we assume a time independent CRe normalisation $n_0$ along the shock normal. \par
Assuming our formalism, we show the total synchrotron spectrum from both models alongside the observed spectrum from \citet{2016MNRAS.455.2402S} in figure \ref{fig:alt_synchro_spectrum}. Both parameter sets do not fit the steepening above 3 GHz as expected. \citet{2010Sci...330..347V}'s model gives a good fit to the data below 3 GHz, while \citet{2014MNRAS.445.1213S} model is too steep and does not fit the data at 150 and 320 MHz, probably because we left $n_0$ constant downstream. See also appendix \ref{app.Mach}.\par

\section{Conclusions}\label{sect.conlusions}

We have proposed to attribute the spectral steepening in radio relics to higher energy cosmic-ray electrons preferentially emitting in lower magnetic fields than the bulk of the CRe. We then presented a first model for a radio relic considering time dependent magnetic fields and adiabatic expansion of the thermal plasma behind the shock. Our model requires a gradual amplification of the magnetic field in a small region downstream of the shock and exceeds expected field strengths at the shock due to compression.  \par
We focused our efforts on the prototype of radio relics, the ''Sausage'' relic in CIZA J2242.8+5301. We developed the formalism to obtain cooling CRe spectra and synchrotron spectra for such models. Considering standard parameters for the northern shock in CIZA J2242.8+5301, we find that
\begin{enumerate}
	\item the standard model of the relic using a constant magnetic field of $5\,\mu\mathrm{G}$ is inconsistent with the observed brightness profile at 150 MHz. The standard downstream speed of 1000 km/s is inconsistent with the spectral index profiles between 150 and 610 MHz. The profiles support a downstream speed of 1200 km/s.
	\item Behind the brightness peak, we find a best fit model where the magnetic field declines exponentially alongside adiabatic expansion of the gas with an e-fold distance of 240 kpc.
	\item the observed radio synchrotron spectrum and the brightness profiles are consistent with a shock Mach number of 4.6 and an exponentially increasing magnetic field from 0.3 to 3.0 $\mu\mathrm{G}$ in the 40 kpc before the brightness peak.  The model predicts a shift in the brightness peak of the relic observing frequencies above 16 GHz. It also predicts that intrinsic polarised intensity depends on the distance from the shock and on observing frequency, which can be used to falsify the model observationally. 
	\item an interpretation of the increase as a turbulent dynamo, caused by a short burst of turbulence, yields an injection scale of about 10 kpc, if 5 \% of the shock energy is converted into turbulence and 5\% of the turbulent energy drives the dynamo. We have also shown that the picture does not appear inconsistent with relevant quantities and time-scales.
	\item relatively large acceleration efficiencies are required by our model provided that electrons are accelerated from the thermal pool.  However, plausible energy densities ($<1\%$) of pre-existing CRe before the shock can at least reduce the required efficiency, due to adiabatic compression by the shock. We find such a scenario well motivated: The expected lifetime of CRe  in the cluster outskirts is very long: $\approx 7\,\mathrm{Gyrs}$ at $100 \,\mathrm{m}_\mathrm{e}\mathrm{c}$. Hence, CRe are able to accumulate  over the lifetime of the cluster in its outskirts as well.  
\end{enumerate}

\section{Acknowledgements}\label{sect.ack}

JD would like to thank T. Jones, R. van Weeren, H. Intema, A. Beck and M.Hoeft for discussions regarding relics, dynamos and this paper. We thank R. van Weeren for providing the GMRT brightness profile. JD acknowledges support from the People Programme (Marie Sklodowska Curie Actions) of the European Union's Eighth Framework Programme H2020 under REA grant agreement no. [658912]. HR acknowledges support from the ERC Advanced Investigator programme NewClusters 321271. 

\bibliographystyle{mn2e} \bibliography{master}

\appendix

\section{Brightness Profiles}

\begin{figure*}
	\centering
	\includegraphics[width=0.4\textwidth]{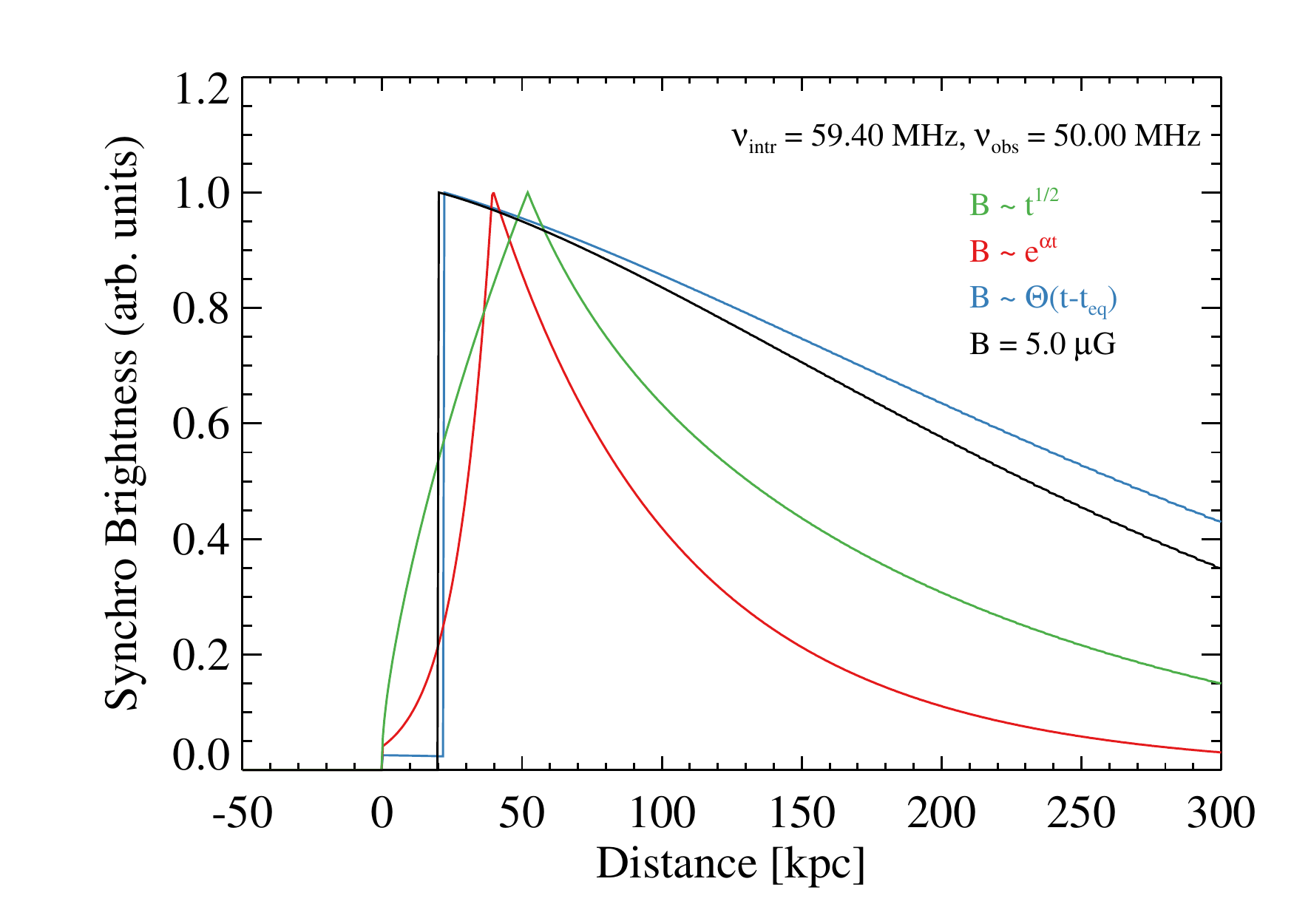}
	\includegraphics[width=0.4\textwidth]{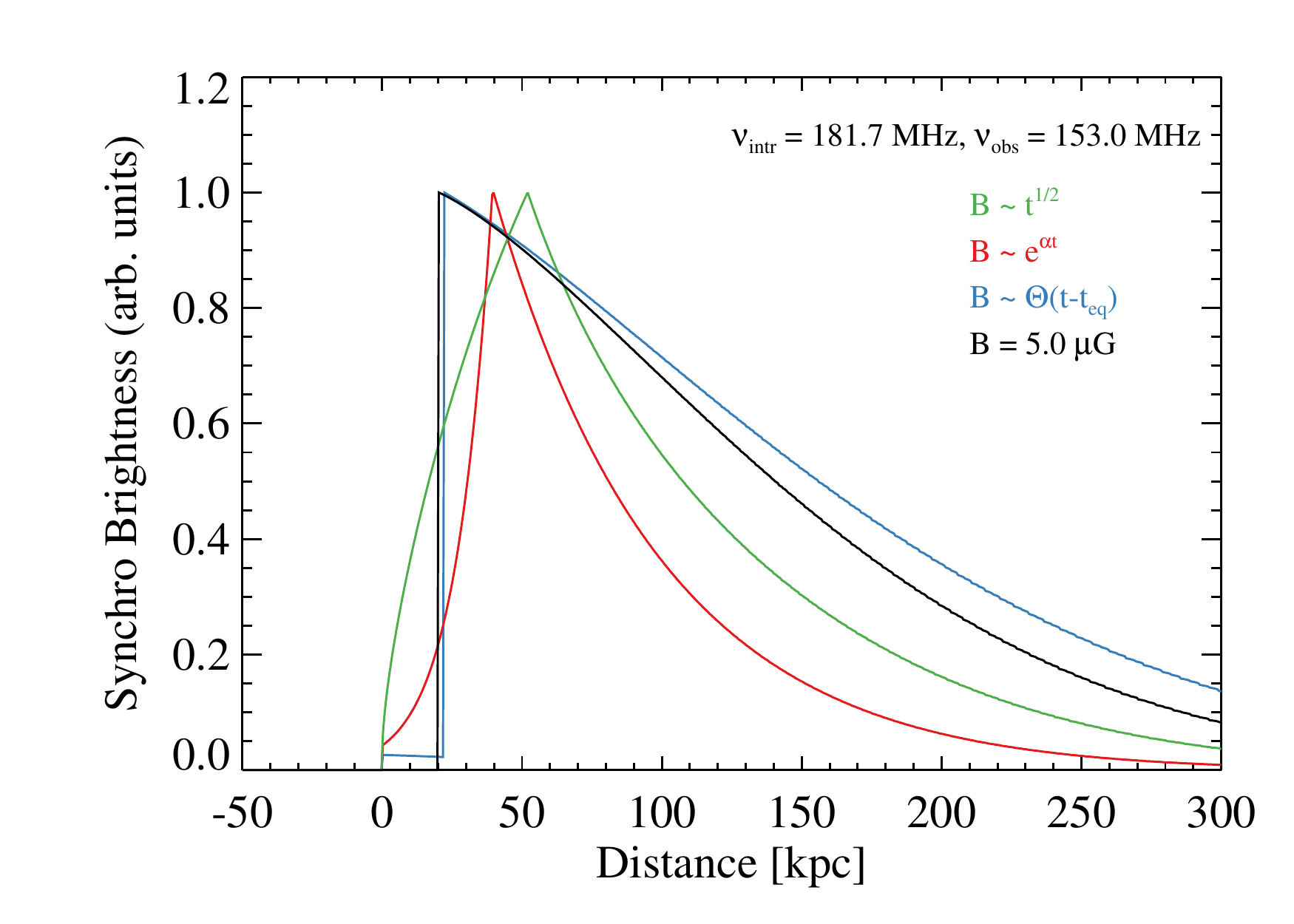}\\
	\includegraphics[width=0.4\textwidth]{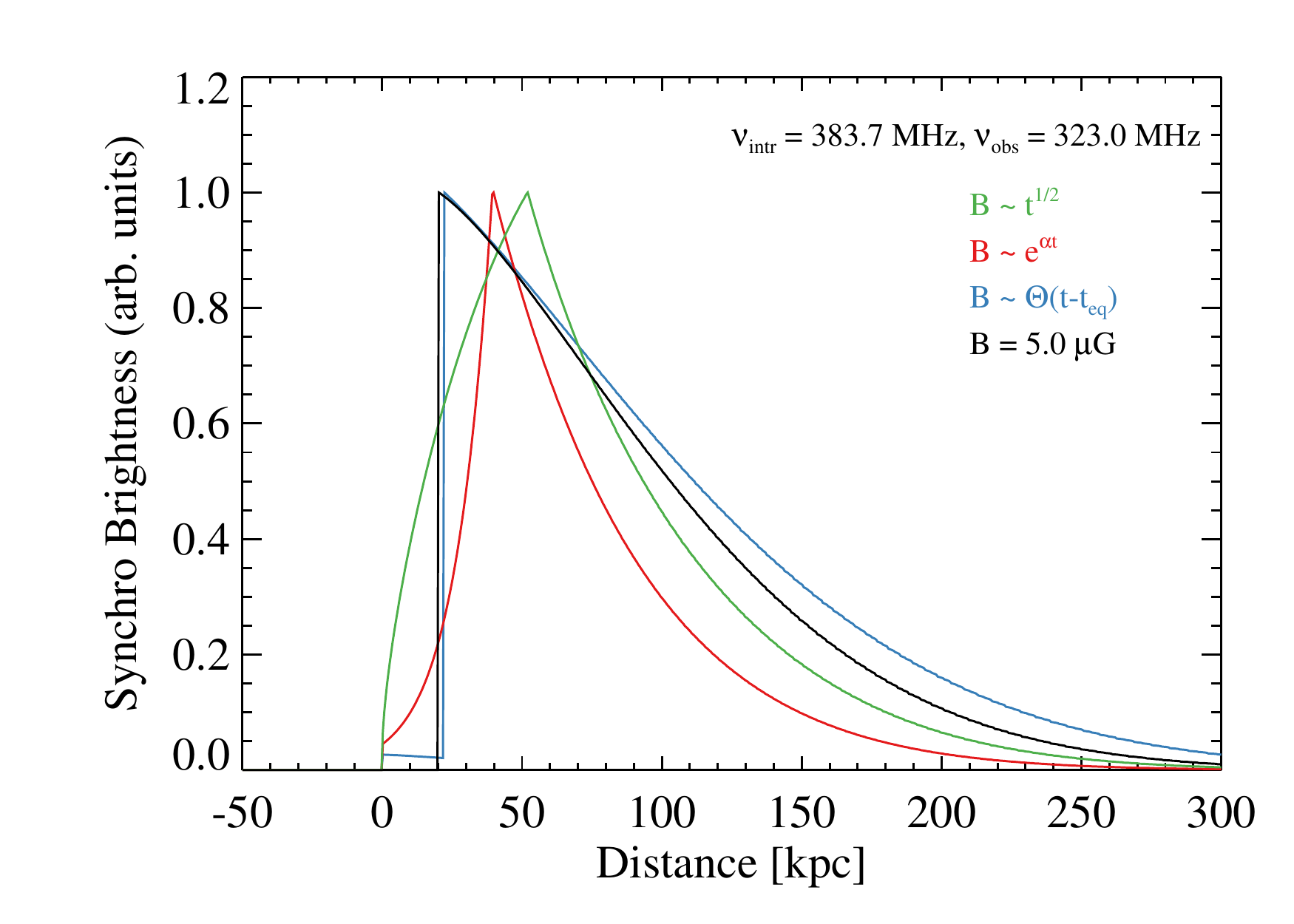}
	\includegraphics[width=0.4\textwidth]{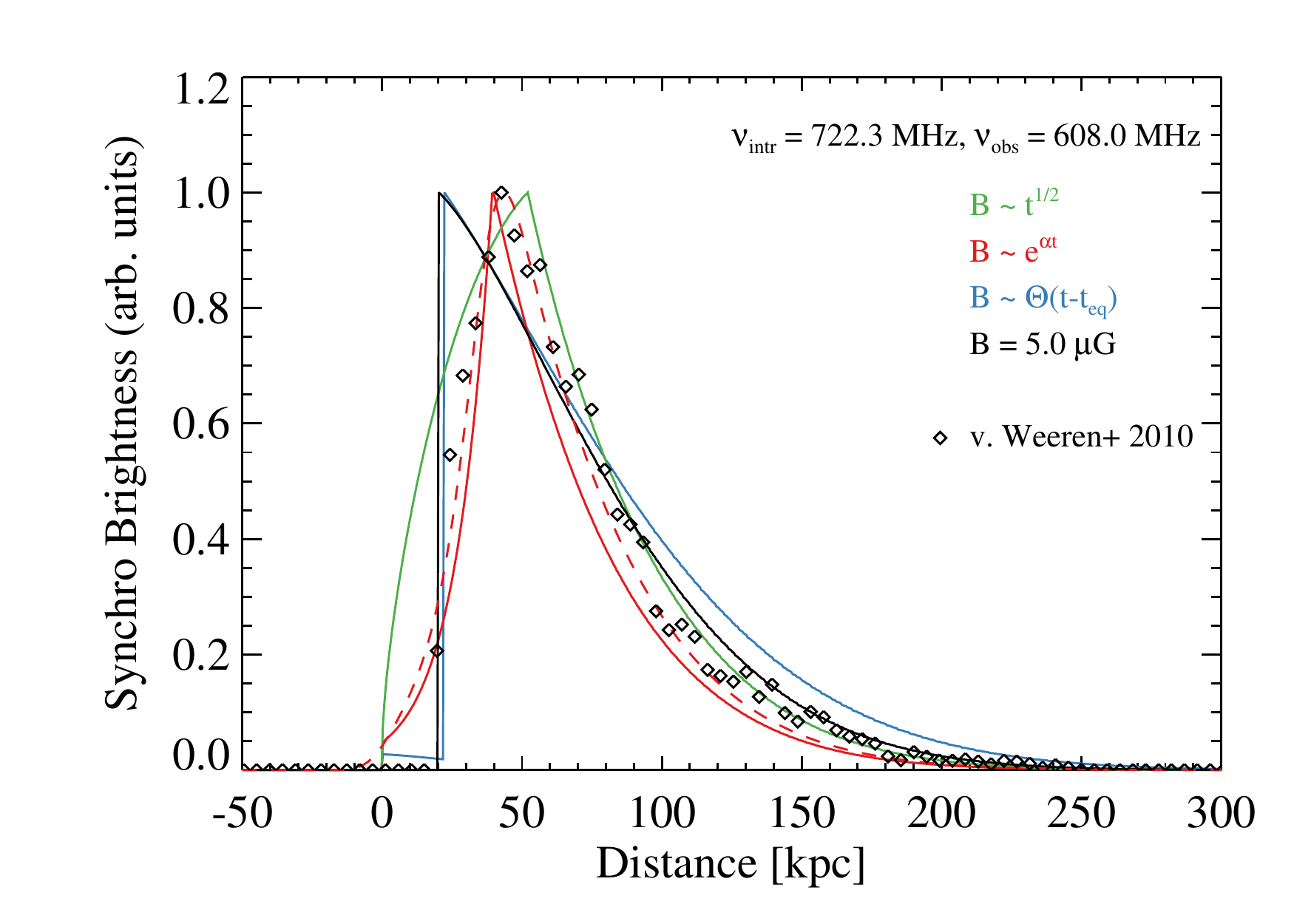}\\
	\includegraphics[width=0.4\textwidth]{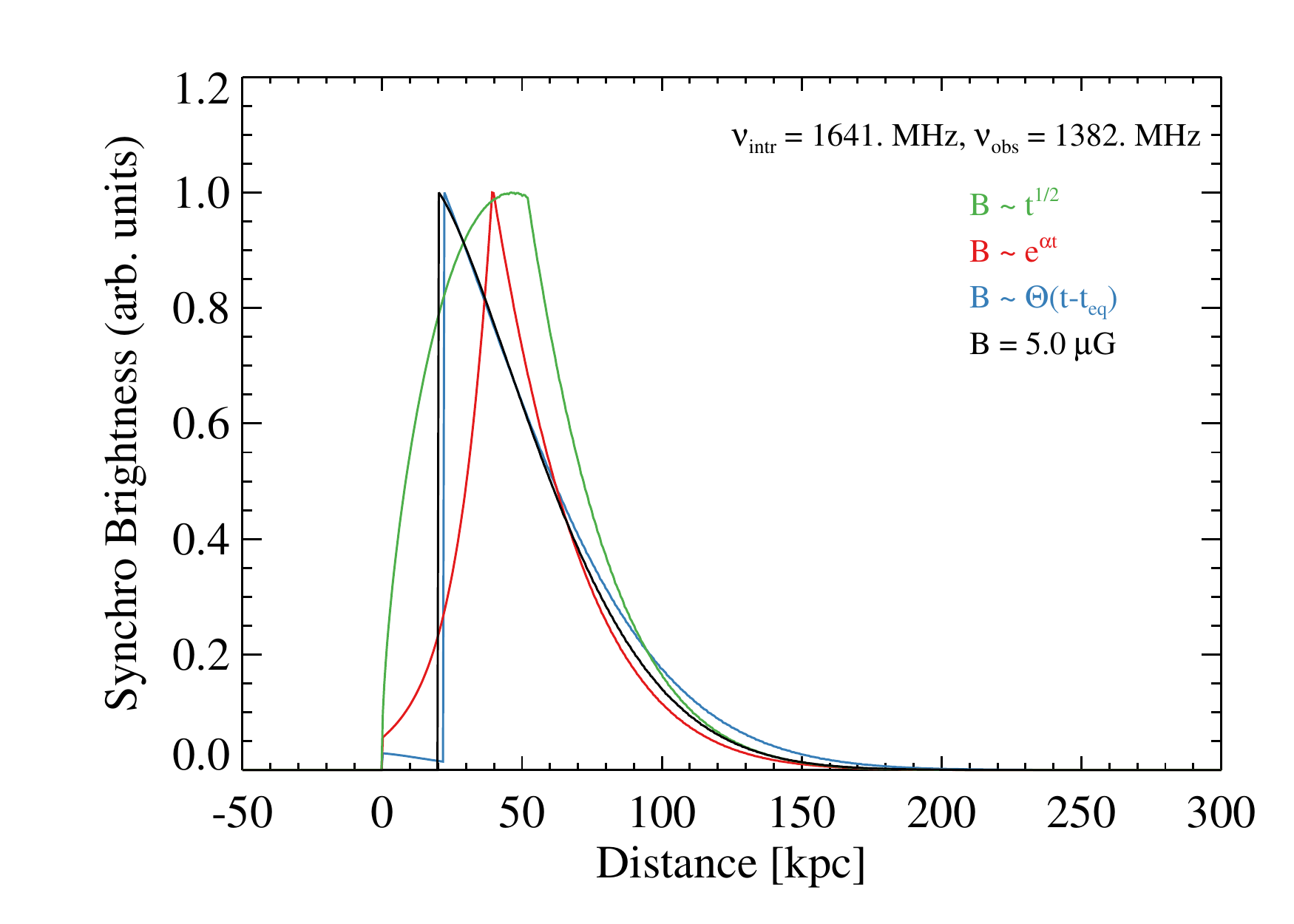}
	\includegraphics[width=0.4\textwidth]{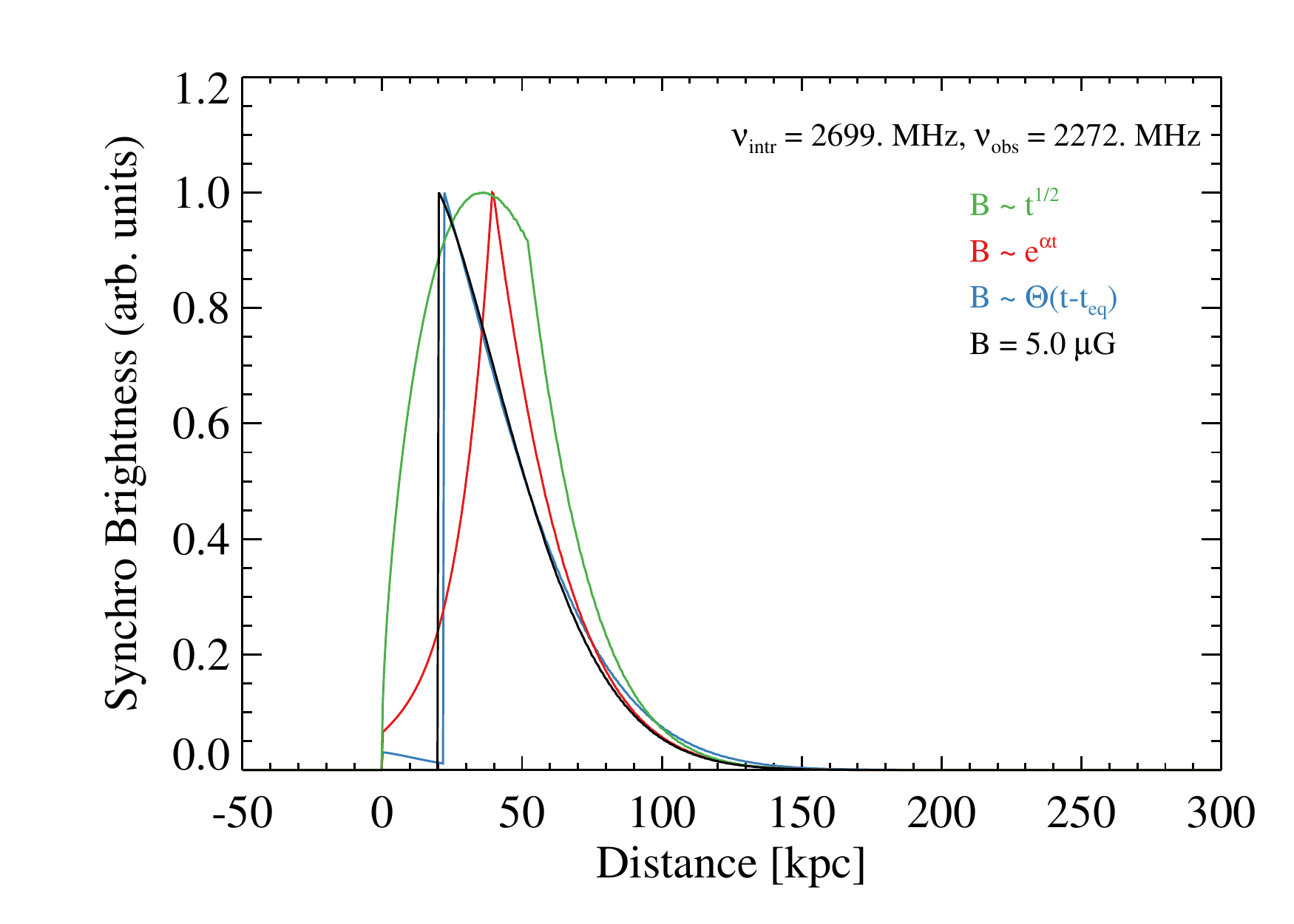}\\
	\includegraphics[width=0.4\textwidth]{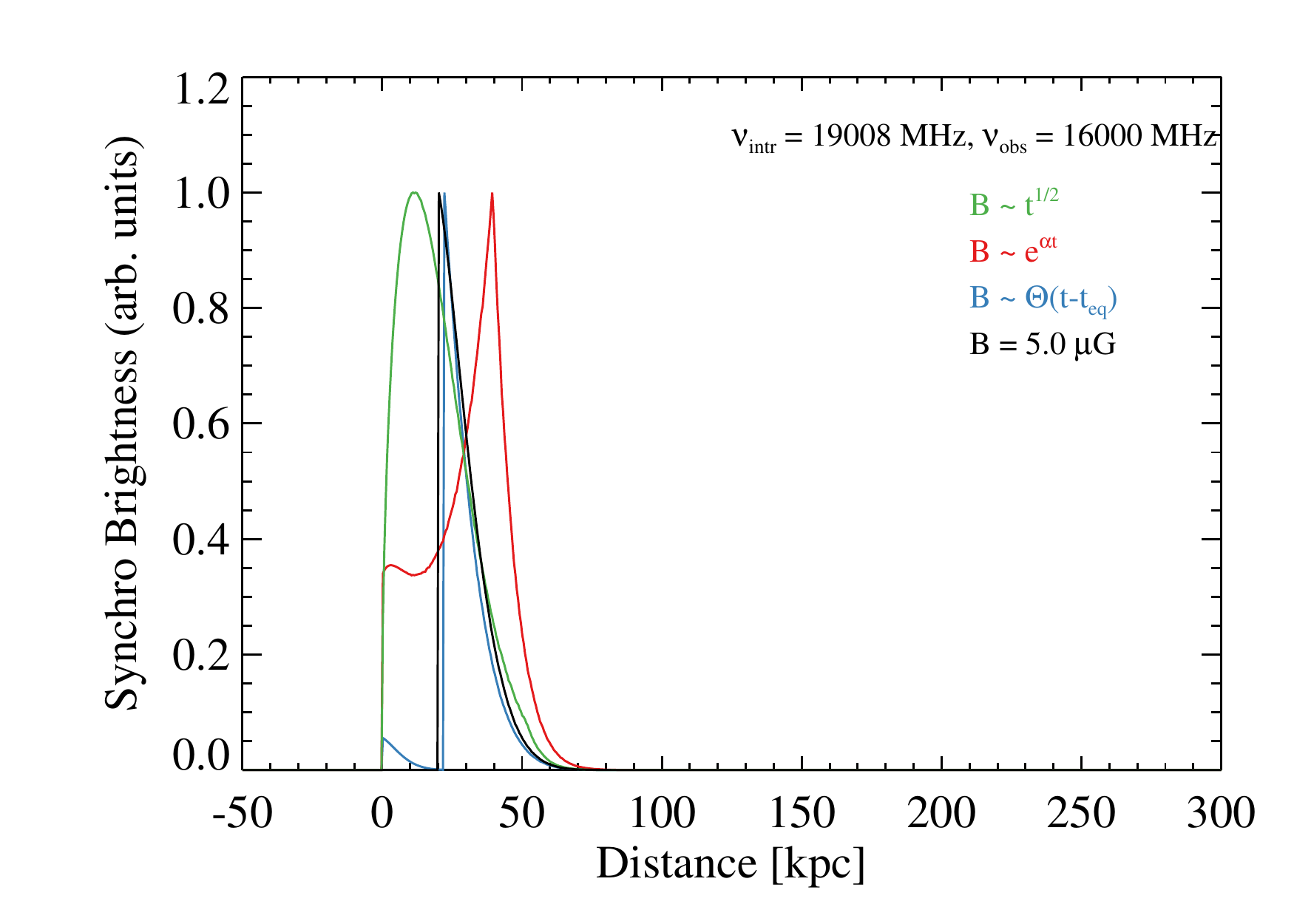}
	\includegraphics[width=0.4\textwidth]{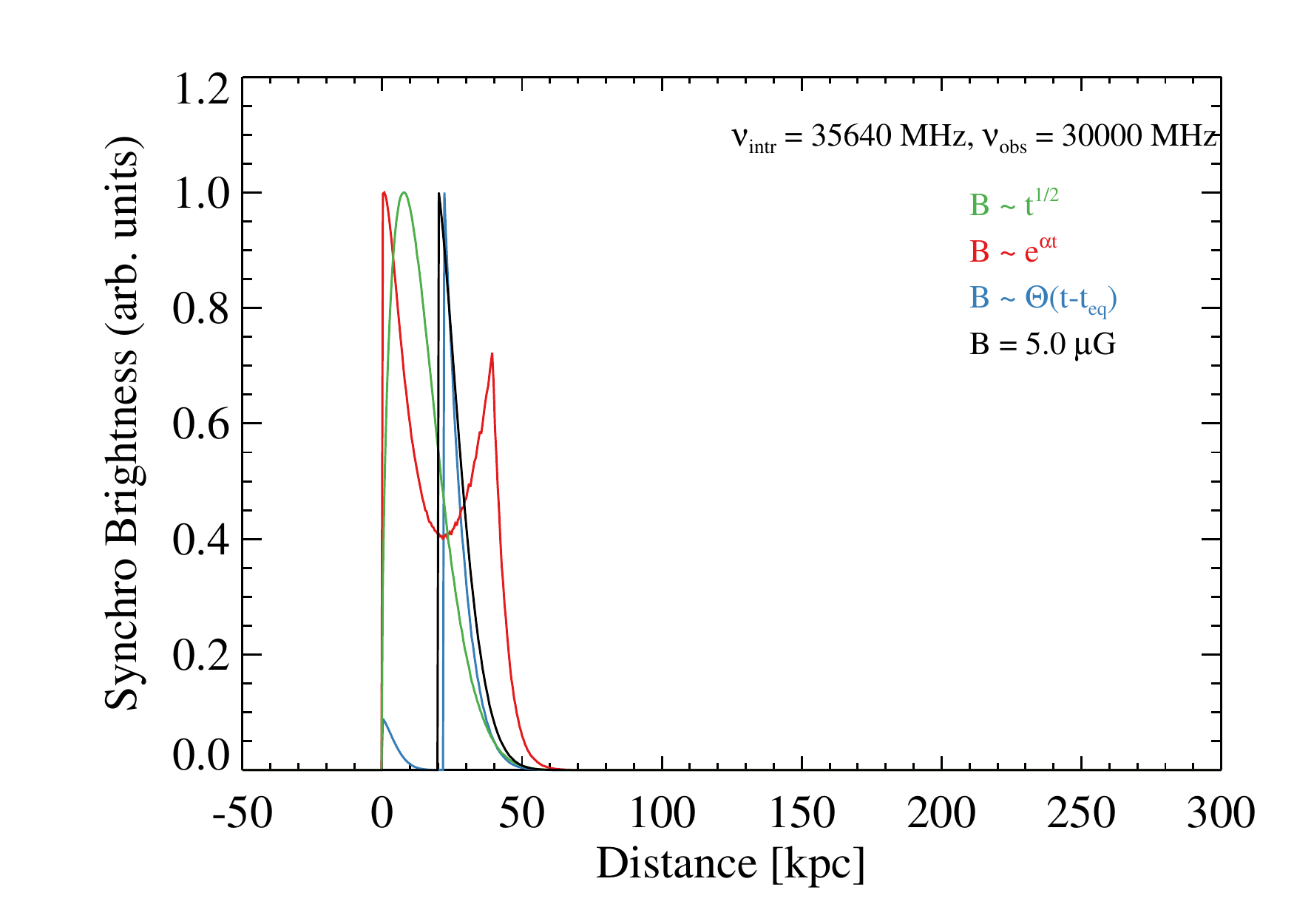}
	\caption{Normalised relic brightness from the models over distance perpendicular to the relic in kpc. The model was computed at eight intrinsic frequencies (top left to bottom right: 59.50, 182, 384.3, 723.5, 1644, 2703, 19040, 35700 MHz), corresponding to observed frequencies: 50, 153, 323, 608, 1382, 2272, 16000, 30000 MHz. The standard model is shown in black with grey errors, the step function model in blue, the exponential model in red and the linear model in green. In the third graph at 723.5 MHz we add the exponential model convolved with a 12 kpc beam as red dashed line. Black diamonds are deconvolved WSRT observations from \citet{2010Sci...330..347V}.  }
	\label{fig:brightness_profiles}
\end{figure*}

For completeness, we report here the brightness profiles reported in figure \ref{fig:conv_brightness_profiles} without convolving with the observed beam. In figure \ref{fig:brightness_profiles}, we show these predicted relic brightness in arbitrary units over distance.
 In the third panel we also show the deconvolved profile observed at 610 MHz from \citet{2010Sci...330..347V} as black diamonds. In that panel we also plot the exponential model convolved with a Gaussian with FWHM of 12 kpc.  We find a reasonable match to the deconvolved observed profile at 610 MHz, especially considering that the deconvolution is probably not perfect. At frequencies above 2 GHz the predicted relic emission becomes only a few ten kpc wide. Additionally, the linear model predicts a shift in the brightness peak at these frequencies compared to frequencies below 2 GHz. This motivates high resolution observations of the Sausage or similar relics to observe the shift in the brightness profile. 

\section{A Model with Lower Mach Number} \label{app.Mach}

\begin{figure*}
	\centering
	\includegraphics[width=0.45\textwidth]{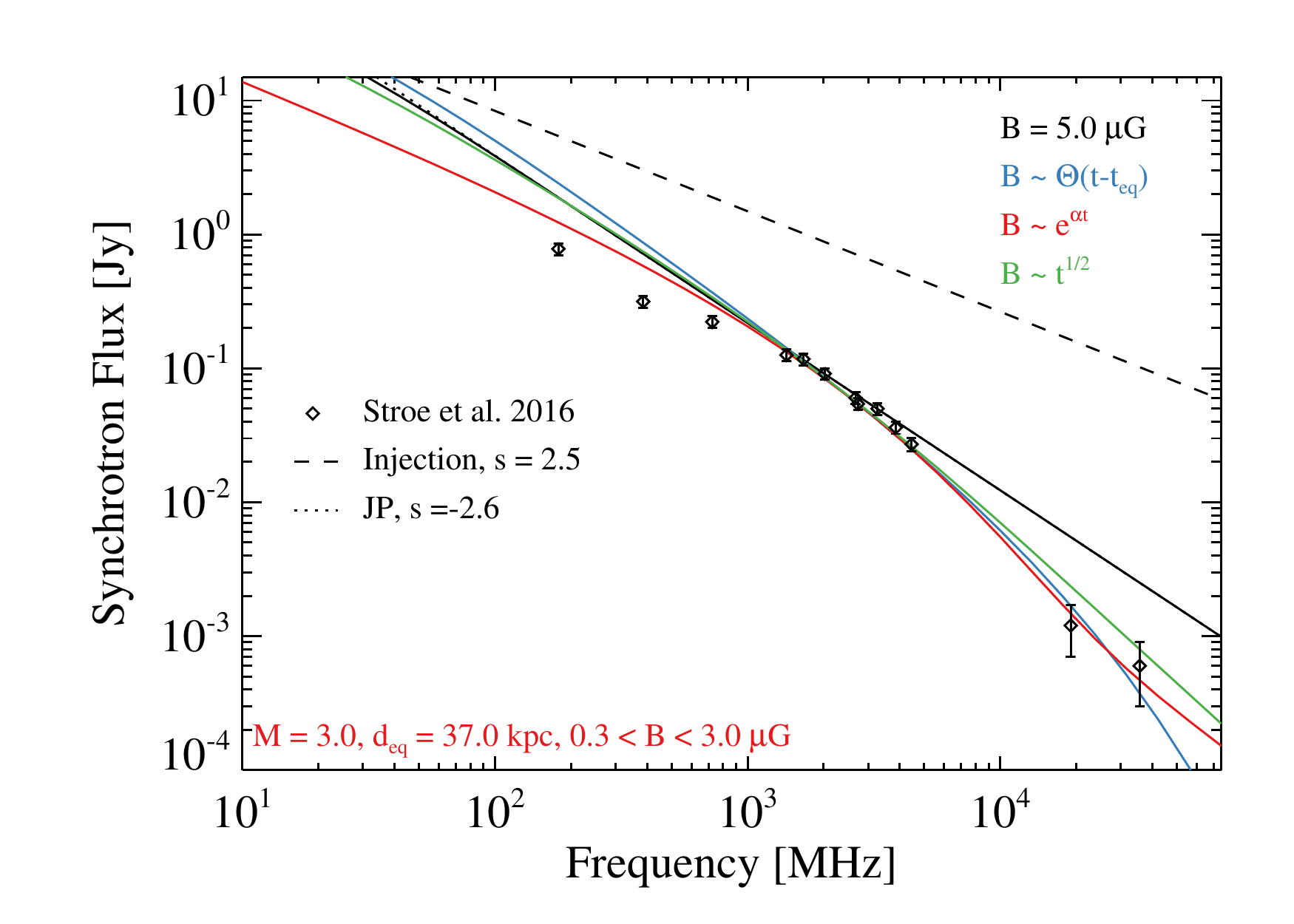}
	\includegraphics[width=0.45\textwidth]{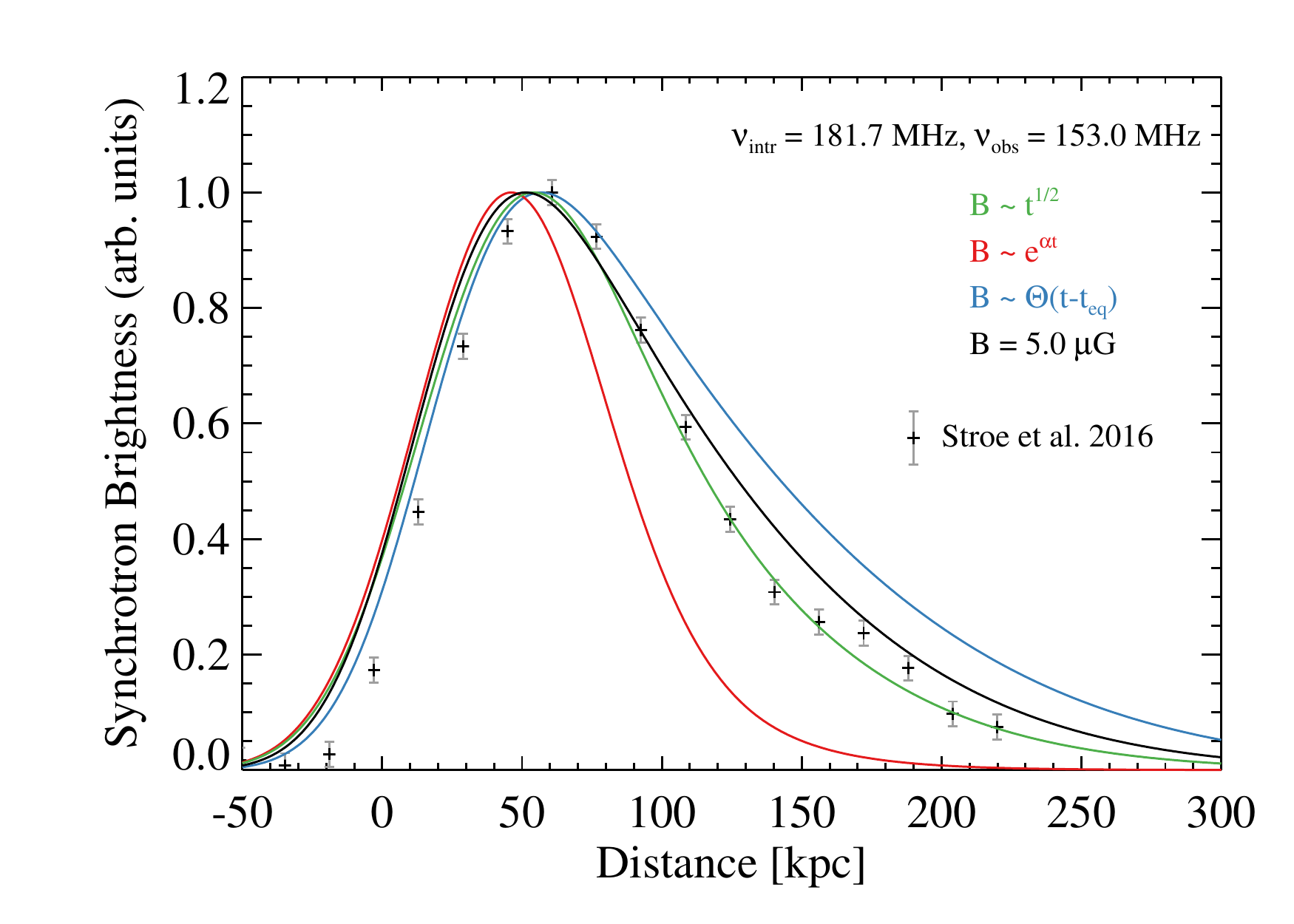} \\
	\includegraphics[width=0.45\textwidth]{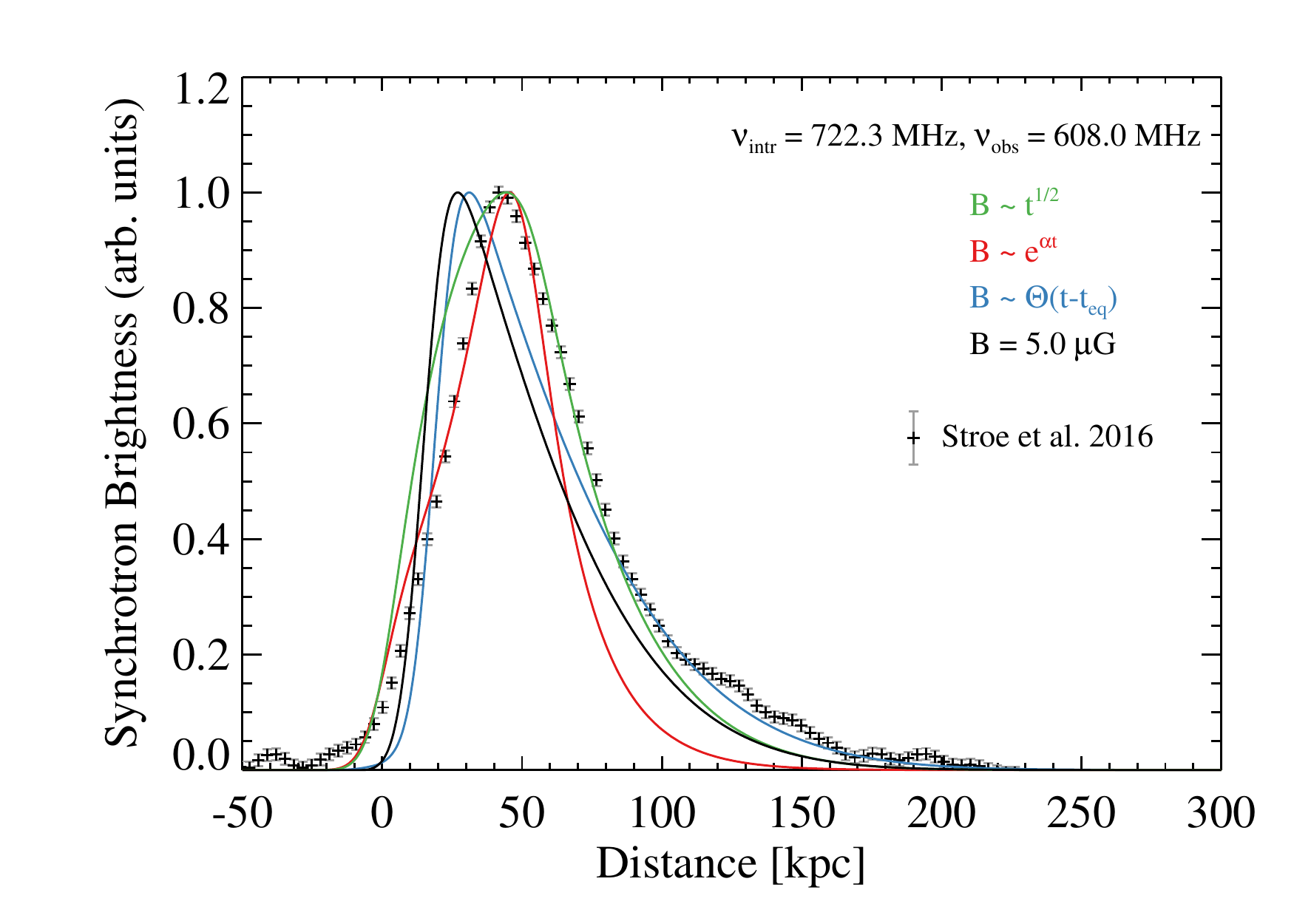}
	\includegraphics[width=0.45\textwidth]{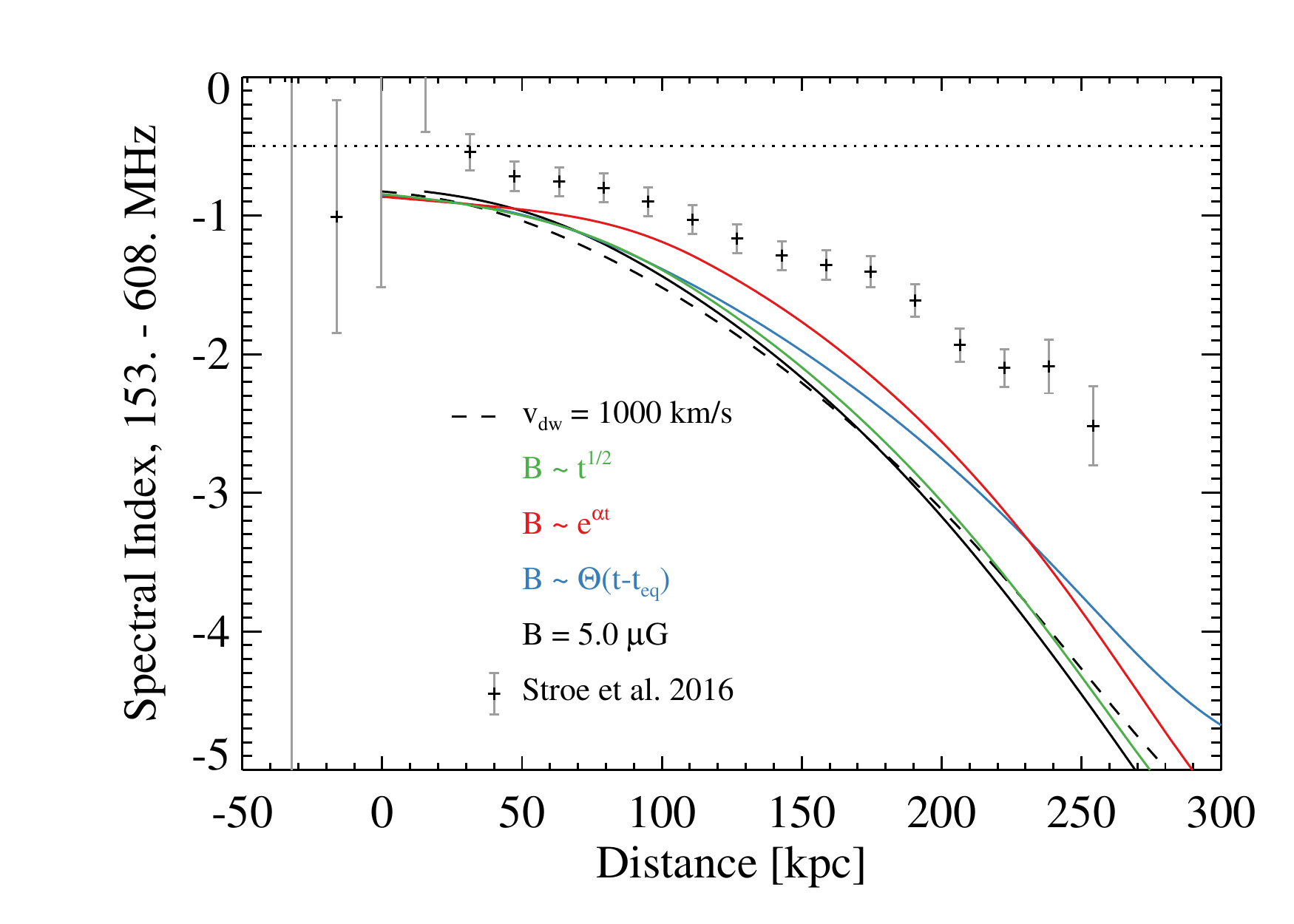}
	\caption{Top left to bottom right: Integrated radio spectrum, beam convolved brightness profiles at 150 MHz and 610 MHz, and spectral index profile between 150 MHz and 610 MHz, for a Mach number of three. All panels are analogous to the same figures show previously.} 
	\label{fig:app_M=3}
\end{figure*}

In section we show results from a model with Mach number of $M=3$. This is motivated by X-ray observations, which suggest a shock weaker than that adopted in our paper, specifically $M \leq 3$ \citep[e.g.][]{2015A&A...582A..87A}.\par
The normalisations ($n_0$) used in this model are  $5.4\times 10^{-4}$, $16\times 10^{-4}$, $160\times 10^{-4}$ and $18\times 10^{-4}$ in the same units\footnote{Note that the energy densities associated with these normalisations are larger than for the $M=4.6$ model, because of the change in spectral index.} as in table \ref{tab:cre_norm}, for the standard, step, exponential and linear energy model, respectively.   \par
In figure \ref{fig:app_M=3}, we show the integrated radio spectrum, the convolved brightness profile at 150 MHz and 610 MHz and the spectral index profile of the models using a Mach number of $M=3$, from top left to bottom right. We employ a magnetic field minimum of $0.3 \,\mu\mathrm{G}$ and a maximum of $3 \,\mu\mathrm{G}$. We require saturation scales of 15, 19, 37 and 52 kpc for the four models, respectively. The values for the field and the saturation scale are partially degenerate, as shown in appendix \ref{app.parameters}. Hence, a lower $B_\mathrm{min}$ does not lead to a better overall fit of the model. \par
The total integrated radio spectrum is fit well above 1 GHz by all but the standard model. However, all models are inconsistent with the observed radio spectrum at low frequencies. For the exponential model we used an e-folding distance of the adiabatic expansion of 80 kpc. We refrain from lowering this parameter even further to ensure $d_\mathrm{eq} \ll t_\mathrm{exp}v_\mathrm{dw}$ \par
The convolved brightness profile at 150 MHz and 610 MHz is compared with model predictions in the top right and bottom left panels. The exponential model declines too quickly due to the combined effect of adiabatic expansion and small downstream velocity. It also shows a significant shift in the brightness peak, not consistent with the observations.\par
Finally the comparison with the spectral index profile is very instructive. All models show a steepening of the spectrum with distance that is stronger than the observed one. The steepening with distance constrains the velocity of the flow downstream, which is only 900 km/s  for a Mach number of $M = 3$. As we explained in sect. \ref{sect.spix} a minimum downstream velocity of $v_\mathrm{dw}= 1200 \,\mathrm{km}/\mathrm{s}$ is required to explain observations. \par
We conclude that a model with a Mach number of three is disfavoured by radio observations, unless additional mechanisms that maintain CR electrons at higher energies for longer times (e.g. re-acceleration) or that enhance the spatial transport of high energy electrons (e.g. diffusion) are considered.

\section{Dependence on Model Parameters}\label{app.parameters}

In this section, we explore the parameter space of the exponential model, to show the importance of the individual input parameters. These are :
\begin{itemize}
	\item the minimum magnetic field value: $B_\mathrm{min}$,
	\item the maximum magnetic field value: $B_\mathrm{max}$,
	\item the saturation distance: $d_\mathrm{eq}$,
	\item the shock speed: $v_\mathrm{dw}$.
\end{itemize}

We conclude that the e-fold distance is degenerate with the downwind speed. Hence, given a shock speed, the e-fold distance can always be found so the model brightness profiles fit the observed profiles. We keep the Mach number fixed at 4.6, a model with $M=3$ was considered above. \par
In what follows, we present the total integrated synchrotron spectrum and the two brightness profiles at 150 MHz and 610 MHz varying one parameter, while leaving the other three fixed. The exponential model with standard values is always shown in black, the lower  value in green and the upper value in blue.

\subsection{Minimum Magnetic Field Value}

\begin{figure*}
	\centering
	\includegraphics[width=0.33\textwidth]{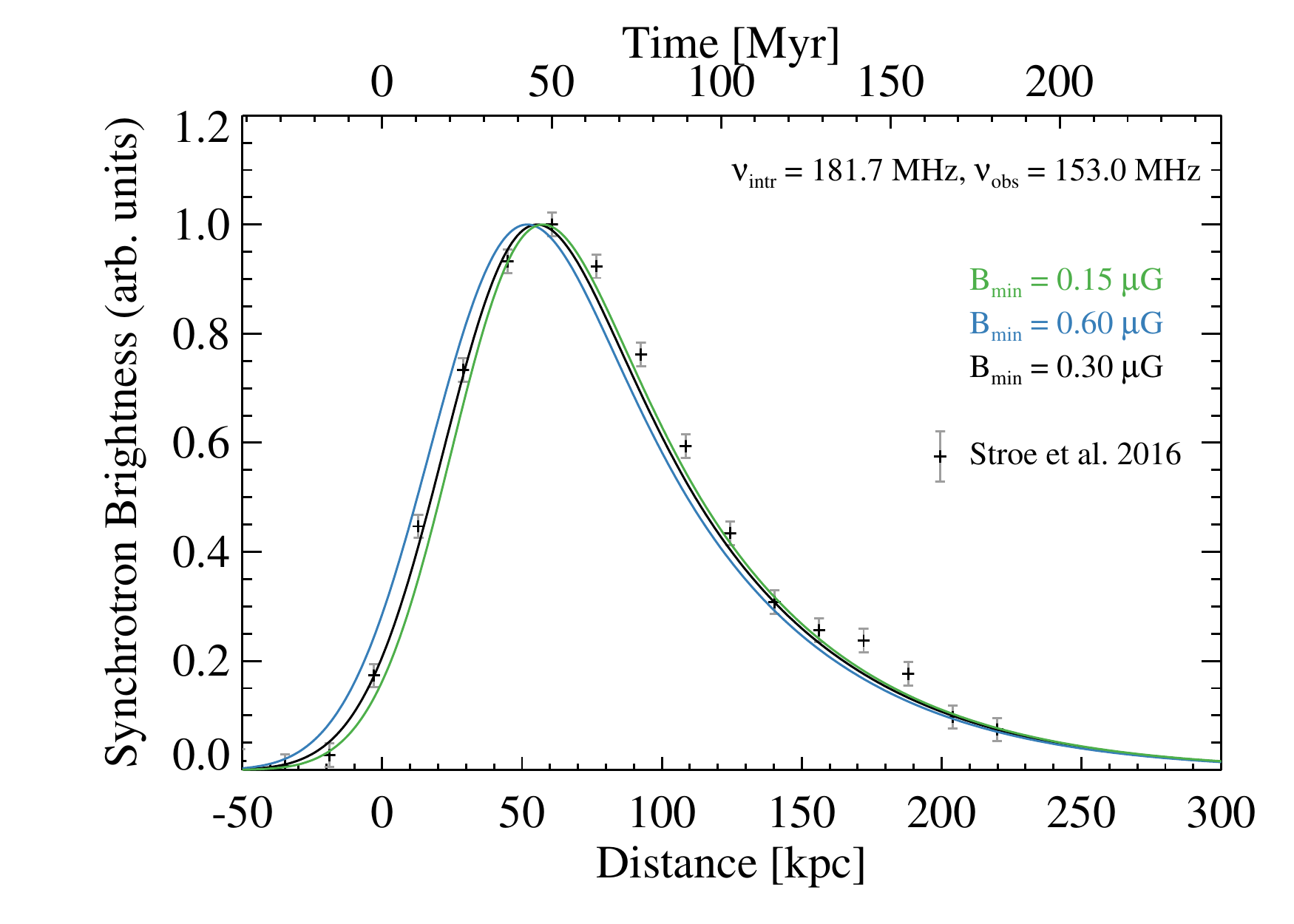}
	\includegraphics[width=0.33\textwidth]{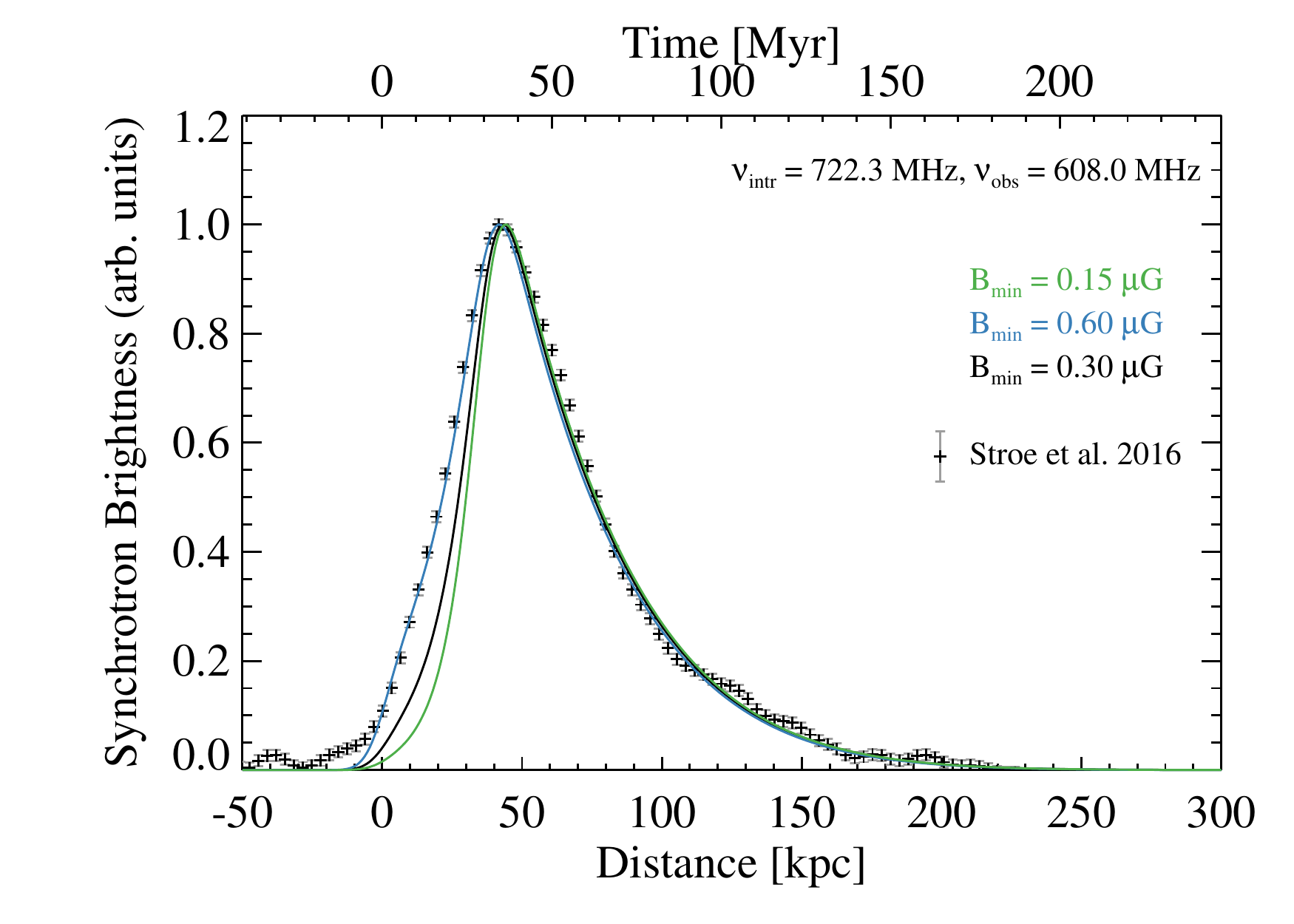} 
	\includegraphics[width=0.33\textwidth]{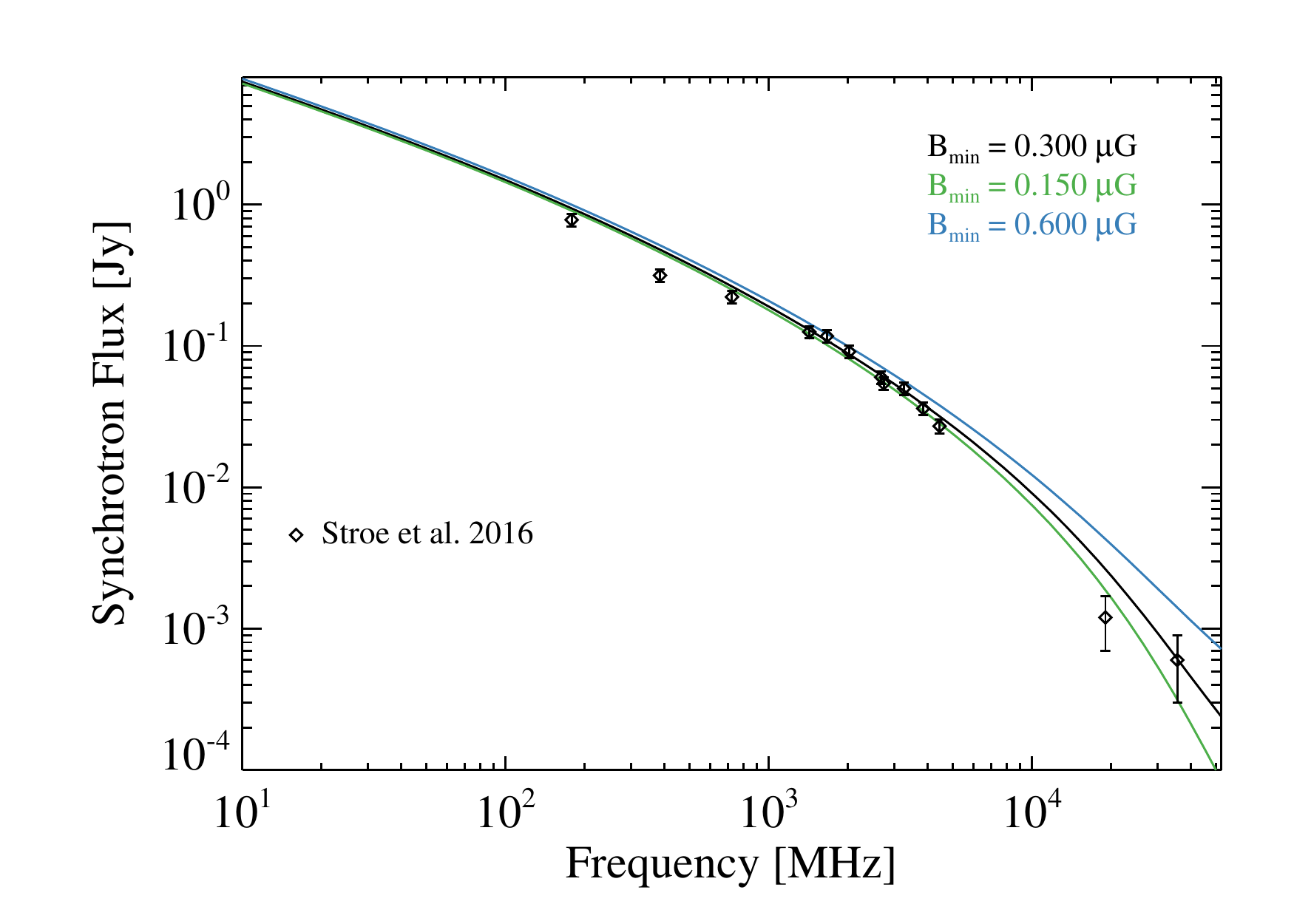}
	\caption{Beam convolved brightness profiles at 150 MHz and 610 MHz and integrated synchrotron spectrum (top left to bottom right) for the exponential model with minimum magnetic field $B_\mathrm{min}$ equal to $0.3\,\mu\mathrm{G}$ (standard model, black), $0.15\,\mu\mathrm{G}$ (green) and $0.6\,\mu\mathrm{G}$ (blue).} 
	\label{fig:app_var_bmin}
\end{figure*}

We first consider the minimum magnetic field strength in the exponential model. For the three values of $B_\mathrm{min}$, 0.15, 0.3 and 0.6 $\mu\mathrm{G}$ we show brightness profiles at two frequencies and the total synchrotron spectrum in figure \ref{fig:app_var_bmin}. This parameter has two effects on the observables: An increase in $B_\mathrm{min}$ increases the brightness at the rising flank of the profile, most significantly at low frequencies. It also decreases the steepening of the spectrum at high radio frequencies. Given the magnetic field values expected at the location of the relic, smaller values than $0.1 \,\mu\mathrm{G}$ are  difficult to justify behind the shock. 

\subsection{Maximum Magnetic Field Value}

\begin{figure*}
	\centering
	\includegraphics[width=0.33\textwidth]{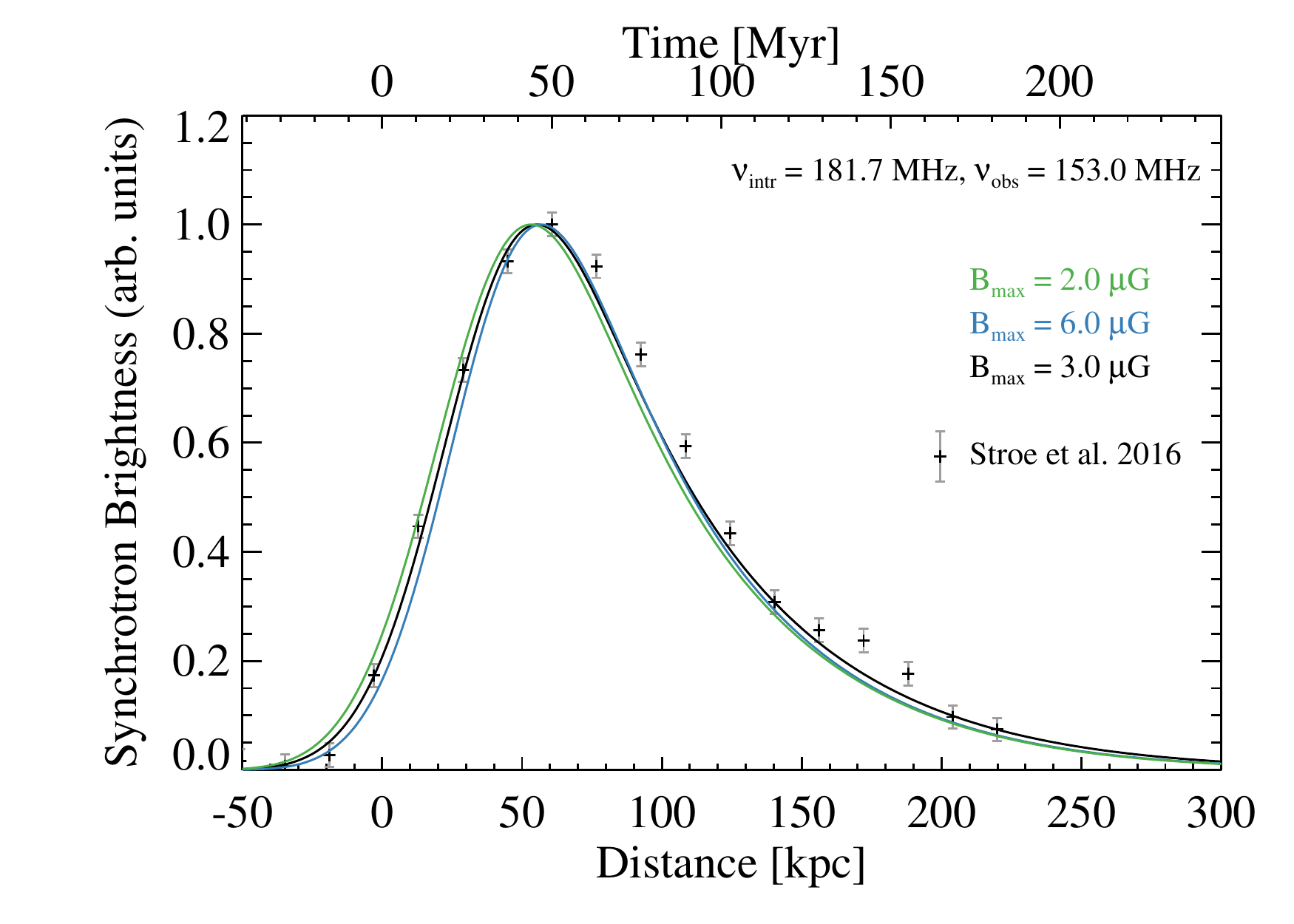} 
	\includegraphics[width=0.33\textwidth]{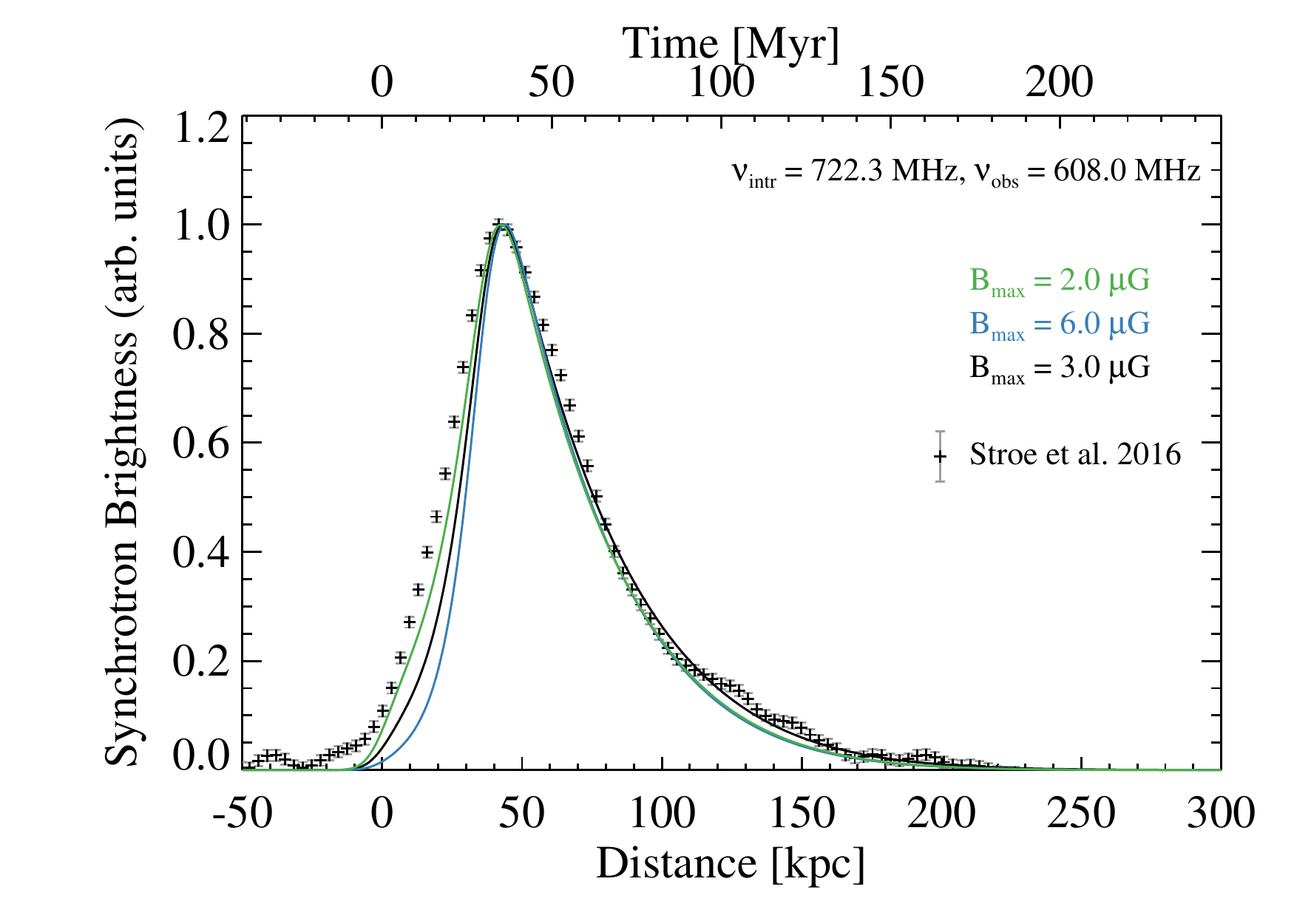} 
	\includegraphics[width=0.33\textwidth]{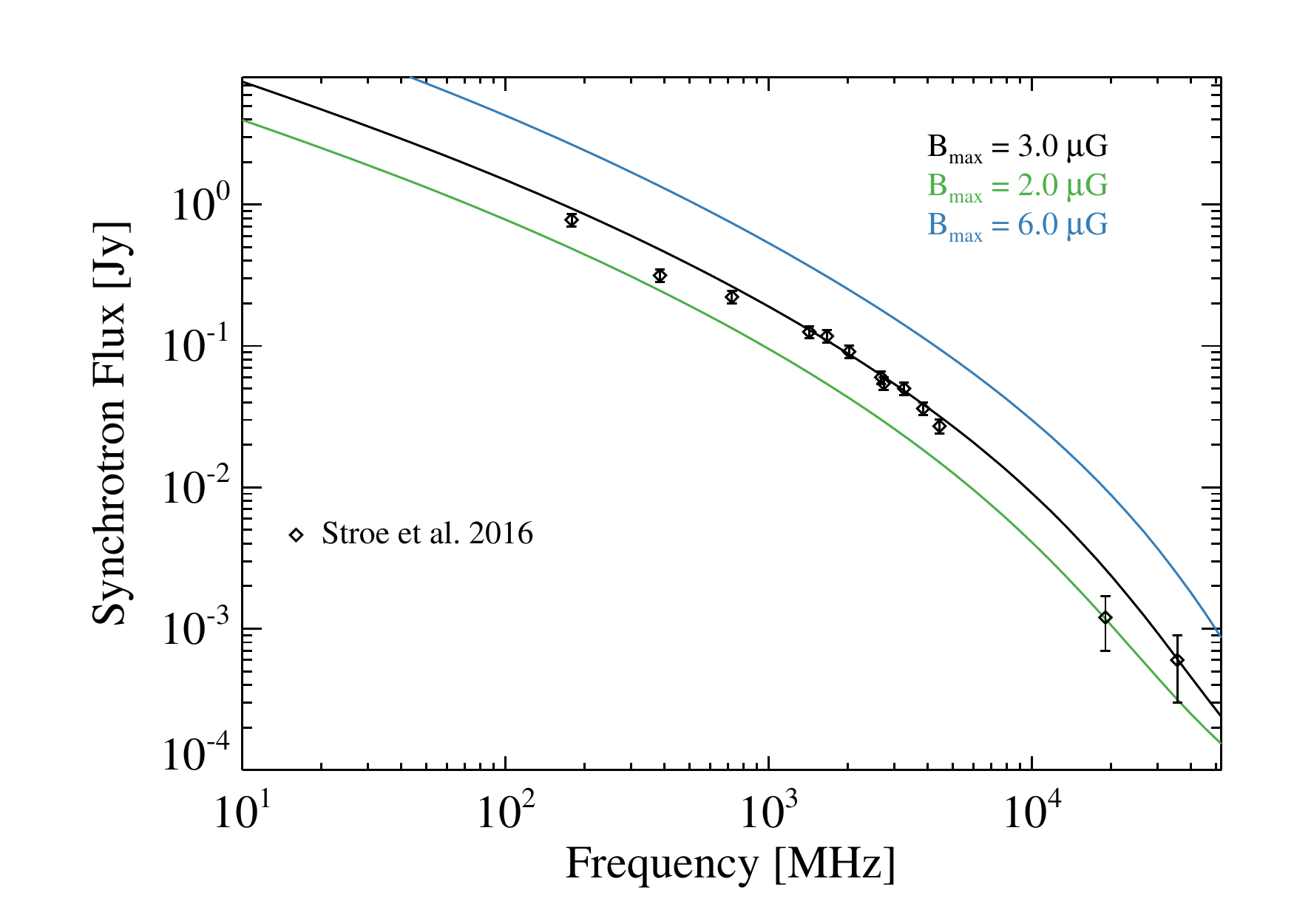}
	\caption{Beam convolved brightness profiles at 150 MHz and 610 MHz and integrated synchrotron spectrum (top left to bottom right) for the exponential model with minimum magnetic field $B_\mathrm{max}$ equal to $3\,\mu\mathrm{G}$ (standard model, black), $2\,\mu\mathrm{G}$ (green) and $6\,\mu\mathrm{G}$ (blue).}
	\label{fig:app_var_bmax}
\end{figure*}

Next we vary the maximum magnetic field strength $B_\mathrm{max}$, which is $3 \,\mu\mathrm{G}$ in the standard model (black), to $2\,\mu\mathrm{G}$ (green) and $6\,\mu\mathrm{G}$ (blue). The lower limit is here set by the inverse Compton limit, the higher value is consistent with the standard model from \citet{2010Sci...330..347V}. \par
The resulting brightness profiles and radio synchrotron spectrum are shown in figure \ref{fig:app_var_bmax}. The major impact of the parameter is visible in the normalisation of the radio synchrotron spectrum, where the low and high models show a shift in the model spectrum. This would reduce (increase) the CRe normalisation and required acceleration efficiencies.The changes to the brightness profiles are minor and roughly within the error bars of the observations. We conclude that the maximum magnetic field strength mostly affects the normalisation of the CRe spectrum and has only minor influence on the brightness profile.

\subsection{Saturation Distance}

\begin{figure*}
	\centering
	\includegraphics[width=0.33\textwidth]{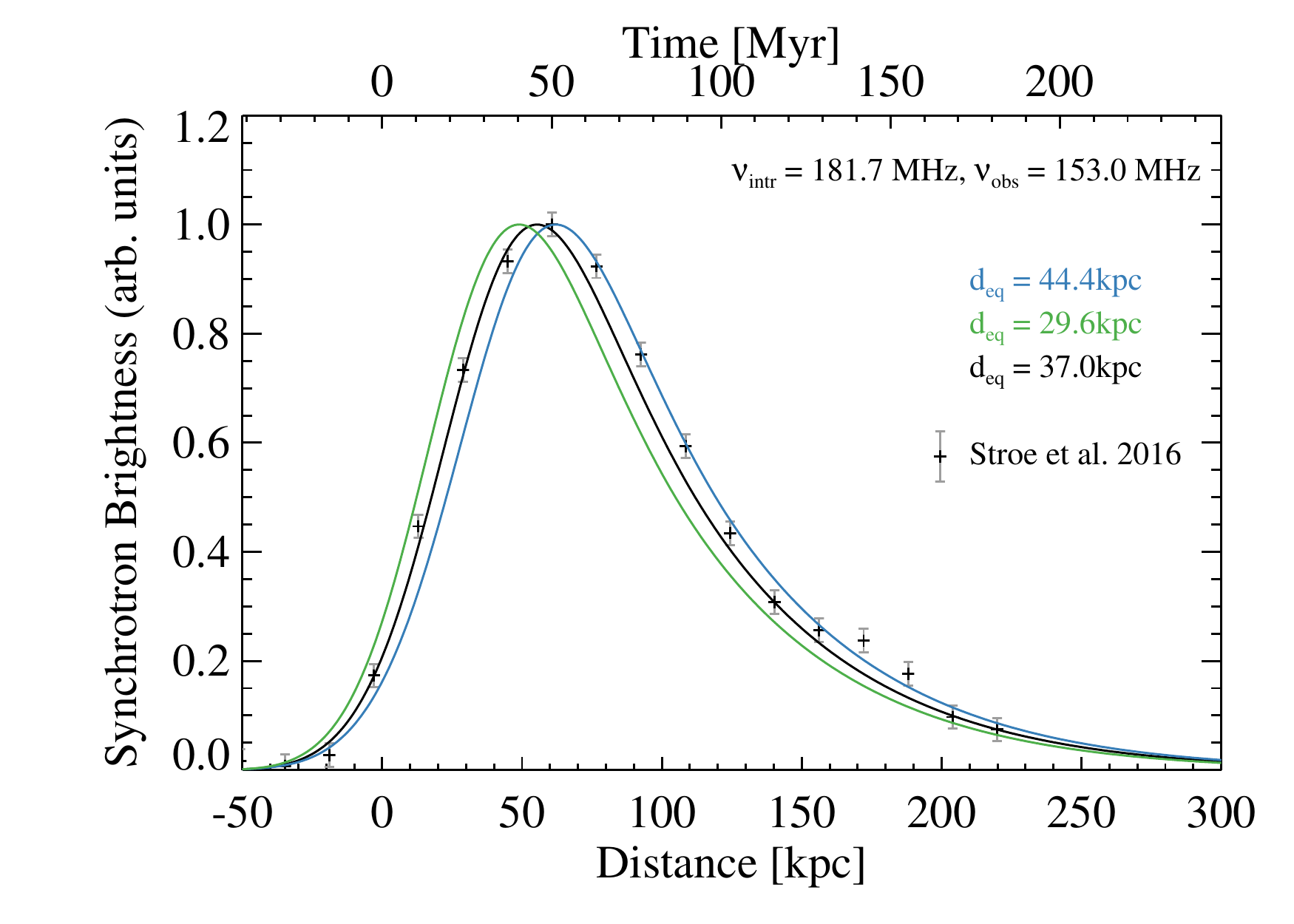} 
	\includegraphics[width=0.33\textwidth]{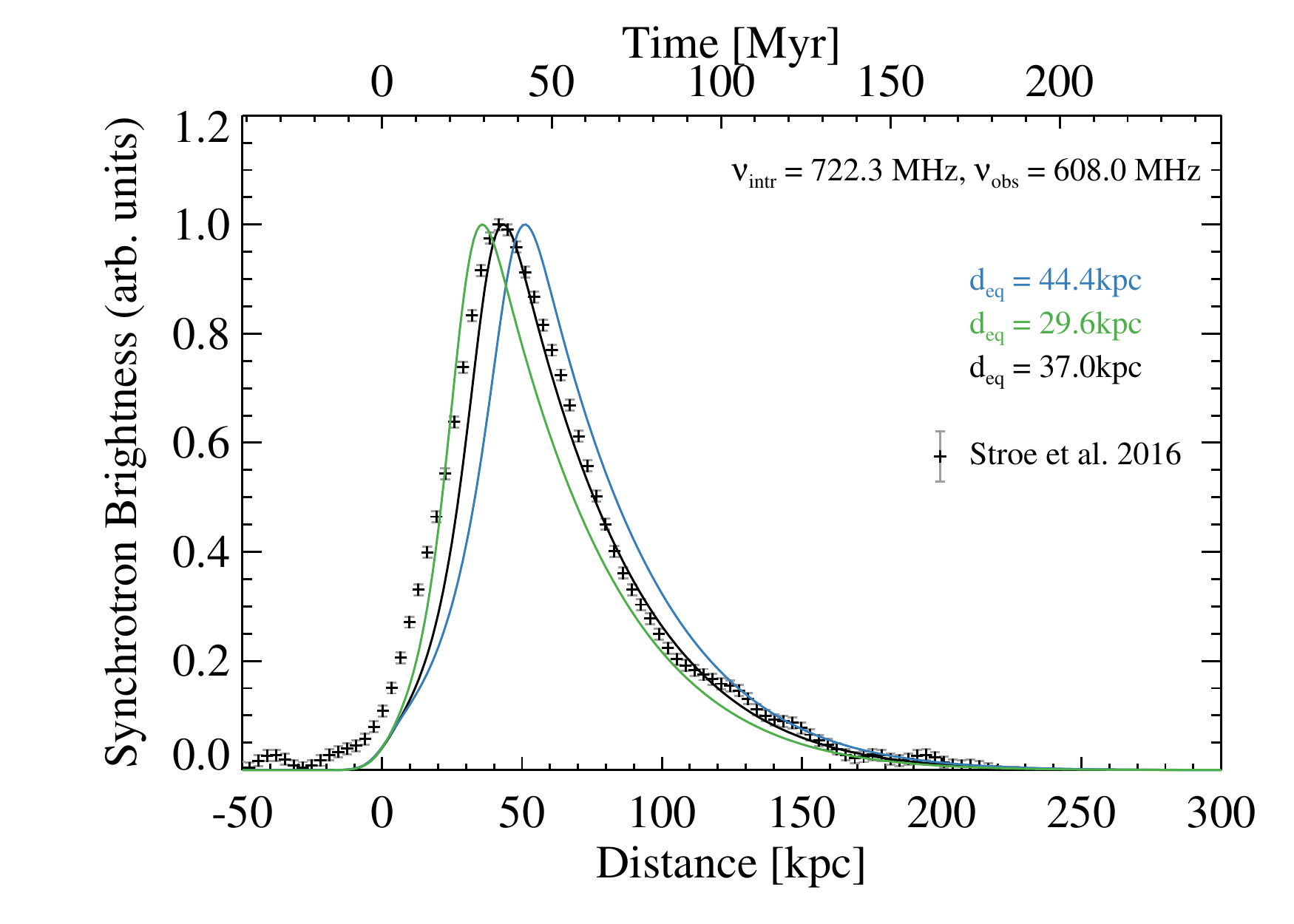} 
	\includegraphics[width=0.33\textwidth]{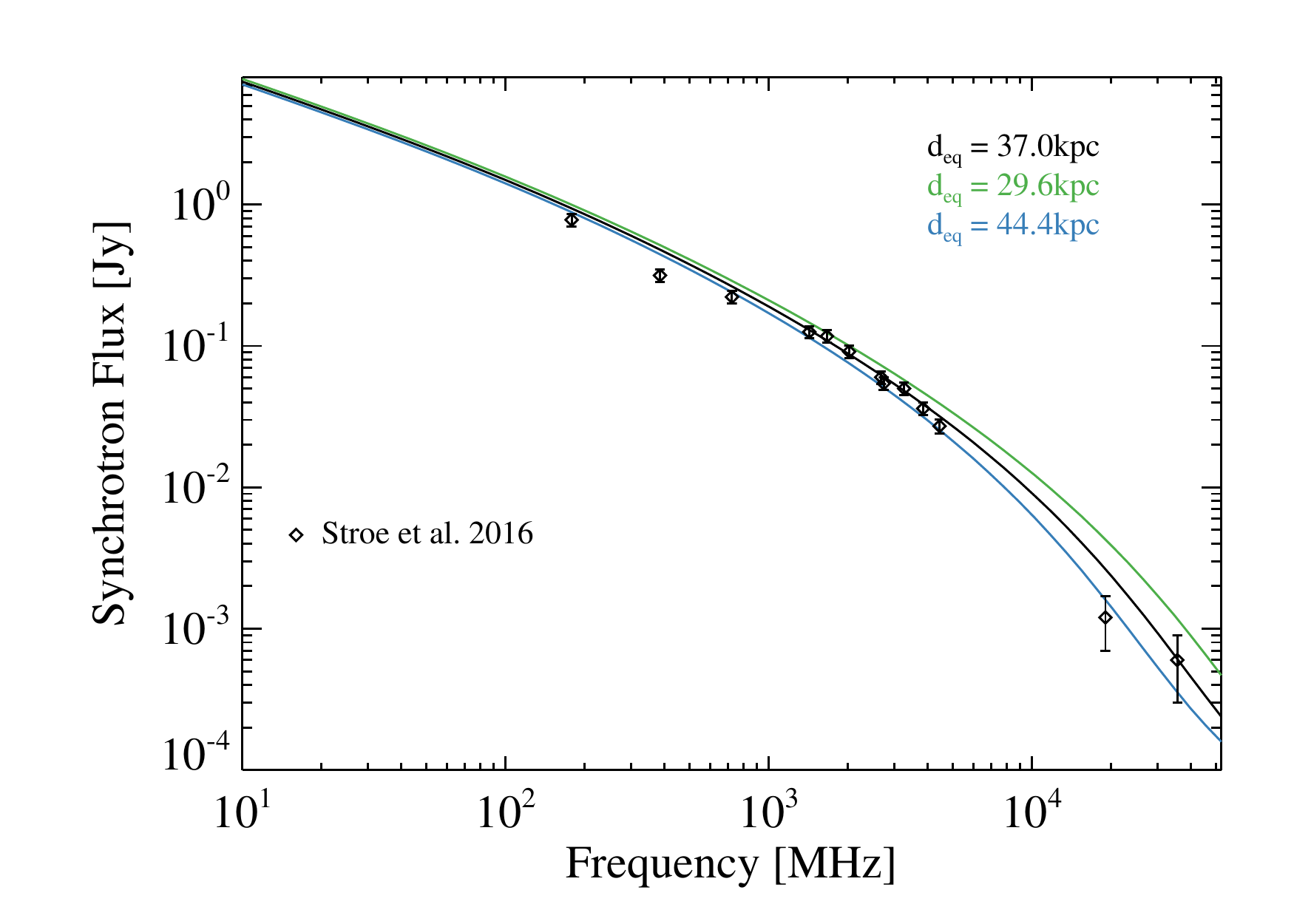}
	\caption{Beam convolved brightness profiles at 150 MHz and 610 MHz and integrated synchrotron spectrum (top left to bottom right) for the exponential model with saturation distance $d_\mathrm{eq}$ equal to $37\,\mathrm{kpc}$ (standard model, black), $29\,\mathrm{kpc}$ (green) and $44\,\mathrm{kpc}$ (blue).}
	\label{fig:app_var_deq}
\end{figure*}

In figure \ref{fig:app_var_deq} we show brightness profiles and synchrotron spectrum, varying the saturation distance $d_\mathrm{eq}$ by $20\%$, from 37 kpc, to 29 kpc (green) and 44 kpc (blue). The change leads to a shift in the peak of the brightness profiles by roughly 10 kpc and an decrease (increase) in the steepening of radio spectrum and high frequencies for smaller (larger) saturation distances. We conclude that this parameter has a rather large impact on the model. In the absence of any projection effects it could be constrained very well by the observed high resolution brightness profiles.

\subsection{Downwind Speed}

\begin{figure*}
	\centering
	\includegraphics[width=0.33\textwidth]{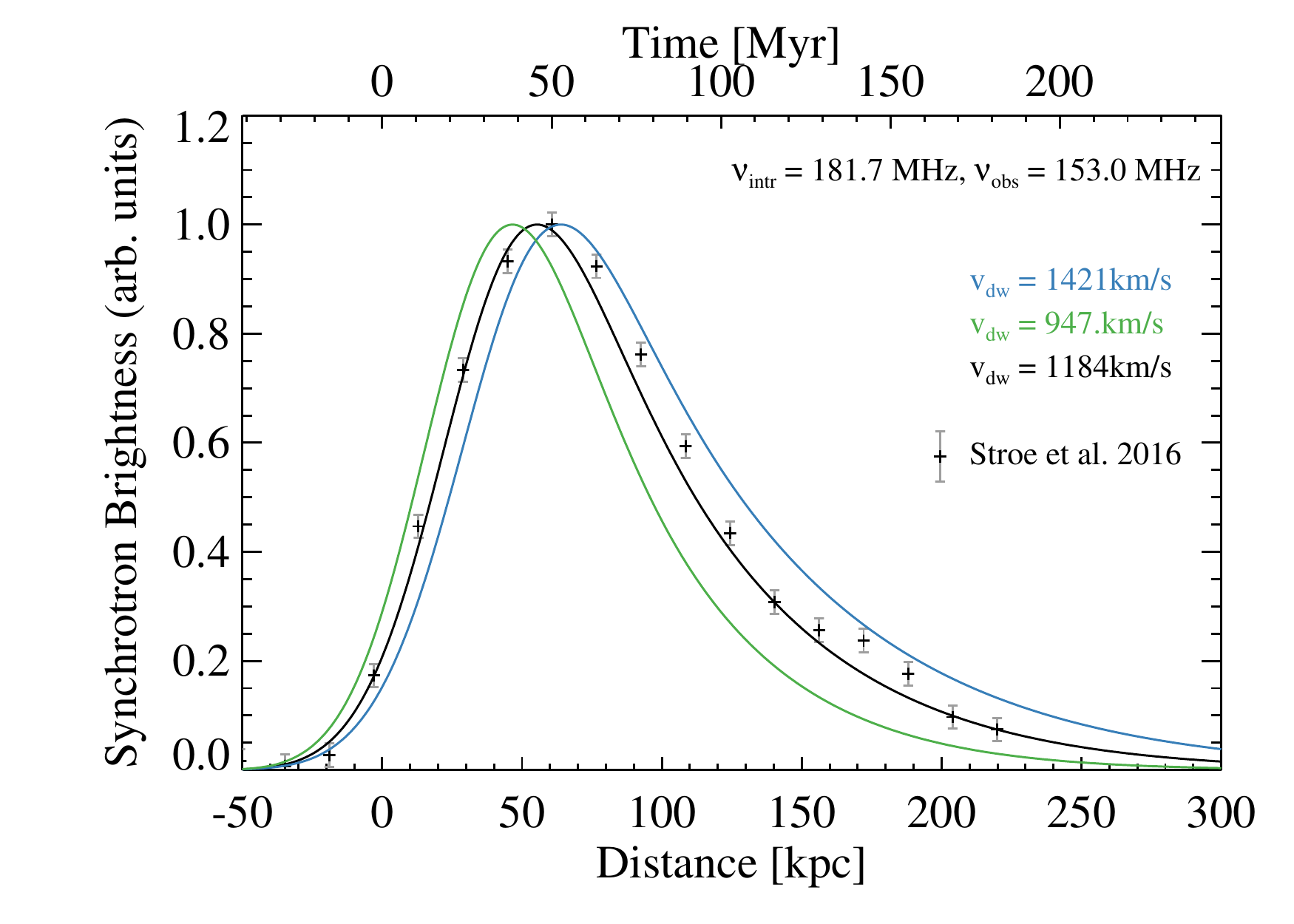} 
	\includegraphics[width=0.33\textwidth]{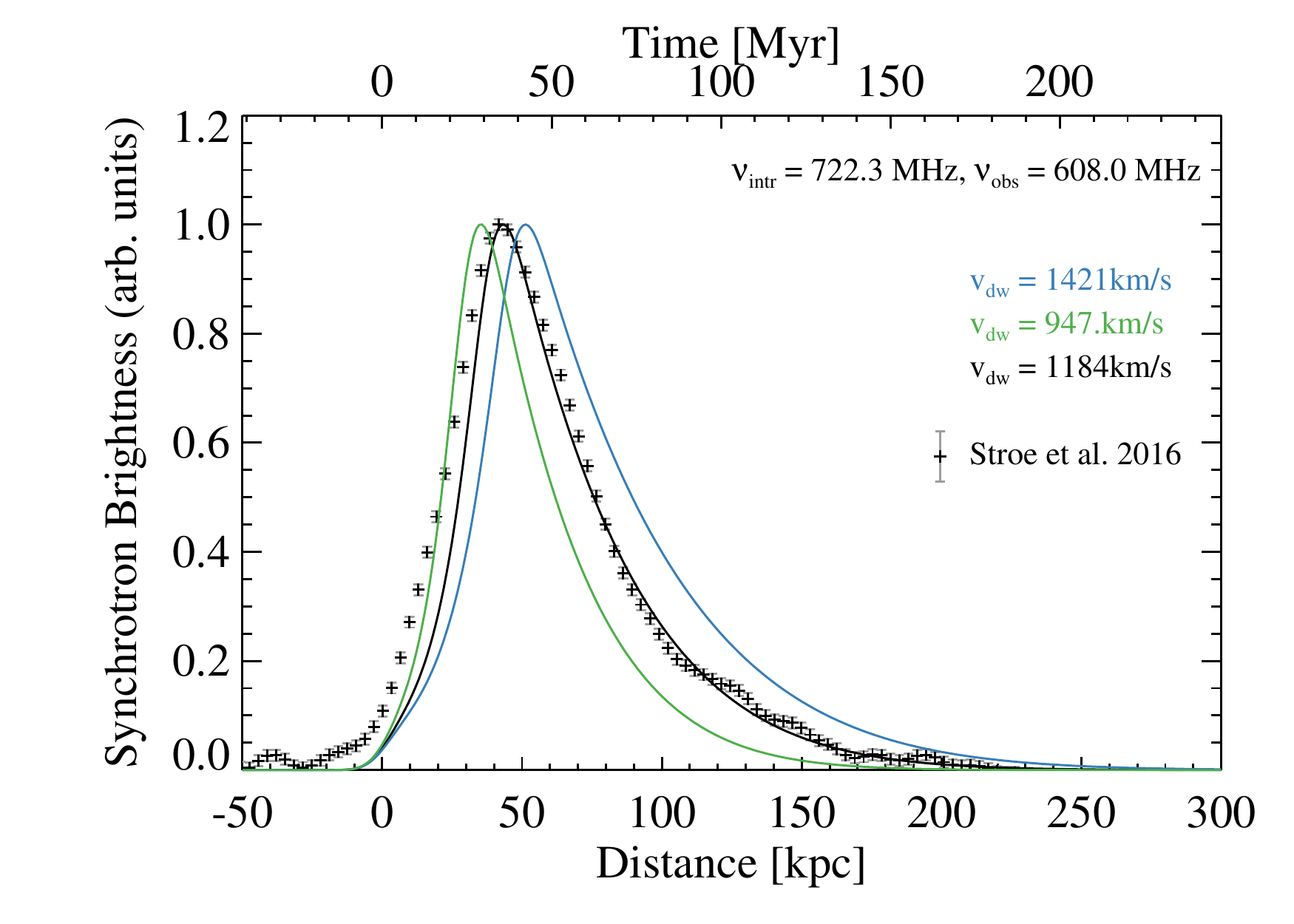}
	\includegraphics[width=0.33\textwidth]{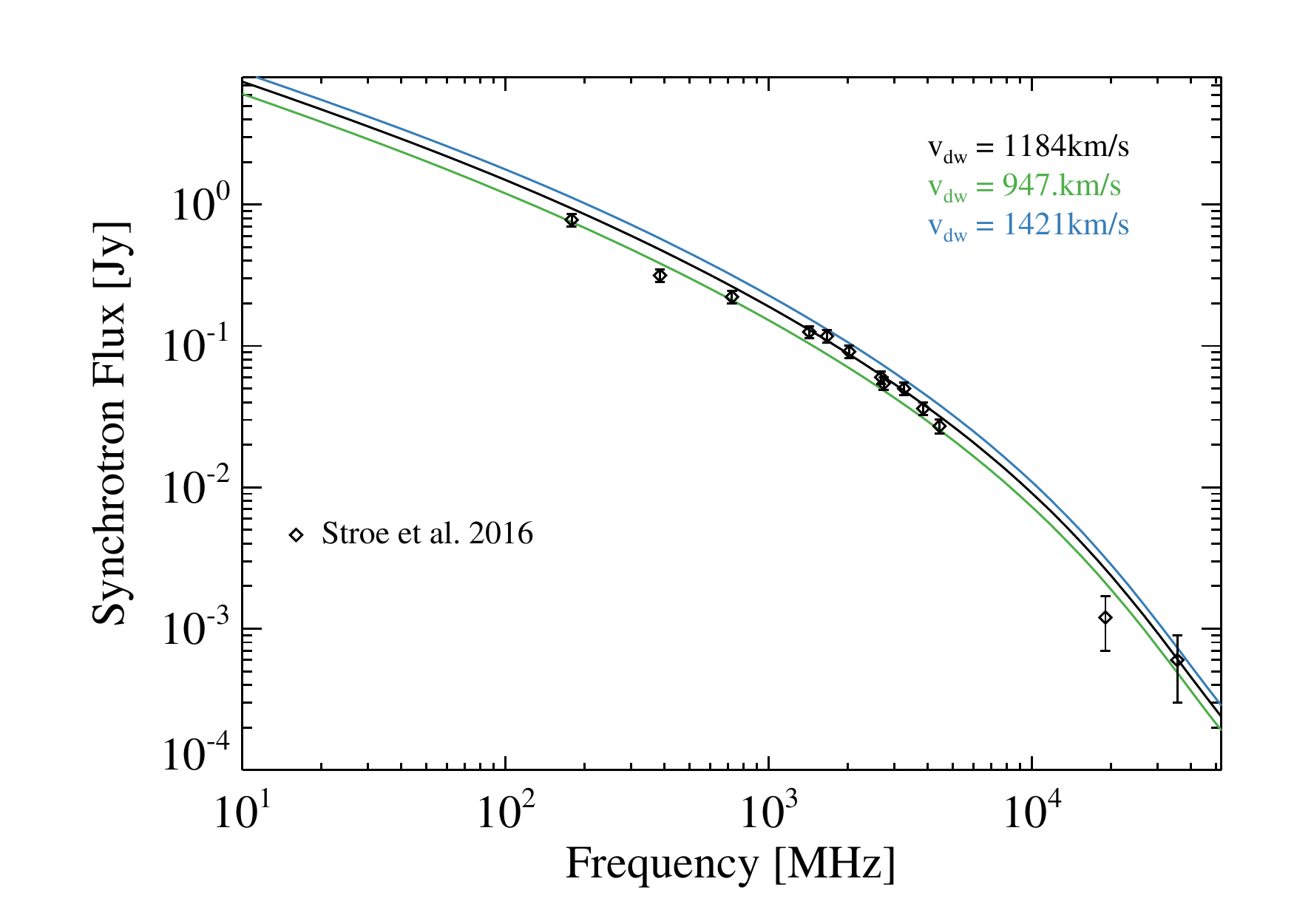}
	\caption{Beam convolved brightness profiles at 150 MHz and 610 MHz and integrated synchrotron spectrum (top left to bottom right) for the exponential model with downwind velocity $v_\mathrm{dw}$ equal to $1000\,\mathrm{km/s}$ (standard model, black), $800\,\mathrm{km/s}$ (green) and $1200\,\mathrm{km/s}$ (blue).}
	\label{fig:app_var_vs}
\end{figure*}

We consider the downstream shock speed $v_\mathrm{dw}$ as the final parameter, again changing it by $20\%$ to 1421 km/s (blue) or 950 km/s (green). The change leads to a shift in the peak position of the brightness profiles to larger (green) or smaller (blue) distances. The change in cooling velocity also leads to an increase (decrease) for smaller (larger) of the spectral steepening, respectively. \par

We conclude that shock speed and saturation distance have the largest impact on the model. The saturation distance can in principle be well constrained by very high resolution observation of the relic at intermediate frequencies. The down stream shock speed is highly uncertain as it depends on the Mach number, the expansion timescale and the upstream sound speed, which in turn depends on the upstream temperatures.  \par

\label{lastpage}

\end{document}